\documentclass[
twocolumn,aps,prb,showpacs,amsmath,superscriptaddress,longbibliography,notitlepage
]{revtex4-2}

\usepackage{amssymb,mathtools}
\usepackage[english]{babel}
\usepackage{mathrsfs}
\usepackage{graphicx}
\usepackage{float}
\usepackage{pbox}
\usepackage[caption=false]{subfig}
\usepackage{dcolumn}
\usepackage{appendix}
\usepackage{tabularx}
\usepackage{txfonts}
\usepackage[normalem]{ulem}
\usepackage{verbatim}
\usepackage{xcolor}
\usepackage{bm}
\usepackage{natbib}
\usepackage{multirow}
\usepackage{soul}

\graphicspath{{fig}}

\begin{document}

\title{Exceptional Points and Braiding Topology in Non-Hermitian Systems with long-range coupling}

\author{S. M. Rafi-Ul-Islam}
\email{rafiul.islam@u.nus.edu}
\affiliation{Department of Electrical and Computer Engineering, National University of Singapore, Singapore 117583, Republic of Singapore}

\author{Zhuo Bin Siu}
\email{elesiuz@nus.edu.sg}
\affiliation{Department of Electrical and Computer Engineering, National University of Singapore, Singapore 117583, Republic of Singapore}

\author{Md. Saddam Hossain Razo}
\email{shrazo@u.nus.edu}
\affiliation{Department of Electrical and Computer Engineering, National University of Singapore, Singapore 117583, Republic of Singapore}

\author{Haydar Sahin}
\email{sahinhaydar@u.nus.edu}
\affiliation{Department of Electrical and Computer Engineering, National University of Singapore, Singapore 117583, Republic of Singapore}
\affiliation{Institute of High Performance Computing, A*STAR, Singapore 138632, Republic of Singapore}

\author{Mansoor B.A. Jalil}
\email{elembaj@nus.edu.sg}
\affiliation{Department of Electrical and Computer Engineering, National University of Singapore, Singapore 117583, Republic of Singapore}

\begin{abstract}
We present a study of complex energy braiding in a 1D non-Hermitian system with $n$th order long range asymmetrical coupling. Our work highlights the emergence of novel topological phenomena in such systems beyond the conventional nearest-neighbor interaction. The modified SSH model displays $n$ distinct knots and links combinations in the complex energy-momentum space under periodic boundary conditions (PBC), which can be controlled by varying the coupling strengths. A topological invariant, namely the braiding index, is introduced to characterize the different complex energy braiding profiles, which depends on the zeros and poles of the characteristic polynomials. Furthermore, we demonstrate that the non-Hermitian skin effect can be localized at one or both ends, signifying conventional or bipolar localization, depending on the sign of the braiding index. Phase transitions between different braiding phases with the same (opposite) sign of the topological invariant occur at Type-1 (Type-2) exceptional points, with Type-1 (Type-2) phase transitions accompanied by single (multiple) exceptional points. We propose an experimental set-up to realize the various braiding schemes based on the RLC circuit framework, which provides an accessible avenue for implementation without recourse to high-dimensional momentum space required in most other platforms.
\end{abstract}

\maketitle

\section{Introduction} 
The study of non-Hermitian systems has a long history, dating back to the early days of quantum mechanics \cite{ashida2020non,bergholtz2021exceptional,gong2018topological,leykam2017edge,rafi2022critical,yokomizo2019non}. However, it has gained renewed interest in recent years, and has been analyzed in various contexts, ranging from optics \cite{el2019dawn,cao2020reservoir,zhang2018non}, photonics \cite{feng2017non,pan2018photonic,longhi2015robust,midya2018non}, metamaterials \cite{ghatak2020observation,hou2020topological,zhou2020non,rosenthal2018topological}, quantum field theory \cite{fring2020pseudo,moiseyev2011non,fring2020goldstone}, topolectrical circuits \cite{rafi2021non,hofmann2020reciprocal,zhang2023anomalous,sahin2023impedance,rafi2020topoelectrical,rafi2020realization,rafi2020anti,rafi2023valley,sahin2023impedance,zhang2023anomalous,rafi2021topological,zou2021observation,helbig2020generalized,rafi2022system,lee2018topolectrical,rafi2022interfacial,rafi2022type} to condensed matter physics \cite{martinez2018topological,ghatak2019new,rafi2020strain,lee2014entanglement,rafi2022critical,lee2019anatomy,li2022non,rafi2022unconventional}. Non-Hermitian systems are usually characterized by non-Bloch theory \cite{yokomizo2019non,kawabata2020non}, and exhibit novel and exotic physical phenomena that are not present in Hermitian systems.  One of the key features of non-Hermitian systems is the presence of complex eigen energies \cite{zhang2020correspondence,liu2019second,hamazaki2019non}, which can have a profound impact on the system's behavior. In particular, the concept of energy braiding \cite{hu2021knots,wang2021topological,lee2020imaging,li2021homotopical,ezawa2019braiding,hu2022knot,zhang2023observation,li2022topological}, i.e., the criss-crossing of two eigen energy braids in complex energy space, has provided key insights into the understanding of non-Hermitian properties, such as (i) the non-Hermitian skin effect \cite{okuma2020topological,siu2023terminal,rafi2024saturation,li2020critical,rafi2024twisted,longhi2019probing,song2019non,kawabata2020higher,zhang2021observation,zhang2022universal},  which localizes the bulk states to the edges of the system under open boundary conditions, and (ii) exceptional points \cite{rafi2021non,kawabata2019classification,minganti2019quantum,hu2017exceptional,parto2021non}, which are special points in the complex energy space where two eigenvalues and their corresponding eigenvectors coalesce. 

Complex energy braiding opens up a new avenue for exploring novel topological phases and their corresponding phase transitions. This is owing to the complex-valued eigenvalues of non-Hermitian systems confers extra degree of freedom in their energy-momentum spectra \cite{rafi2022valley,rafi2023conductance,rafi2024chiral,sun2019field,sun2020spin}. Energy braiding in non-Hermitian systems has been realized in a variety of platforms, both theoretically and experimentally. On the theoretical side, the concept of complex energy braiding has been studied using various models such as the Kitaev chain \cite{malciu2018braiding,homeier2021z,sekania2017braiding,ezawa2020non,backens2017emulating}, and the topological superconductor \cite{pahomi2020braiding,levin2012braiding}. Experimentally, energy braiding has been realized in a variety of platforms such as photonic systems \cite{wang2021topological,noh2020braiding,iadecola2016non}, metamaterial systems \cite{barlas2020topological}, and electronic circuits \cite{ezawa2020non,nayak2008non}. In photonic systems, complex energy braiding has been observed in wave guide arrays and resonator arrays with asymmetrical couplings \cite{wang2021topological,wang2021generating}. A single-band system with long-range coupling was experimentally realized in an optical resonator system \cite{wang2021generating}. In mechanical systems, the braiding of energy eigenvalues has been demonstrated in the motion of coupled pendulums \cite{patil2022measuring}, while in electronic circuits, it has been realized using  LC (Inductor-Capacitor) circuits. 
Finally,  one-band \cite{zhang2021acoustic} and two-band systems \cite{zhang2023observation} with long-range coupling have been experimentally realized in acoustic systems.

Despite the above plethora of studies on complex admittance energy braiding in non-Hermitian systems, several key questions remain unanswered:

\begin{enumerate}
  \item Is it possible to extend the braiding concepts to multiple knots and links in a generalized 1D system?
  \item Is there a correlation between the localization of the  non-Hermitian skin effect and the handedness of the energy braiding? 
  \item Can generalized braiding circuits provide a general approach to trace the locus of the exceptional points which is not tied to any  specific model?
  \item Can an extended topological braiding index be defined to describe non-Hermitian systems with arbitrary long-distance coupling?
\end{enumerate}

In this letter, we aim to address the above-mentioned questions by reference to the complex energy braiding of a model 1D non-Hermitian system which incorporates  long-range unidirectional coupling. Our model which is based on the bipartite modified non-Hermitian SSH model \cite{zhang2019partial},  allows for an arbitrary number of braids or knots to be engineered in the energy spectrum. Depending on the degree of long-range coupling in the model, the eigen energy strands of in the complex energy spectra  can tangle with one another and form an arbitrary number of loops. In general, an $n$th-order asymmetrical inter-chain coupling allows for $n$ different configurations of complex energy braiding. The braid configurations are characterized by a new topological invariant, the knot/braiding index, which depends on the number of zeros and poles of the characteristic polynomial of the system's Hamiltonian. Interestingly, the sign of the braiding index would determine the localization of the non-Hermitian skin effect (NHSE) at either or both ends of the 1D chain. This is unlike the conventional NHSE which is governed by the relative coupling strength within the chain, and is restricted to unipolar (localized at one end) as opposed to bipolar localization. Furthermore, distinct topological phases with different numbers of knots can be transformed from one to the other via exceptional points, which can be classified as Type-1 or Type-2 depending on the sign of the braiding index across the phase boundary. Finally, we propose an experimental framework for realizing these braiding schemes in an $RLC$ circuit, which offers a simple accessible route to the realization of braiding in 1D non-Hermitian systems. 

\section{Model with arbitrary number of knots}
To realize various complex energy-momenta knots via 1D non-reciprocal lattice, we consider a bipartite  model with tunable long-range coupling with two types of sublattice sites (A and B) per unit cell.  The model Hamiltonian is given by  

\begin{equation}
	H_1(k) = \begin{pmatrix} 0 &  C_{\mathrm{AB}, 0} + C_{\mathrm{AB}, -m} e^{-i k m}  \\ 
	C_{\mathrm{AB}, 0} + C_{\mathrm{BA}, n} e^{i k n} & 0 \end{pmatrix}.  \label{H1}
\end{equation}

\begin{figure*}[htp]
\centering
\includegraphics[width=0.8\textwidth]{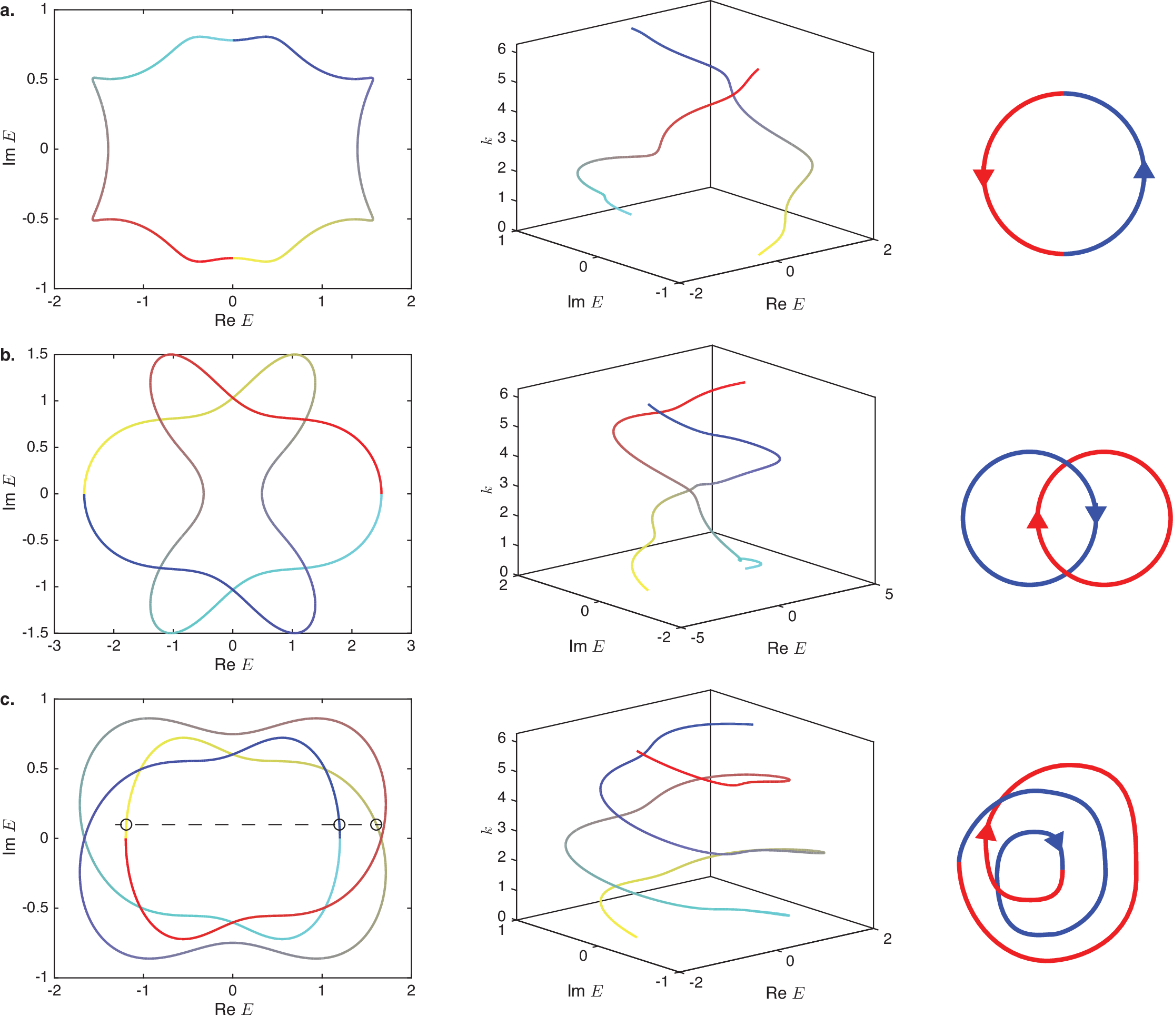}\caption{ \textbf{Complex energy braiding with arbitrary number of knots in a 1D two-band modified SSH chain with long-range coupling} From left to right : complex energy plane projection, 3D complex energy--$k$ braiding, and schematic representation of braiding for $m=3$, $n=1$ and $C_{\mathrm{AB}, 0}=1$ in (a) an Unknot system with one knot and $C_{\mathrm{AB}, -m}=0.22$, $C_{\mathrm{BA}, n}=-1.5$, (b) Hopf link system with two knots and $C_{\mathrm{AB}, -m}=1.4$, $C_{\mathrm{BA}, n}=1.6$, and (c) trefoil system with three knots and $C_{\mathrm{AB}, 0}=1.4$, $C_{\mathrm{BA}, n}= -0.4$. The dotted line and circles denote the crossing points of the yellow-blue line with a line slightly displaced above the $\mathrm{Re} E$ axis mentioned in the text.  The yellow-blue and gray-red lines in the left and middle plots represent the eigen energies of the two distinct bands with darker colors (i.e., red and blue) representing larger values of $k$ for $0 \leq k < 2\pi$.   }
\label{gFig1}
\end{figure*}

Eq. \eqref{H1} represents a two-band model with long-range unidirectional couplings of $n$ unit cells towards the right and $m$ unit cells towards the left. For a given value of momentum $k$, the corresponding eigenvalues of $H_1(k)$ are  given by 
\begin{equation}
	E_\pm (k) = \pm \sqrt{(C_{\mathrm{AB}, 0} + C_{\mathrm{AB}, -m} e^{-i k m})(C_{\mathrm{AB}, 0} + C_{\mathrm{BA}, n} e^{i k n})}. \label{Epm}
\end{equation}
As the momentum $k$ varies from 0 to $2\pi$ within the Brillouin zone, the trajectories of the two complex eigen energies may form knots in the 3D $(\mathrm{Re}E, \mathrm{Im} E, k)$  complex energy--momentum space. If we assume, without loss of generality that $m>n$, then the two eigen energy bands may tangle with each other for up to $m$ times  (e.g., if we set $m=3$, the energy bands may tangle for zero, one, two, or three times) depending on the coupling parameters.  Fig. \ref{gFig1} shows exemplary knot systems realized by the Hamiltonian in Eq. \ref{H1} with $m=3$ and $n=1$ in which the bands tangle once (Fig. \ref{gFig1}a), twice (Fig. \ref{gFig1}b), and three times (Fig. \ref{gFig1}c). These correspond to the Unknot, Hopf link, and trefoil configurations, respectively.  We can systematically describe these configurations via the Artins notation \cite{rehren1989einstein,hu2022knot},where a complex braiding with $p$ knots follows a sequence of band crossings, each of which can be described by the braiding word $\tau_1^{\pm p}$. Here, $\tau_1^1$ ($\tau_1^{-1}$) denotes the case where the trajectory of the first energy strand crosses that of the second energy strand from the left (right) in the projection of the 3D trajectory onto the $\mathrm{Im} E=\infty$ plane. The systems in Fig. \ref{gFig1}a -- c can thus be described by the braid words $\tau^1_1$, $\tau^{-2}_1$, and $\tau^{-3}_1$, respectively. 
	
Furthermore, the  different types of knot configurations can be characterized by an integer topological number known as the braiding index $\xi$, which is defined as \cite{wang2021topological}
\begin{equation}
	\xi \equiv  \int \frac{\mathrm{d}k}{2\pi i} \partial_k \mathrm{log} \left| H(k) - \frac{1}{2}\mathrm{Tr} H(k)\mathbf{I}_2 \right| \label{xi0} 
\end{equation}
where $\mathbf{I}_2$ is the two-by-two identity matrix. 

The magnitude of $\xi$ indicates the number of times that the eigen energy strands of $H(k)$ wind or braid around each other in $(\mathrm{Re}(E), \mathrm{Im}(E), k)$ space, while its sign denotes the net handedness of the  braiding. Using the fact that the determinant of a square matrix is equal to the product of its eigenvalues, introducing $\tilde{H}(k) \equiv H(k) - \frac{1}{2}\mathrm{Tr}H(k)\mathbf{I}_2$, and denoting the two eigenvalues of $\tilde{H}(k)$ as $\tilde{E}_\pm(k)$,  $\xi$ can be rewritten as
\begin{equation}
	\xi =  \int \frac{\mathrm{d}k}{2\pi}\ \partial_k\left( \mathrm{Arg} \tilde{E}_+(k) + \mathrm{Arg} \tilde{E}_-(k) \right).  \label{xiArg} 
\end{equation}

Because the Hamiltonian Eq. \ref{H1} is traceless, $\xi$ for this Hamiltonian can be interpreted as counting the number of times that its eigenvalues wind around the complex energy origin. For example, the leftmost plot in Fig. \ref{gFig1}a shows that for this Unknot system, the two energy strands join together to form a single loop that winds in the counter-clockwise direction as $k$ increases (as is evident from the color coding of the energy strings where darker colors correspond to larger $k$ values), and hence corresponds to a braiding index of +1. For the Hopf link in Fig. \ref{gFig1}b, the gray-red and yellow-blue bands each winds around the origin once in a clockwise manner, and hence collectively give rise to an overall braiding  index of -2. For the trefoil in Fig. \ref{gFig1}c, the gray-red and yellow-blue bands each winds around the origin for one and a half times in a counter-clockwise manner. (The fact that each band winds around the origin for one and a half times can be seen from, e.g., the yellow-blue line crossing a line slightly displaced above the  real energy axis only once on the negative side but twice on the positive side, with the crossing points depicted by open circles in the left panel of  Fig. \ref{gFig1}c. ) The two bands therefore collectively give rise to a braiding index of -3. Interestingly, for two-band systems, each of the two energy strings begins and ends at the same value of complex energy at $k=0$ and $k=2\pi$, respectively, when $\xi$ is even, but when $\xi$ is odd, the start-point of one band coincides with the end-point of the other, and vice versa (so that the trajectory of the two bands collectively form a closed loop).

\begin{figure*}[htp]
\centering
\includegraphics[width=0.8\textwidth]{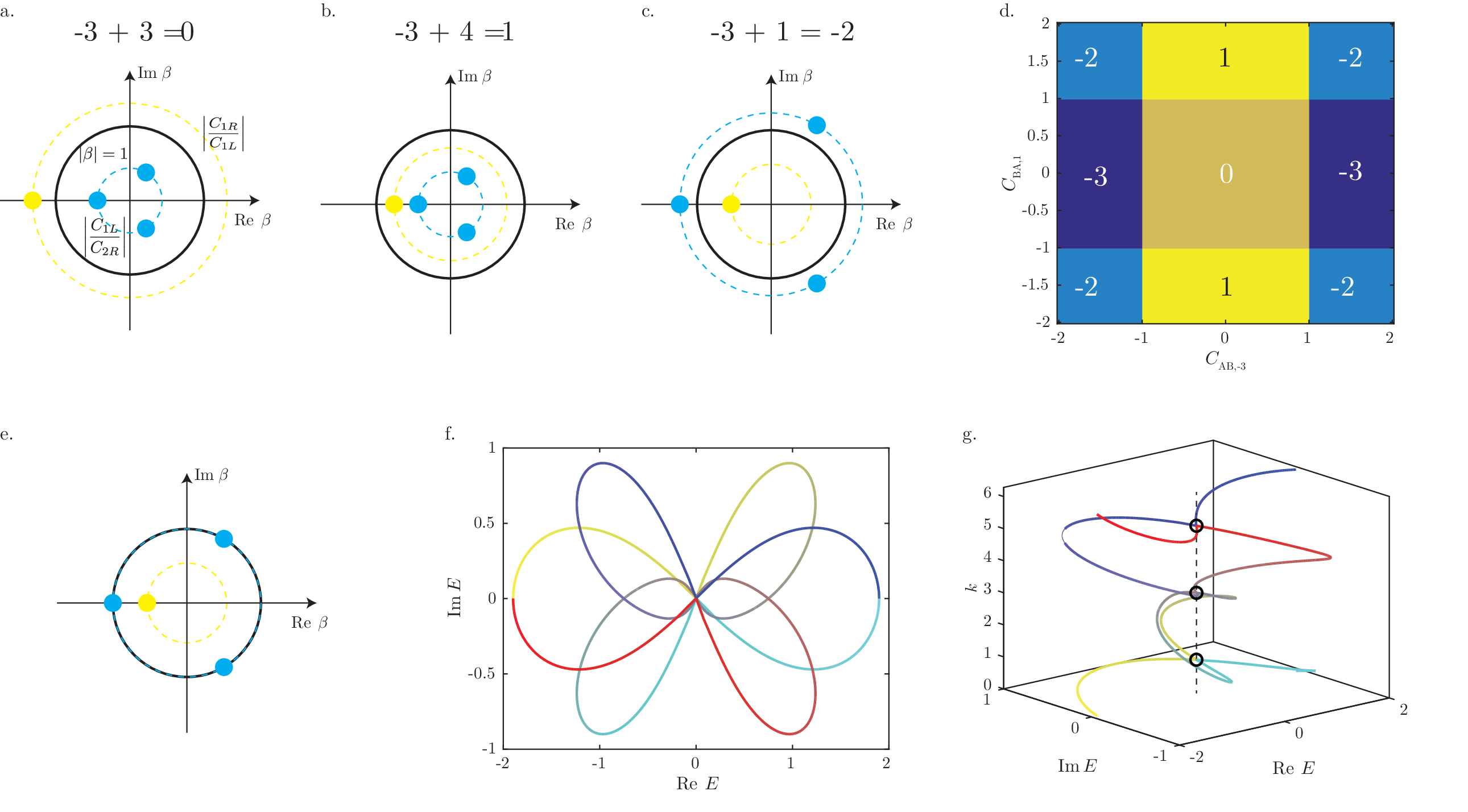}\caption{ \textbf{Braiding index phase diagram }  (a) - (c) Schematic illustration of distribution of $\beta$s at which $E=0$ on complex $\beta$ plane at (a) $\xi=0$, (b) $\xi=1$, and (c) $\xi=-2$ for a system described by Eq. \eqref{H1} for $m=3$ and $n=1$. The yellow circle denotes the value of $\beta$ due to the $n=1$ term and the blue circles the values of $\beta$ due to the $m=3$ term. (d) $\xi$ phase diagram of system in (a) with $C_{\mathrm{AB}, 0}=1$.  (e) Schematic illustration of distribution of $\beta$s  at which $E=0$ on complex $\beta$ plane on transition line between $\xi=-2$ and $\xi=1$.  (f) Complex energy projection and (g) $(\mathrm{Re}(E), \mathrm{Im}(E), k)$-space trajectory of energy strings at transition in (e). The black circles in (g) denote the string-touching points at $E=0$. }
\label{gFig2}
\end{figure*}	

Beyond counting the number of times that the eigen energy strands wind around the origin on the complex energy plane, the braiding index has an alternative mathematical interpretation. Eq. \eqref{xi0} can be recast into the form of a contour integral over the complex unit circle in $\beta\equiv \exp(ik)$ space:
\begin{equation}
	\xi = \frac{1}{2\pi i} \oint_{|\beta|=1} \frac{\partial_k \tilde{E_+}(\beta)}{\tilde{E_+}(\beta)}. \label{xiBeta} 
\end{equation}
Eq. \eqref{xiBeta} can be evaluated using the argument principle as $\xi = N - P$, where $N (P)$ is the number of zeros (poles) of $E_+$ enclosed inside the contour integral. From Eq. \ref{Epm}, $\tilde{E}_+ = \sqrt{(C_{\mathrm{AB}, 0} + C_{\mathrm{AB}, -m}\beta^{-m})(C_{\mathrm{AB}, 0} + C_{\mathrm{BA}, n}\beta^n)}$. There are therefore $m$ poles of $E_+$ at $\beta=0$ within the complex unit circle due to the $\beta^{-m}$ term. The zeros of $\beta$ occur when $\beta^m = -C_{\mathrm{AB}, -m}/C_{\mathrm{AB}, 0}$, the solutions of which are  given by the following equation
\begin{equation}
\beta = \beta_{1,f} = |C_{\mathrm{AB}, -m}/C_{\mathrm{AB}, 0}|^{1/m}e^{(i(\mathrm{Arg}(C_{\mathrm{AB}, -m})-\mathrm{Arg}(C_{\mathrm{AB}, 0})+ (2f+1)\pi /m))},
\end{equation}  
where, $f \in (1, ..., m)$, and when $\beta^n = -C_{\mathrm{AB}, 0}/C_{\mathrm{BA}, n}$, i.e., the solutions of which are given by
\begin{equation}
\beta = \beta_{2,g} = |C_{\mathrm{AB}, 0}/C_{\mathrm{BA}, n}|^{1/n}e^{(i(\mathrm{Arg}(C_{\mathrm{AB}, 0})-\mathrm{Arg}(C_{\mathrm{BA}, n})+ (2g+1)\pi /n))},
\end{equation}
where, $g \in (1, ..., n)$.  Depending on the relative magnitudes of $C_{\mathrm{AB}, 0}$, $C_{\mathrm{AB}, -m}$, and $C_{\mathrm{BA}, n}$, the $m$ ($n$) values of $\beta_{1,f}$ ($\beta_{2,g}$) may fall inside, on, or outside the complex unit circle. Neglecting for now the cases where the $\beta_{1,f}$s or $\beta_{2,g}$s lie exactly on the complex unit circle, the possible values of $\xi$ are thus $-m$, when $|C_{\mathrm{AB}, -m}/C_{\mathrm{AB}, 0}| > 1$, $|C_{\mathrm{AB}, 0}/C_{\mathrm{BA}, n}| > 1$ so that  the $\beta_{1,f}$s and $\beta_{2,g}$s all lie outside the unit circle; 0, when $|C_{\mathrm{AB}, -m}/C_{\mathrm{AB}, 0}| < 1$,  $|C_{\mathrm{AB}, 0}/C_{\mathrm{BA}, n}| > 1$ so that the $m$ $\beta_{1,f}$s all lie within the unit circle while the $n$ $\beta_{2,g}$s all lie outside  (Fig. \ref{gFig2}a);  $n$, when  $|C_{\mathrm{AB}, -m}/C_{\mathrm{AB}, 0}| < 1$,  $|C_{\mathrm{AB}, 0}/C_{\mathrm{BA}, n}| < 1$, so that the $m$ $\beta_{1,f}$s  and $n$ $\beta_{2,g}$s all lie inside the unit circle (Fig. \ref{gFig2}b) ; and $(n-m)$, so that the $n$ $\beta_{2,g}$s all lie within the unit circle while the $m$ $\beta_{1,f}$s  all lie outside (Fig. \ref{gFig2}c).  This dependence of $\xi$ on the relative magnitudes of $C_{\mathrm{AB}, 0}$, $C_{2L}$, and $C_{\mathrm{BA}, n}$ results in the phase diagram shown in Fig. \ref{gFig2}d where the boundaries between different values of the braiding indices are defined by the equations: $C_{\mathrm{AB}, -m}= \pm C_{\mathrm{AB}, 0}$ and $C_{\mathrm{BA}, n} = \pm C_{\mathrm{AB}, 0}$.

Let us now consider the case where the $m$ values of $\beta_{1,f}$s (or $n$ values of $\beta_{2,g}$s) fall on the unit circle on the complex $\beta$ plane (Fig.  \ref{gFig2}e), for which  the value of $C_{\mathrm{AB}, -m}$ ($C_{\mathrm{BA}, n}$) must necessarily fall on one of the phase boundaries between different values of $\xi$. This is because the $\beta_{1,f}$s  and/or $\beta_{2,g}$s have to pass through the unit circle, for the number of $\beta$s   solutions to $E_+=0$ that lie within the unit circle to change.  The correspondence between the complex unit circle on the $\beta$ plane to real values of $k$, i.e., $\beta\equiv\exp(ik)$ means that there are $m$, $n$, or $(m+n)$ values of $k$ corresponding to zero energy on the PBC spectrum when only $|C_{\mathrm{AB},-m}/C_{\mathrm{AB},0}| = 1$, only $|C_{\mathrm{AB},0}/C_{\mathrm{BA},n}| = 1$, or $|C_{\mathrm{AB},-m}/C_{\mathrm{AB},0}| = |C_{\mathrm{AB},0}/C_{\mathrm{BA},n}| = 1$, respectively.    Figs. \ref{gFig2}f and \ref{gFig2}g illustrate this  for the specific case of  $m=3$. For the particular model in Eq. \eqref{H1}, the zeros of the $E_+$ also happen to be exceptional points because these zeros occur only when $H_1(k)$ is rank-deficient and all the matrix elements in the left or right columns are zero. 

\begin{figure*}[htp]
\centering
\includegraphics[width=0.8\textwidth]{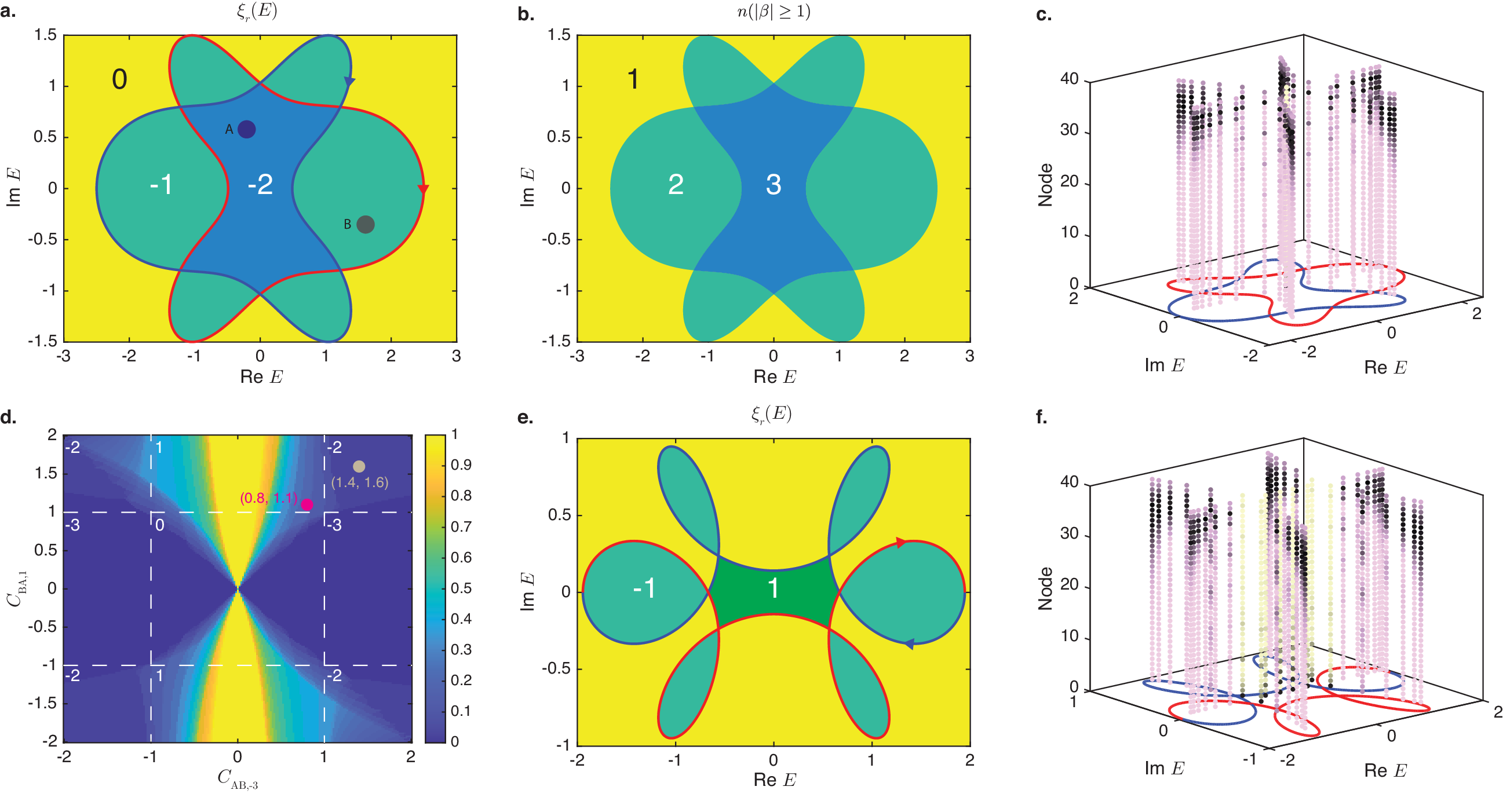}\caption{ \textbf{Braiding number and NHSE}  (a) Distribution of $\xi_r(E)$ on the complex energy plane and PBC spectrum for the Hopf link system of Fig. \ref{gFig1}b with $m=3$, $n=1$, $C_{\mathrm{AB}, 0}=1$, $C_{\mathrm{AB}, -m}=1.4$, and $C_{\mathrm{BA}, n}=1.6$. The two energy bands are indicated by the blue and red lines, and the arrows on the lines indicate the direction of increasing $k$. The blue dot A and red dot B denote values of $E$ for which $\xi_r(E)$ takes the values of -2 and -1, respectively. (b) Distribution of the number of distinct eigenstates with $|\beta|>1$ on the complex energy plane for the system in (a). (c) PBC spectra for the Hopf link system in (a), and the OBC spectra and spatial density distribution for its eigenstates in a finite system with 40 nodes. Darker dots indicate a higher density distribution. The lines comprising faint blue (green) dots denote eigenstates localized towards the right (left) with larger (smaller) node indices. (The green dots in (c) are located at $E=0$.)  (d) Variation of proportion of OBC eigenstates in finite systems comprising 40 nodes localized nearer the left edge with $C_{\mathrm{AB}, -m}$ and $C_{\mathrm{BA}, n}$. The dotted white lines denote the partitions between phase regions with different values of the braiding index $\xi$, which are indicated at the top left corner  of the regions. The values of $(C_{\mathrm{AB}, -m},C_{\mathrm{BA}, n})=(1.4,1.6)$ plotted in (a) -- (c), and  $(C_{\mathrm{AB}, -m},C_{\mathrm{BA}, n})=(0.8,1.1)$ plotted in (e) and (f) are indicated. (e) Distribution of $\xi_r(E)$ on the complex energy plane for a system with with $m=3$, $n=1$, $C_{\mathrm{AB}, 0}=1$, $C_{\mathrm{AB}, -m}=0.8$, and $C_{\mathrm{BA}, n}=1.1$. (f) PBC spectrum, and OBC spectrum and eigenstate spatial distribution for a finite system containing 40 nodes for the system in (e). }
\label{gFig3}
\end{figure*}

\section{GBZ and NHSE variation in complex braiding configurations}
We comment briefly on the relationship between the braiding index and the non-Hermitian skin effect (NHSE). In 1D systems, a winding number (or braiding index) $\xi_r(E)$ \cite{PRL124_086801} may be defined with respect to a reference energy $E$ as  
\begin{equation}
	\xi_r(E) =  \int \frac{\mathrm{d}k}{2\pi i} \partial_k \mathrm{log} \left| H(k) - E \right|.
\end{equation}
The braiding index $\xi$ in Eq. \eqref{xi0} may be interpreted as a special case of the winding number $\xi_r(E)$ where $E$ is set to $\frac{1}{2} \mathrm{Tr}(H)$, which for the model in Eq. \eqref{H1} is equal to 0.  The winding number $\xi_r(E)$ may be visually determined from the PBC spectrum of a system provided that the individual bands can be separately identified and the direction of increasing $k$ indicated.

 For example, in Fig. \ref{gFig3}a the winding number at point A. $ \xi_r(i) = -2$ because the bands indicated by the red and blue lines each winds around point A in a complete circle along the clockwise direction. In contrast, the winding number at point B, $\xi_r(B) = -1$ because point B falls completely outside the loop enclosed by the blue band and is enclosed only the red band, which winds around the point in the counter-clockwise direction.

The magnitude of the winding number $\xi_r(E)$ is associated with the number of bulk NHSE states that exist in a semi-infinite system at $E$, as explained shortly, while the sign of the winding number denotes whether it is a semi-infinite system that extends from 0 to $+\infty$ (positive $\xi_r(E)$) or a semi-infinite system that extends from $-\infty$ to 0 that hosts such bulk NHSE states (\cite{PRX8_031079})  The braiding index $\xi=\xi_r\left(-\frac{1}{2}\mathrm{Tr}(H)\right)$ therefore indicates the number and localization of  NHSE states that exist at $E=-\frac{1}{2}\mathrm{Tr}(H)$ in the  semi-infinite system.

To further elucidate the relationship between the winding number and semi-infinite eigenstates, we examine these eigenstates more carefully. We consider first the eigenstates for a semi-infinite system that extends from $x=-\infty$ to $x=0$. We introduce the surrogate Hamiltonian $H_1(\beta)$  \cite{li2020critical} to the Hamiltonian in Eq. \eqref{H1} as 
\begin{equation}
	H_1(\beta) = \begin{pmatrix} 0 &  C_{\mathrm{AB}, 0} + C_{\mathrm{AB}, -m}\beta^{-m}  \\ 
	C_{\mathrm{AB}, 0} + C_{\mathrm{BA}, n}\beta^n & 0 \end{pmatrix},  \label{H1b}
\end{equation}
where $\beta\equiv \exp(ik)$ and $k$ is now allowed to be complex. If an eigenstate of the semi-infinite system at a given energy $E$ exists, the eigenstate can be generically written as a linear combination 
\begin{equation}
	\psi (x) = \sum^{a=n+1}_{a=1, |\beta_{a;E}|\geq 1} |\phi_{a;E}\rangle c_a \beta_{a;E}^x  \label{psiLeft}
\end{equation}
where $\beta = \beta_{a; E}$ is the $a$th value of $\beta$ for which one of the eigenvalues of $H_1(\beta = \beta_{a; E})$ is $E$ and $|\phi_a\rangle_E$ is the corresponding eigenvector. The summation of $a$ from 1 to $n+1$ in Eq. \eqref{psiLeft} is due to the following considerations: The coupling between the B node of a unit cell and the A node of the $n$th neighbor to its right represented by the $\beta^n$ term in Eq. \eqref{H1b} requires the A node component of $\psi(x)$, i.e. $\langle A|\psi(x)$ to satisfy the boundary conditions $\langle A|\psi(1)=\langle A|\psi(2)=... = \langle A|\psi(n) =0$ beyond the rightmost unit cell of the semi-infinite system at $x=0$. A linear combination of $n+1$ linearly independent eigenvectors is therefore required in Eq. \eqref{psiLeft} to ensure that these $n$ constraints are always satisfied. Moreover, the corresponding $\beta$ values for these eigenvectors are required to satisfy $|\beta_{a;E}|\geq 1$ to ensure that $\psi(x)$ remains bounded as $x\rightarrow-\infty$.

Whether such right-localized eigenstates exist in a semi-infinite system extending from $x=-\infty$ to $x=0$ for a given $E$ therefore hinges on whether there are at least $n+1$ linearly independent eigenvectors of $H_1(\beta)$ satisfying $|\beta_{a;E}|\geq 1$ at that given $E$. The characteristic equation for $|H_1(\beta)-E\mathbf{I}_2| = 0$ can be cast into the form of a $(m+n)$th order polynomial $E^2\beta^m - (C_{\mathrm{AB}, 0}\beta^m + C_{\mathrm{AB}, -m})(C_{\mathrm{AB}, 0} + C_{\mathrm{BA}, n}\beta^n) = 0$, which in general has $(m+n)$ roots for the $\beta$s, corresponding to $(m+n)$ linearly independent eigenvectors.  

Now recall  that the braiding index is $\xi=N-P$ where $N$ is the number of zeros of $E_+$ and $P=m$ is the number of poles of $E_+$ that lie within $|\beta|<1$ on the complex $\beta$ plane.As before, for simplicity, we first set aside the cases where any of the $\beta$ values lies exactly on the unit circle. The zeros of $E_+$ on the complex $\beta$ plane, are merely the $\beta_{a;E=0}$ values corresponding to $E=0$. The number of $\beta_{a;E=0}$s that lie within the complex unit circle  on the $\beta$ plane (excluding the $\beta=0$ term),  i.e. $N$, is therefore $m+\xi$ because $\xi=N-m$. Conversely, the number of $\beta$s that lie outside the complex unit circle on the $\beta$ plane is $m+n-(m + \xi) = n - \xi$. In particular, $\xi=-1$ implies that there are $n+1$ values of $\beta$ that lie outside the complex plane with $n+1$ corresponding linearly independent eigenvectors. This is exactly the number of linearly independent eignvectors with $|\beta|\geq 1$ required to constitute a right-localized semi-infinite NHSE eigenstate. A more negative value of $\xi$ implies that there is an excess number of $|\beta|\geq 1$ eigenstates that may be used to construct a right-localized semi-infinite NHSE eigenstate. The fact that only $(n+1)$  such $|\beta|\geq 1$ eigenstates are required out of the $n+|\xi|$ available ones implies that we can obtain $C^{n+|\xi|}_{n+1}$ linearly independent right-localized NHSE eigenstates in a semi-infinite system extending from $x=-\infty$ to $x=0$ at $E=0$.  Here, $C^a_b$ denotes the number of different combinations of picking $a$ items out of a set of $b$ items. (Note that this differs from the interpretation in \cite{PRX8_031079} where $|\xi|$ itself is interpreted to be the number of degenerate semi-infinite NHSE eigenstates.) In a similar manner, a positive value of $\xi$ implies that there are $C^{m+\xi}_{m+1}$ linearly independent left-localized NHSE eigenstates in a semi-infinite system extending from $x=0$ to $x=\infty$ at $E=0$.

The above arguments that $ \xi = \xi_r (E=0)$ is correlated to the number of semi-infinite NHSE eigenstates at $E = 0$ can be readily extended to the case of arbitary $E$. Fig. \ref{gFig3}b shows $n(|\beta|>1)$, i.e., the number of $\beta$s for which $|\beta|>1$ corresponding to  each value of $E$ in the parameter set in Fig. \ref{gFig3}a, for which $m=3$ and $n=1$. It can be readily verified by a comparison between Fig. \ref{gFig3}a and \ref{gFig3}b that the relation between the negative values of $\xi$ in Fig. \ref{gFig3}a  to $n(|\beta| \geq 1)$ in Fig. \ref{gFig3}b is given by $n(|\beta| \geq 1) = \xi_r + m$ as discussed above. Further, the PBC spectrum coincides with the boundary lines between different values of $\xi_r(E)$. This is because changes in $\xi_r(E)$ correspond to changes in the number of $\beta$s within the complex unit circle on the $\beta$ plane, which are necessarily accompanied by $\beta_{a,E}$s crossing the complex unit circle as $E$ is varied. A $\beta_{a,E}$ values that lies on the complex unit circle corresponds to a real value of $k=-i\mathrm{ln} (\beta_{a,E})$ in Eq. \eqref{H1} and therefore lies on the PBC spectrum.  

Let us label the $\beta_{a, E}$s such that $|\beta_{1,E}| \leq |\beta_{2,E}| \leq ... |\beta_{n+m, E}|$. Then, the general Brillouin zone (GBZ) of Eq. \eqref{H1} for a system with open boundary conditions (OBC) in the thermodynamic limit is given by the loci of $E$ for which $|\beta_{m,E}|=|\beta_{m+1,E}|$. Recall from the above discussion that the number of $\beta_{a;E}$s that lie within the complex unit circle on the $\beta$ plane at energy $E$ is $m + \xi_r(E)$. For a value of $E$ that lies on the GBZ, a positive finite value of $\xi_r(E)$ therefore implies that $|\beta_{m,E}|=|\beta_{m+1,E}|\leq 1$, which corresponds to a left-localized OBC NHSE state. Conversely, a negative value of $\xi_r(E)$ in turn corresponds to a right-localized OBC NHSE state.  Although strictly speaking, the GBZ condition $|\beta_{m,E}|=|\beta_{m+1,E}|$ applies only in the thermodynamic limit,  the NHSE localization direction of a sufficiently long \textit{finite}-sized chain follows that of the  chain in the thermodynamic limit. In particular, a positive (negative) value of the braiding index $\xi = \xi_r(0)$ implies that OBC NHSE states within the central area around $E=0$ bounded by the PBC spectrum are localized at the left (right) edge of the system. This is illustrated in Fig. \ref{gFig3}c, which shows the eigenenergy and spatial density distribution for the eigenstates of a 40-node long finite system described by the Hamiltonian with the parameter set of Fig. \ref{gFig3}a, b. As predicted by the negative signs of $\xi_r(E)$ shown in Fig. \ref{gFig3}a, most of the eigenstates are localized towards the right edge (large $N$) of the system, except for two topological zero modes (TZMs) that are localized towards the left edge  (these TZMs are not described by  the GBZ theory). 

Apart from the TZMs, the NHSE localization direction may differ from that predicted by $\xi$ outside the central area on the complex energy plane around $E=0$ bounded by the PBC spectrum. Fig. \ref{gFig3}d shows $f_{\mathrm{L}}$, the fraction of the 40 eigenstates of a 40 node-long system described by Eq. \eqref{H1} localized closer to the left edge,  as a function of $C_{\mathrm{AB}, -m}$ and $C_{\mathrm{BA}, n}$. A $f_{\mathrm{L}}$ value of 1 indicates that all of the eigenstates are localized closer to the left edge while a proportion of 0 indicates that all of the eigenstates are localized closer to the right edge. In the $\xi_r(E)=-2$ regions at the four corners of the plot in Fig. \ref{gFig3}d, there are regions at which $f_{\mathrm{L}}$ deviates very slightly from 0. These slight deviations are due to TZMs that are localized at the left edge opposite to that of the NHSE eigenstates, as shown in Fig. \ref{gFig3}c. Apart from the $\xi_r(E)=-3$ regions where all the eigenstates are localized to the right, bipolar localization, in which the NHSE eigenstates at different eigenenergies are localized at different edges of the system, occurs throughout much of the $C_{\mathrm{AB}, -m}$--$C_{\mathrm{BA}, n}$ parameter space. One example at $(C_{\mathrm{AB}, -m}, C_{\mathrm{BA}, n})=(0.8, 1.1)$ is shown in Fig. \ref{gFig3}e and f where Fig. \ref{gFig3}f shows that the eigenstates in the central complex energy region around $E=0$ are localized to the left, consistent with $\xi_r(E)=1$, whereas the more numerous eigenstates in the complex $E$ side-lobes are localized towards the right, consistent with $\xi_r(E)=-1$ in these lobes, as shown in Fig. \ref{gFig3}e.

\section{Continuous jumps in braiding index via intra-unit cell coupling and onsite coupling}
\begin{figure*}[htp]
\centering
\includegraphics[width=0.8\textwidth]{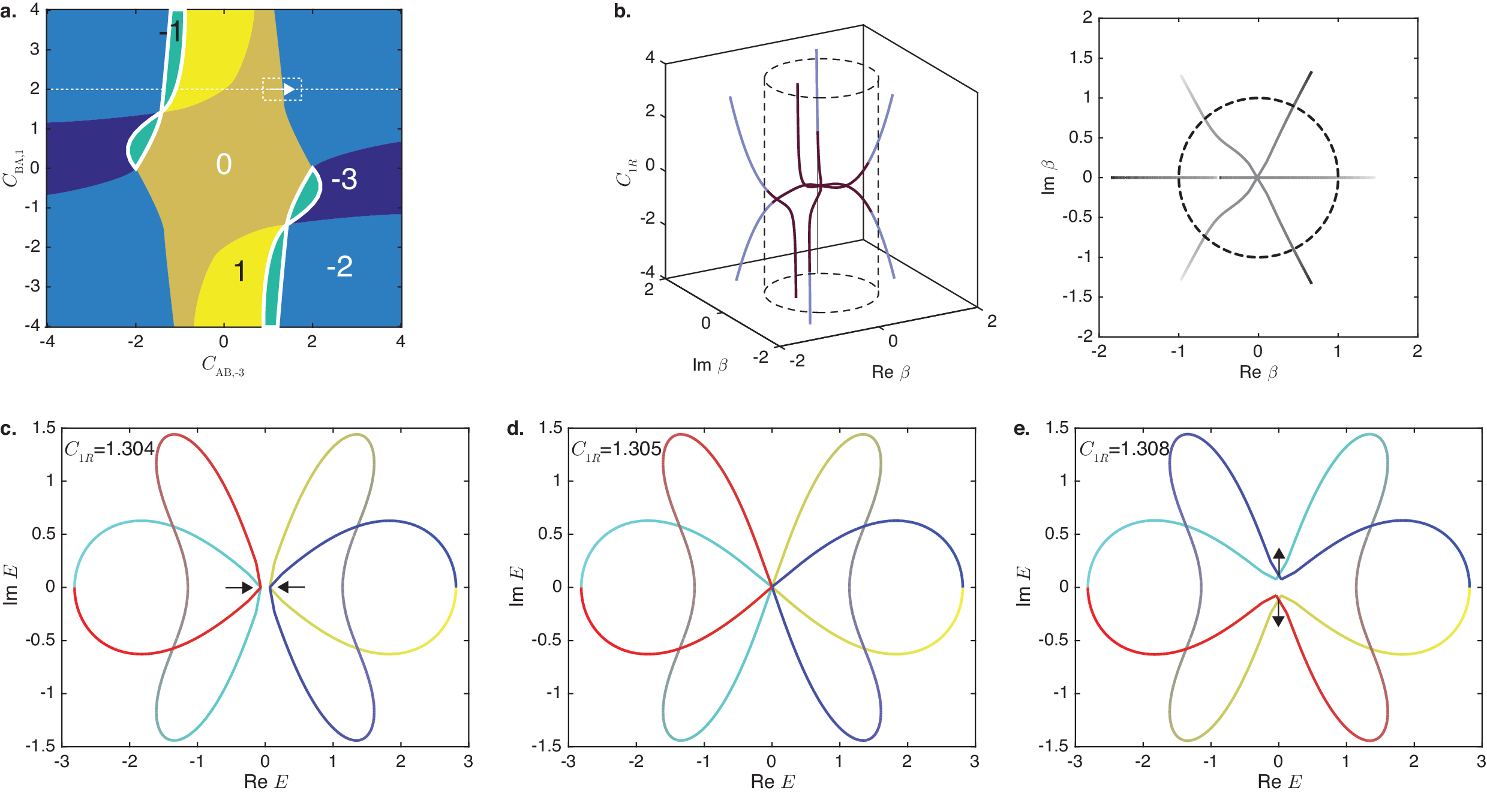}\caption{ \textbf{Continuous jumps in $\xi$ via intra-unit cell coupling }  (a) $\xi$ phase diagram of system with surrogate Hamiltonian given by Eq. \eqref{H2b} with $C_{\mathrm{AB}, 0}=C_i=1$, $m=3$, and $n=1$. The $\xi=-1$ regions, which are absent in Fig. \ref{gFig3}d, are highlighted in white. (b) (Left) Variation of complex $\beta$s for which $E=0 $ with $C_{\mathrm{AB}, -m}$ at $C_{\mathrm{BA}, n}=2$, as denoted by the white dotted line in (a).  The values of $\beta$ for which $|\beta|>1$ ($|\beta|<1$) are marked in black (blue). (Right) Projection of graph on left onto complex $\beta$ plane. The points with lighter (darker) colors correspond to smaller (larger) values of $C_{\mathrm{AB}, -m}$. (c), (d), (e) show the evolution of the PBC spectrum with $C_{\mathrm{AB}, -m}$ in the vicinity of the $\xi=0$ to $\xi=-2$ transition denoted by the white arrow in (a). The black arrows in (c) and (e) show the movement direction of the energy bands as $C_{\mathrm{AB}, -m}$ is increased across the transition. }
\label{gFig4}
\end{figure*}

In the model in Eq. \ref{H1}, there is a constraint in the braiding index $\xi$ in that it changes only in steps of $\pm m$ or $\pm n$ . This is because the $m$ ($n$) roots of $C_{\mathrm{AB}, 0}+C_{\mathrm{AB}, -m}\exp(-ikm)=0$ ($C_{\mathrm{AB}, 0}+C_{\mathrm{BA}, n}\exp(ikn)=0$) in Eq. \eqref{Epm} all have the same absolute values. We can remove this constraint by modifying the Hamiltonian with the introduction of a finite on-site potential, i.e.,  
\begin{equation}
	H_2(\beta) = \begin{pmatrix}  -C_i &  C_{\mathrm{AB}, 0} + C_{\mathrm{AB}, -m}\beta^{-m} \\ C_{\mathrm{AB}, 0} + C_{\mathrm{BA}, n}\beta^n & C_i \end{pmatrix}, \label{H2b} 
\end{equation}
in which case the eigenenrgies $E$ now satisfy
\begin{equation}
	C_i^2 - (C_{\mathrm{AB}, 0} + C_{\mathrm{BA}, n}\beta^n)(C_{\mathrm{AB}, 0}+C_{\mathrm{AB}, -m}\beta^{-m}) = E^2.  \label{E2}
\end{equation}
This results in a phase diagram for which all values of $\xi$ across the entire range of $\xi=-m$ to $\xi=n$ appear, as shown in Fig. \ref{gFig4}a. Due to the offset $C_i^2$ in Eq. \ref{E2},  the $(m+n)$ values of $\beta$ that satisfy the $E=0$ condition will no longer have the same absolute values in general (Fig. \ref{gFig4}b), thus removing the restriction on the change in $\xi$ to steps of $\pm m$ or $\pm n$ as was the case when $C_i=0$. As mentioned previously, a change in $\xi$ will be accompanied by the appearance of a zero eigenenergy on the PBC spectrum since $\xi$ varies with the number of $\beta$ values at which the energy eigenvalues are 0 and $|\beta|<1$. From Eq. \ref{E2}, it can be seen that $E=0$ corresponds to an eigenenergy at which the two energy bands coincide with each other (Fig. \ref{gFig3}d). As $C_{\mathrm{AB}, -m}$ in Fig. \ref{gFig4})c--e is varied towards the boundary between two $\xi$ values (along the dotted line shown in Fig. \ref{gFig4}a), the two bands begin to approach each other along the real energy axis (Fig. \ref{gFig4}c). They then touch each other at $E=0$ at the transition point corresponding to the $\xi$ boundary line (Fig. \ref{gFig4}d), and then drift apart along the imaginary energy axis (Fig. \ref{gFig4}e) as $C_{AB,-m}$ is increased further. As the energy bands touch at the transition point, they swap partners, which in turn results in a change in the connectivity between the energy bands and accordingly, the number of times the two bands wind around $E=0$. Consider the series of figures  Fig. \ref{gFig4}c--e. Before the transition (Fig. \ref{gFig4}c), $\mathrm{Re}(E^2)$ is less than 0 at the minimum value of $|E|$, which results in the two bands being separated by a finite gap along the real energy axis; $\mathrm{Re}(E^2)$ becomes 0 at the transition point, which results in the bands touching (Fig. \ref{gFig4}d); and finally $\mathrm{Re}(E^2)$ becomes positive after the transition point in Fig. \ref{gFig4}e, which results in the two bands being separated by a finite gap along the imaginary energy axis.

\section{ Impact of bidirectional asymmetric long range coupling on braiding profile}
In the models introduced so far,  the inter-unit cell A to B sublattice site coupling $\langle \mathrm{A}|H(\beta)|\mathrm{B}\rangle$ occurs only to the left ( $C_{\mathrm{AB}, -m}\beta^{-m}$) while the B to A sublattice coupling $\langle \mathrm{B}|H(\beta)|\mathrm{A}\rangle$  occurs only to the right ($C_{\mathrm{AB}, 0}\beta^n$) term. This implies that the maximum value of $|\xi|$ that can occur is $\mathrm{max}([m,n])$, i.e., the largest distance of the inter-unit cell coupling either along the left or right directions. We can further generalize the model to extend the range of possible $\xi$ values and attain a larger maximum value of $|\xi|$ for a given inter-cell coupling distance  by incorporating inter-unit cell bidirectional coupling of both directions in both the A to B sublattice sites and vice versa. The corresponding Hamiltonian in the presence of this bidirectional coupling is given by

\begin{widetext}
\begin{equation}
	H_3(\beta) = \begin{pmatrix}  -C_i &  \sum_{a=1}^{m_{\mathrm{AB}}} C_{\mathrm{AB}; -a}\beta^{-a} +  \sum_{b=0}^{n_{\mathrm{AB}}} C_{\mathrm{AB}; b}\beta^b   \\ \sum_{c=1}^{m_{\mathrm{BA}}} C_{\mathrm{BA}; -c}\beta^{-c} +  \sum_{d=0}^{n_{\mathrm{BA}}} C_{\mathrm{BA}; d}\beta^{d} & C_i \end{pmatrix} \label{H3b}.
\end{equation}
\end{widetext}

\begin{figure*}[htp]
\centering
\includegraphics[width=0.8\textwidth]{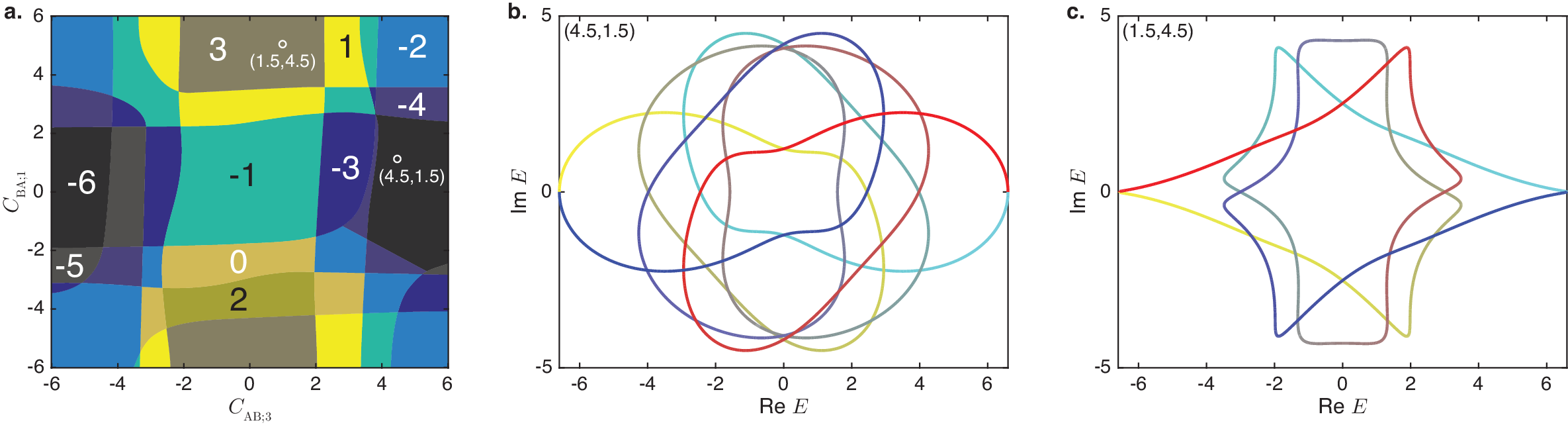}
\caption{ \textbf{Extended  values of $\xi$ in the presence of bi-directional coupling}  (a) $\xi$ phase diagram of system with surrogate Hamiltonian on the $C_{\mathrm{AB}; 3}$--$C_{\mathrm{BA}; 1}$ plane at $C_{\mathrm{AB}; 0}=C_{\mathrm{BA}; 0}=C_i=1$, and $C_{\mathrm{AB};2}=C_{\mathrm{BA};-3}= 3$.   (b), (c) PBC spectra of system in (a) at (b)  $(C_{\mathrm{AB}; 3}, C_{\mathrm{BA}; 1} = (4.5,1.5)$ with $\xi=-6$ and (c) $(C_{\mathrm{AB}; 3}, C_{\mathrm{BA}; 1} = (1.5,4.5)$ with $\xi=3$ . }
\label{gFig5}
\end{figure*}	

Fig. \ref{gFig5} shows an exemplary phase diagram for a system represented by Eq. \eqref{H3b} where $m_{\mathrm{AB}}=m_{\mathrm{BA}}=3$, $n_{\mathrm{AB}} = 2$, and $n_{\mathrm{BA}} = 1$ and two PBC spectra from the two extreme (positive and negative) values of $\xi=n_{\mathrm{AB}} + n_{\mathrm{BA}} = 3$ and $\xi=-m_{\mathrm{AB}}-m_{\mathrm{BA}}=-6$ obtained when the $\beta$ terms with the largest positive and negative exponents in the upper right element of Eq. \eqref{H3b} are multplied with their counterparts in the lower left element. The presence of coupling along both the left and right directions in the two off-diagonal terms of $H$ implies that the exponents of the $\beta$ terms sum up in the characteristic equation for the eigenenergy , which is given by
\begin{widetext} 
\begin{equation}
	E^2 - C_i^2 + \left( \sum_{a=1}^{m_{\mathrm{AB}}} C_{\mathrm{AB}; -a}\beta^{-a} +  \sum_{b=0}^{n_{\mathrm{AB}}} C_{\mathrm{AB}; b}\beta^b \right)\left( \sum_{c=1}^{m_{\mathrm{BA}}} C_{\mathrm{BA}; -c}\beta^{-c} +  \sum_{d=0}^{n_{\mathrm{BA}}} C_{\mathrm{BA}; d}\beta^{d} \right)=0.
\end{equation}
\end{widetext} 
In the characteristic equation,  the exponents of $\beta$ now range from $-(m_{\mathrm{AB}}+m_{\mathrm{BA}})$ to $n_{\mathrm{AB}}+n_{\mathrm{BA}}$. Correspondingly, for the example depicted in Fig. \ref{gFig5}, the values of $\xi$ range from $-(m_{\mathrm{AB}}+m_{\mathrm{BA}}) = -6$ when all of the ($m_{\mathrm{AB}}+m_{\mathrm{BA}}+n_{\mathrm{AB}}+n_{\mathrm{BA}}$) $|\beta|$ values at $E=0$ fall outside the unit circle   on the complex $\beta$ plane (Fig. \ref{gFig5}b) to $n_{\mathrm{AB}} + n_{\mathrm{BA}} = 3$ when all of the $|\beta|$ values fall within the unit circle. Compared to the earlier models with unidirectional couplings depicted in Figs. \ref{gFig2} to \ref{gFig4} for which $m=3$ resulted in the maximum value of $|\xi|$ being only 3,the presence of bidirectional coupling results in the doubling of the maximum value of $|\xi|$ to 6 even though the furthest inter-unit cell coupling is still maintained  to the third neighbor. 

\section{Topological phase transition through exceptional curves}
Two topologically distinct phases associated with different braiding index cannot be continuously transformed into one another without transiting through a vanishing bandgap state, at which point the braiding index changes its value. This transition signifies the presence of exceptional points (EPs) lying on the boundary between the two distinct phases. Specifically, all points lying on the phase boundaries between two distinguished knot configurations(see phase diagram of Fig. \ref{gFig2}d and \ref{gFig4}a) constitute EPs which are characterized by vanishing bandgap with real solution of $k$.

To illustrate the phase transition via EPs, let us consider Eq. \ref{H1} and set $n=1$ without loss of generality. The modified Laplacian then reads 
\begin{equation}
H(k, m) = (C_{\mathrm{AB}, 0}+ C_{\mathrm{AB}, -m} e^{-i m k}) \sigma^{+} + (C_{\mathrm{AB},0}+C_{\mathrm{BA}, n} e^{i k}) \sigma^{-}
\label{Eqmat8}
\end{equation}
where, $\sigma^{\pm}= \frac{1}{2} (\sigma_x \pm i \sigma_y)$.Note that the band-gap of the two band non-Hermitian model can vanish in two possible ways. First, if the transition points lie at the phase boundary between two opposite handedness (i.e., the sign of the braiding index changes from positive sign to negative sign or vice versa) configurations, the two strings with undercross ($\tau_1^m$) configuration transforms into an overcross ($\tau_1^{-m}$) configuration or vice versa, this sort of phase transition is referred as Type-2 phase transition which   is accompanied by $m$ number of EPs (here we vary $k_x$ from $-\pi$ to $\pi$ for simplicity). On the other hand,  a second type of phase transition occurs when two strings change its braiding index number at the phase transition points while maintaining the same sign, so that  the handedness of the braiding and the NHSE localization remains at the same edge. This is the so-called Type-1 phase transition, where an undercross (overcross) configuration remains undercross (overcross) in the braiding configuration. Type-1 transition is accompanied by a single EP. 

\begin{figure*}[htp]
    \centering
    \includegraphics[width=0.8\textwidth]{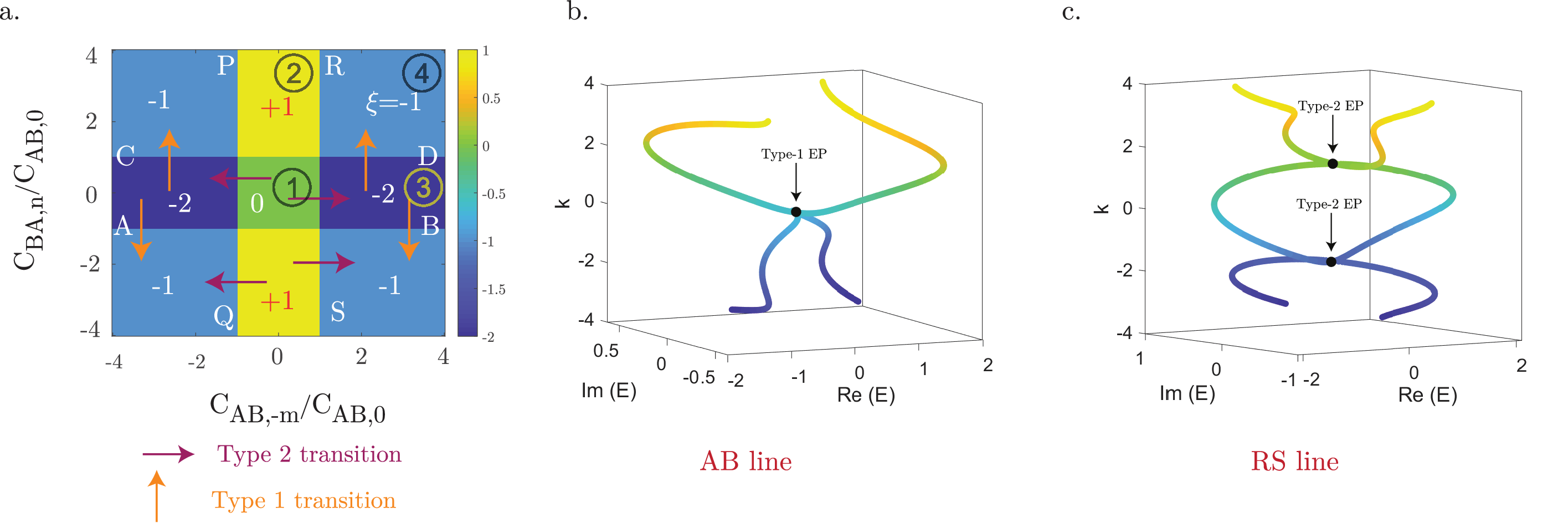}
    \caption{\textbf{Exceptional points evolution in the various braided non-Hermitian bands } (a) Phase diagram and the phase boundaries with various types and number of EPs in the two band system defined in Eq. \ref{Eqmat8} with second order long range coupling (i.e. $m=2$). Type-1 and Type-2 EPs are defined at the phase boundaries between different  knot  index with same sign and opposite sign, respectively and shown in the vertical and horizontal arrows, respectively. (b-c) Various type EPs in the complex energy-momentum space. Type-1 transition is accompanied by single EP located at $k=0$ in (b). Type-2 transition is accompanied by two EPs located at $k= \pm \pi/2$ in (c).}
    \label{fig4} 
\end{figure*}
As an illustrative example, we consider a system with second-order long range coupling (i.e. setting $m=2$ in Eq. \ref{Eqmat8}), and plot the corresponding phase diagram in Fig. \ref{fig4}. Clearly, the phase boundaries coincide with the $C_{\mathrm{AB}, -m}/C_{\mathrm{AB}, 0}=\pm 1$ and $C_{\mathrm{BA}, n}/C_{\mathrm{AB}, 0}=\pm 1$ lines. Interestingly, the band diagrams corresponding to points lying on the  lines PQ and RS (corresponding to $C_{\mathrm{AB}, -m}=\pm C_{\mathrm{AB}, 0} $ ) host two exceptional points with Type-2 transition as there are  only two real values of $k$ (i.e., $k=\pm \pi/2$ for RS line, see Fig. \ref{fig4}c and $k=0, \pi$ for PQ line) which result in zero bandgap of Eq. \ref{Eqmat8}. On the contrary, the band diagrams corresponding to lines AB and CD (corresponding to  $C_{\mathrm{BA}, n}=\pm C_{\mathrm{AB}, 0} $) represents the Type-1 phase transition with a single exceptional point  as there exist only a single real value of $k$ (i.e., $k=0$ for AB line, see Fig. \ref{fig4}b and $k=\pi$ for \underline{EF line}) which satisfies the zero bandgap condition in Eq. \ref{Eqmat8}. Thus a transition from unlink ($|\xi|=0$) to Hopf link with $|\xi|=2$  (unknot with $|\xi|=1$) phase constitutes a Type-2 (Type-1) transition (see Fig. \ref{fig4}a). 


In general, Type-2 transition host $m$ number of EPs, for a system with long-range coupling of the order of $m$, since there are  $m$ real solutions of $k$ corresponding to zero bandgap. On the other hand,  the Type-1 phase transition hosts only a single EP, regardless of the degree of the long-range coupling (i.e. $m$).  We summarized the locations of   $m$ Type-2 EPs and single Type-1 EP in the $k$ space for several example of long-range order in  Table  \ref{tab:table2} (note that for simplicity we consider the values of  $k$ within the range of  $-\pi$ to $\pi$).

\begin{table*}[htp]
\caption{\label{tab:table2}
Location and number of various Type of EPs for different order of long range coupling for the 1D non-Hermitian system defined in Eq. \ref{Eqmat8}. Type-1 and Type-2 EPs emerge when phase boundary lie between two distinct phases with two knot index of same sign and opposite sign, respectively. Type-1 phase transition hosts only single EPs irrespective to the choice of long range order while the number of EPs lying on the Type-2 phase boundary is equal to the order of the long range coupling. We used second order to sixth order long range coupling here. Note that we vary $k$ from $-\pi$ to $\pi$ for simplicity. }
\begin{ruledtabular}
\begin{tabular}{ccccc}
 \vtop{\hbox{\strut Order of long}\hbox{\strut range coupling}} & \vtop{\hbox{\strut Type-1 transition}\hbox{\strut Transition through  \textbf{AB} line}} & \vtop{\hbox{\strut Type-1 transition}\hbox{\strut Transition through  \textbf{EF} line}}&  \vtop{\hbox{\strut Type-2 transition}\hbox{\strut Transition through  \textbf{PQ} line}} & \vtop{\hbox{\strut Type-2 transition}\hbox{\strut Transition through  \textbf{RS} line}}\\
\hline
$m=2$ & $k =0$ & $k=\pi$ & $k= 0,\pi$ & $k=\pm \frac{\pi}{2}$ \\
$m=3$ & $k =0$ & $k=\pi$ & $k= 0,\pm \frac{2 \pi}{3}$ & $k=\pm \frac{\pi}{3}, \pi $ \\
$m=4$ & $k =0$ & $k=\pi$ & $k= 0,\pm \frac{2 \pi}{4}, \pi $ & $k=\pm \frac{\pi}{4}, \pm (\pi -\frac{\pi}{4})$ \\
$m=5$ & $k =0$ & $k=\pi$ & $k= 0,\pm \frac{2 \pi}{5}, \pm (\pi -\frac{\pi}{5})$ & $k=\pm \frac{\pi}{5}, \pm (\pi -\frac{\pi}{5}), \pi $ \\

$m=6$ & $k =0$ & $k=\pi$ & $k= 0,\pm \frac{2 \pi}{6}, \pm (\pi -\frac{2\pi}{6}), \pi $ & $k=\pm \frac{\pi}{6}, \pm (\pi -\frac{\pi}{6}), \pm \frac{\pi}{2}$ \\
\end{tabular}
\end{ruledtabular}
\end{table*}

\section{Proposal for Topological Circuit Realization}

\begin{figure}[htp]
\centering
\includegraphics[width=0.48\textwidth]{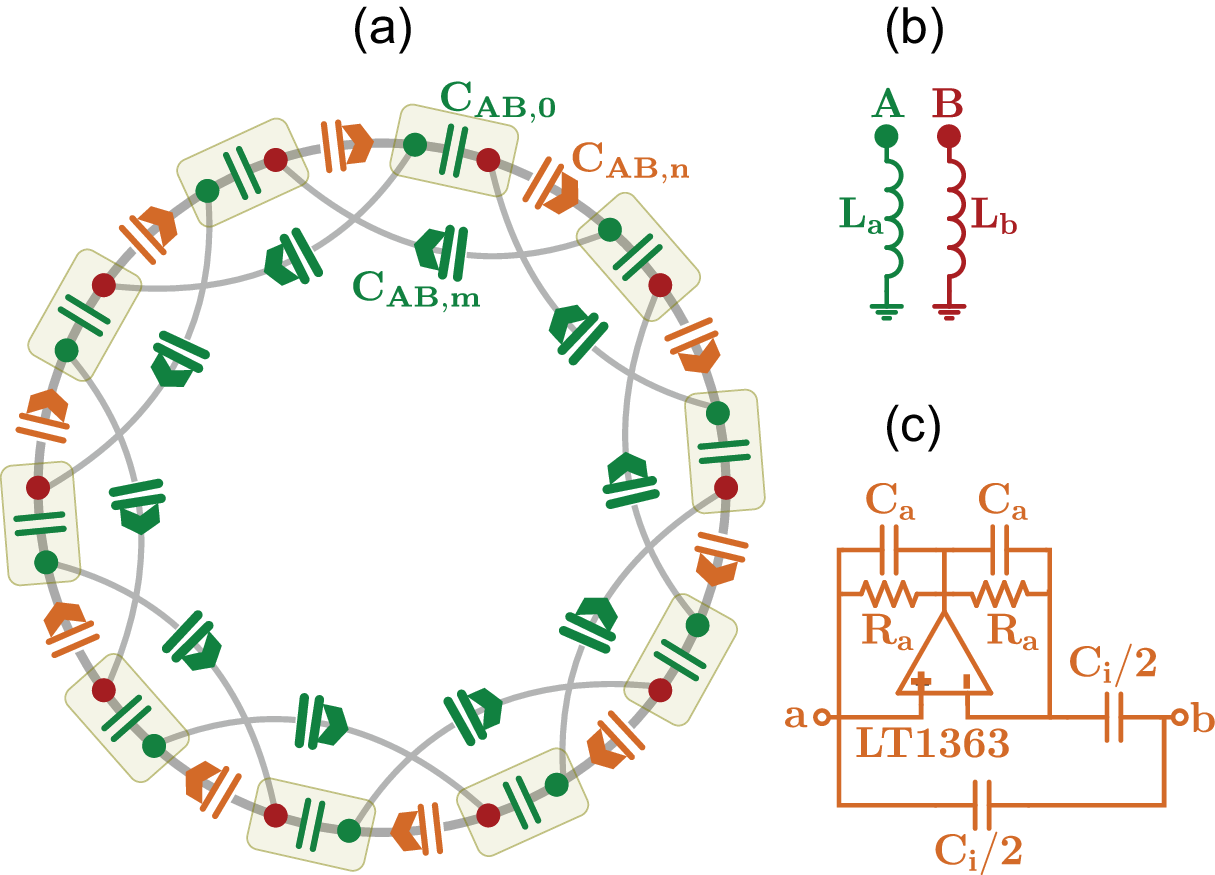}
	\caption{(a) Circuit realization of the non-Hermitian braiding system with long-range coupling in periodic boundary conditions. There are 10 unit cells (one representative unit cell is highlighted with a light yellow rectangle) with A (green dot) and B (red dot) sub-lattice sites. Every node is grounded via an inductor as shown in (b). A and B sub-lattice nodes are connected with a bi-directional capacitor $C_{AB,0}$ within each unit cell. For this demonstration, we selected long-range coupling $m=2$ and $n=1$. So, each A node of the unit cell is connected to the B node of the second unit cell to its left via a unidirectional capacitor $C_{AB,m}$ and similarly, the B node of the cell is connected to the A node of the first unit cell to its right via another unidirectional capacitor $C_{AB,n}$. All these long-range couplings are realized through a uni-directional capacitor designed with the current-inversion negative impedance converters (INIC) as shown in (c). $C_{a}$ and $R_{a}$ are incorporated into the INIC to stabilize the outputs of the LT1363. This will generate unidirectional capacitance of $C_{i}$ from node B to A.} \label{fig_circuitsetup} 
\end{figure}

The intricate energy braiding described by the model in Eq. \ref{H1} can be implemented in a practical electrical circuit using an RLC configuration, as depicted in Fig. \ref{fig_circuitsetup}, for the specific parameter values of $m=2$ and $n=1$ under PBC configuration. For the corresponding OBC setup, the terminal couplings should be replaced with appropriate capacitors to ensure consistent total node admittance across all nodes (refer to Appendix A for detailed information). The behavior of this system is governed by Kirchhoff's law, as expressed in Eq. \eqref{eqn:KCL}.
\begin{equation}
	I_{i}(\omega)=\sum_{j}^ {}J_{ij}(\omega)V_{j}(\omega)\label{eqn:KCL}
\end{equation}

Here, $I_{i}(\omega)$ represents the current flowing into the $i^{th}$ node, $J_{ij}$ denotes the circuit admittance between the $i^{th}$ and $j^{th}$ nodes, and $V_{j}(\omega)$ is the voltage at the $j^{th}$ node. The circuit Laplacian, $J_{ij}(\omega)$, is given by:

\begin{equation}
	J_{ij}(\omega)=i\omega\left[\sum_{m}C_{ij,m}-\sum_{n}\frac{1}{\omega^{2}L_{i,n}}\right]\label{eqn:502}
\end{equation}

Where $C_{ij,m}$ represents $m^{th}$ capacitor connected between node $i$ and $j$, and $L_{i,n}$ are the grounding $n^{th}$ inductor connected to node $i$. Applying Fourier transformation yields the $k-$space representation of the circuit Laplacian, as seen in Eq. \eqref{eq:J_k}:
\begin{widetext} 
	\begin{equation}
	J(k)  =  i\omega\begin{pmatrix}C_{AB,0}+C_{AB,n}-\frac{1}{\omega^{2}L_{a}} & -C_{AB,0}-C_{AB,-m}e^{-i\,m\,k}\\
	-C_{AB,0}-C_{AB,n}e^{i\,n\,k} & C_{AB,0}+C_{AB,-m}-\frac{1}{\omega^{2}L_{b}}
	\end{pmatrix},\label{eq:J_k}
	\end{equation}
\end{widetext} 
Sub-lattice node A and B are grounded through inductors $L_a$ and $L_b$ respectively. $C_{AB,0}$ is a bidirectional capacitor between node A and B, $C_{AB,n}$ is an unidirectional capacitor from node A towards B and $C_{AB,-m}$ is an unidirectional capacitor from node B towards A (Fig. \ref{fig_circuitsetup}). 

To realize any complex braiding phase, the system needs to be driven at the resonant frequency $\omega_{r}=\frac{1}{\sqrt{L_{a}(C_{AB,0}+C_{AB,n})}}=\frac{1}{\sqrt{L_{b}(C_{AB,0}+C_{AB,-m})}}$.

The real-space matrix formation of $J(k)$ for both PBC and OBC can be found in Appendix \ref{sec:Appx_Laplacian}, supplemented by detailed derivations and circuit diagrams. Current-inversion negative impedance converters (INIC) topology, incorporating LT1363 Op-Amps, are employed to realize unidirectional capacitors $C_{AB,m}, C_{AB,n}$ (as shown in Fig \ref{fig_circuitsetup}(c)). To stabilize the Op-Amp output, a small capacitance ($C_{a}$) in the pico Farad range and a low-resistance ($R_{a}$) less than 1 k$\Omega$ are connected in parallel. Since achieving a perfect unidirectional capacitance is challenging in practice, $C_{a}$ and $R_{a}$ are meticulously tuned and tested to attain the best possible unidirectional capacitance for the capacitors used in this study. Detailed results of the INIC optimization can be found in Appendix \ref{sec:Appx_INIC}.

To illustrate the practical implementation of our proposed braiding scheme, four distinct braiding phases—referred to as phases 1 through 4 in Fig \ref{fig4}—have been selected for execution in the electrical circuit set-up. In the circuit implementation of the phases, we dynamically adjust the circuit parameters to demonstrate the versatility and adaptability of our proposed approach. More specifically, as the circuit is exposed to complex admittance eigen values with both positive and negative amplitudes, it is vulnerable to non-causal instabilities\cite{helbig2020generalized}. To prevent the system going unstable, we grounded each of the nodes with additional resistor $R_0$ to shift up the whole spectrum to have complex eigen values with no negative imaginary parts. This shifting, however, has no effect in realizing the braiding topologies this study is concerned about. Moreover, this ensures converged transient circuit simulation results as well. Careful numerical analysis was carried out to get the optimum value of $R_0$ which was chosen as $20 \Omega$. Also, parasitic resistance of $100 m\Omega$ was added to each inductors to mimic more real-world system. Table \ref{tab:Circuit-Parameters} lists all the chosen parameters for the four different phases.

\begin{table}
	\caption{\label{tab:Circuit-Parameters}Circuit Parameters to realize the four different braiding phases}
	\begin{centering}
		\begin{tabular}{|c|c|c|c|c|c|}
			\hline 
			Phase & $C_{AB,0}$ & $C_{AB, m}$ & $C_{AB, n}$ & $L_{a}$ & $L_{b}$\tabularnewline
			\hline 
			\hline 
			1 & 4.7 nF & 0.94 nF & 2 nF & 94.52 $\mu$H & 112.28 $\mu$H\tabularnewline
			\hline 
			2 & 4.7 nF & 0.94 nF & 20 nF & 25.64 $\mu$H & 112.28 $\mu$H \tabularnewline
			\hline 
			3 & 4.7 nF & 20 nF & 2 nF & 94.52 $\mu$H & 25.64 $\mu$H\tabularnewline
			\hline 
			4 & 4.7 nF & 20 nF & 20 nF & 25.64 $\mu$H & 25.64 $\mu$H\tabularnewline
			\hline 
		\end{tabular}
		\par\end{centering}
\end{table}

\begin{figure*}[htp]
	\begin{centering}
		\includegraphics[width=0.9\textwidth]{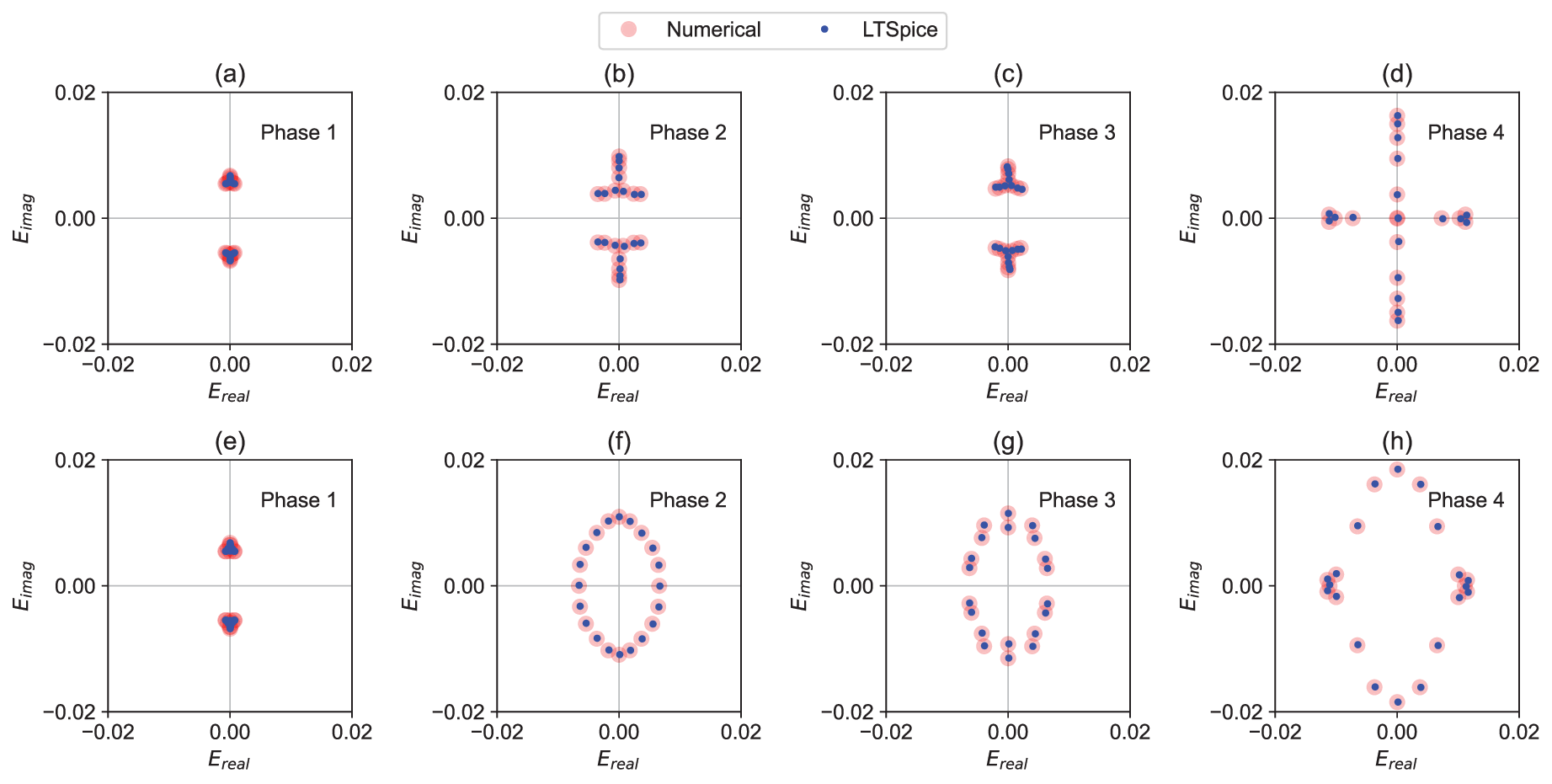}
		\par\end{centering}
	\caption{Admittance energy spectrum -comparison between numerical results and LTSpice simulation of the circuit shown in Fig. \ref{fig_circuitsetup}. Red and blue dots represent numerical and LTSpice simulation results, respectively. (a-d) show the OBC admittance spectrum, while (e-h) show the PBC admittance spectrum. All these results were obtained with 10-unit cells. In LTSpice, we added 500m$\Omega$ ESR(Effective Series Resistance) to each inductor. Other circuit parameters used in realizing different phases are outlined in Table \ref{tab:Circuit-Parameters}.}
	\label{figComparison}
\end{figure*}

Interestingly, the complex energy admittance spectrum can be reconstructed from the circuit by applying current at each node and measuring the node voltages. This allows the construction of the circuit Green's function. Subsequently,  the inversion of this Green's function provides the resulting circuit Laplacian. From the constructed circuit Laplacian, eigenvalues and eigen modes corresponding to the admittance spectrum and node voltage distribution can be calculated. Fig. \ref{figComparison} shows the complex admittance spectrum corresponding to the circuit of Fig. \ref{fig_circuitsetup}. The numerical results (red dots) are compared with LTSpice simulation results (blue dots), revealing that the electrical circuit setup can almost perfectly realize all the distinct phases of the complex braiding system. Although there are slight discrepancies between the numerical results and the circuit simulations, these mainly arise from the inclusion of parasitic resistance in the inductors and the imperfect unidirectional capacitance realized through the INIC, as discussed in detail in Appendices C-E. The necessary precautions for choosing parameters related to the INIC are thoroughly covered in these appendices to ensure the transient analysis of the circuit converges without causing the OpAmps to saturate. Furthermore, the effects of parasitics and disorder on the braiding configurations are discussed in Appendix D.



\section{Conclusion}
In conclusion, we present a comprehensive analysis of complex energy braiding in 1D non-Hermitian systems with a general $m$th order long-range asymmetrical coupling. Our work provides a new perspective on the emergence of novel topological phenomena induced by long-range coupling beyond the conventional nearest-neighbor interaction. By varying the coupling strengths, we demonstrate that the modified SSH model displays $m$ distinct knots and links topologies in the complex energy-momentum space under periodic boundary conditions (PBC). These topologies can be characterized by a new topological invariant, the braiding index. We establish a correlation between the braiding index and the zeros and poles of the characteristic polynomials,  providing a novel insight to the topology of complex energy braiding.

Additionally, our work shows a deep connection between the braiding topology and eigen state localization due to  the non-Hermitian skin effect (NHSE). We find that the NHSE can be localized at one or both terminals of the system, signifying the conventional or bipolar skin localization respectively, depending on the sign of the braiding index. Furthermore, we find that different complex energy braiding configurations with the same sign (opposite sign) can be transformed continuously through Type-1 (Type-2) phase transitions which traverse through single (multiple) exceptional points (EPs). 

Finally, we propose a realistic experimental setup for realizing various braiding schemes in an RLC circuit. This  circuit-based platform to realize braiding topology is readily implementable using basic circuit components and avoids the requirement for high-dimensional momentum space unlike most other proposed platforms \cite{wang2021topological,li2022topological}. The proposed electrical circuit setup provides a promising platform for exploring the rich physics of complex energy braiding in non-Hermitian systems with non-reciprocal long-range coupling.


\subsubsection*{Acknowledgments}
This work is supported by the Ministry of Education (MOE) Tier-II Grant MOE-T2EP50121-0014 (NUS Grant No. A-8000086-01-00), and MOE Tier-I FRC Grant (NUS Grant No. A-8000195-01-00).

\nocite{*}


\begin{thebibliography}{92}%
\makeatletter
\providecommand \@ifxundefined [1]{%
 \@ifx{#1\undefined}
}%
\providecommand \@ifnum [1]{%
 \ifnum #1\expandafter \@firstoftwo
 \else \expandafter \@secondoftwo
 \fi
}%
\providecommand \@ifx [1]{%
 \ifx #1\expandafter \@firstoftwo
 \else \expandafter \@secondoftwo
 \fi
}%
\providecommand \natexlab [1]{#1}%
\providecommand \enquote  [1]{``#1''}%
\providecommand \bibnamefont  [1]{#1}%
\providecommand \bibfnamefont [1]{#1}%
\providecommand \citenamefont [1]{#1}%
\providecommand \href@noop [0]{\@secondoftwo}%
\providecommand \href [0]{\begingroup \@sanitize@url \@href}%
\providecommand \@href[1]{\@@startlink{#1}\@@href}%
\providecommand \@@href[1]{\endgroup#1\@@endlink}%
\providecommand \@sanitize@url [0]{\catcode `\\12\catcode `\$12\catcode
  `\&12\catcode `\#12\catcode `\^12\catcode `\_12\catcode `\%12\relax}%
\providecommand \@@startlink[1]{}%
\providecommand \@@endlink[0]{}%
\providecommand \url  [0]{\begingroup\@sanitize@url \@url }%
\providecommand \@url [1]{\endgroup\@href {#1}{\urlprefix }}%
\providecommand \urlprefix  [0]{URL }%
\providecommand \Eprint [0]{\href }%
\providecommand \doibase [0]{https://doi.org/}%
\providecommand \selectlanguage [0]{\@gobble}%
\providecommand \bibinfo  [0]{\@secondoftwo}%
\providecommand \bibfield  [0]{\@secondoftwo}%
\providecommand \translation [1]{[#1]}%
\providecommand \BibitemOpen [0]{}%
\providecommand \bibitemStop [0]{}%
\providecommand \bibitemNoStop [0]{.\EOS\space}%
\providecommand \EOS [0]{\spacefactor3000\relax}%
\providecommand \BibitemShut  [1]{\csname bibitem#1\endcsname}%
\let\auto@bib@innerbib\@empty
\bibitem [{\citenamefont {Ashida}\ \emph {et~al.}(2020)\citenamefont {Ashida},
  \citenamefont {Gong},\ and\ \citenamefont {Ueda}}]{ashida2020non}%
  \BibitemOpen
  \bibfield  {author} {\bibinfo {author} {\bibfnamefont {Y.}~\bibnamefont
  {Ashida}}, \bibinfo {author} {\bibfnamefont {Z.}~\bibnamefont {Gong}},\ and\
  \bibinfo {author} {\bibfnamefont {M.}~\bibnamefont {Ueda}},\ }\bibfield
  {title} {\bibinfo {title} {Non-hermitian physics},\ }\href@noop {} {\bibfield
   {journal} {\bibinfo  {journal} {Adv. Phys.}\ }\textbf {\bibinfo {volume}
  {69}},\ \bibinfo {pages} {249} (\bibinfo {year} {2020})}\BibitemShut
  {NoStop}%
\bibitem [{\citenamefont {Bergholtz}\ \emph {et~al.}(2021)\citenamefont
  {Bergholtz}, \citenamefont {Budich},\ and\ \citenamefont
  {Kunst}}]{bergholtz2021exceptional}%
  \BibitemOpen
  \bibfield  {author} {\bibinfo {author} {\bibfnamefont {E.~J.}\ \bibnamefont
  {Bergholtz}}, \bibinfo {author} {\bibfnamefont {J.~C.}\ \bibnamefont
  {Budich}},\ and\ \bibinfo {author} {\bibfnamefont {F.~K.}\ \bibnamefont
  {Kunst}},\ }\bibfield  {title} {\bibinfo {title} {Exceptional topology of
  non-hermitian systems},\ }\href@noop {} {\bibfield  {journal} {\bibinfo
  {journal} {Rev. Mod. Phys.}\ }\textbf {\bibinfo {volume} {93}},\ \bibinfo
  {pages} {015005} (\bibinfo {year} {2021})}\BibitemShut {NoStop}%
\bibitem [{\citenamefont {Gong}\ \emph
  {et~al.}(2018{\natexlab{a}})\citenamefont {Gong}, \citenamefont {Ashida},
  \citenamefont {Kawabata}, \citenamefont {Takasan}, \citenamefont
  {Higashikawa},\ and\ \citenamefont {Ueda}}]{gong2018topological}%
  \BibitemOpen
  \bibfield  {author} {\bibinfo {author} {\bibfnamefont {Z.}~\bibnamefont
  {Gong}}, \bibinfo {author} {\bibfnamefont {Y.}~\bibnamefont {Ashida}},
  \bibinfo {author} {\bibfnamefont {K.}~\bibnamefont {Kawabata}}, \bibinfo
  {author} {\bibfnamefont {K.}~\bibnamefont {Takasan}}, \bibinfo {author}
  {\bibfnamefont {S.}~\bibnamefont {Higashikawa}},\ and\ \bibinfo {author}
  {\bibfnamefont {M.}~\bibnamefont {Ueda}},\ }\bibfield  {title} {\bibinfo
  {title} {Topological phases of non-hermitian systems},\ }\href@noop {}
  {\bibfield  {journal} {\bibinfo  {journal} {Phys. Rev. X}\ }\textbf {\bibinfo
  {volume} {8}},\ \bibinfo {pages} {031079} (\bibinfo {year}
  {2018}{\natexlab{a}})}\BibitemShut {NoStop}%
\bibitem [{\citenamefont {Leykam}\ \emph {et~al.}(2017)\citenamefont {Leykam},
  \citenamefont {Bliokh}, \citenamefont {Huang}, \citenamefont {Chong},\ and\
  \citenamefont {Nori}}]{leykam2017edge}%
  \BibitemOpen
  \bibfield  {author} {\bibinfo {author} {\bibfnamefont {D.}~\bibnamefont
  {Leykam}}, \bibinfo {author} {\bibfnamefont {K.~Y.}\ \bibnamefont {Bliokh}},
  \bibinfo {author} {\bibfnamefont {C.}~\bibnamefont {Huang}}, \bibinfo
  {author} {\bibfnamefont {Y.~D.}\ \bibnamefont {Chong}},\ and\ \bibinfo
  {author} {\bibfnamefont {F.}~\bibnamefont {Nori}},\ }\bibfield  {title}
  {\bibinfo {title} {Edge modes, degeneracies, and topological numbers in
  non-hermitian systems},\ }\href@noop {} {\bibfield  {journal} {\bibinfo
  {journal} {Phys. Rev. lett.}\ }\textbf {\bibinfo {volume} {118}},\ \bibinfo
  {pages} {040401} (\bibinfo {year} {2017})}\BibitemShut {NoStop}%
\bibitem [{\citenamefont {Rafi-Ul-Islam}\ \emph
  {et~al.}(2022{\natexlab{a}})\citenamefont {Rafi-Ul-Islam}, \citenamefont
  {Siu}, \citenamefont {Sahin}, \citenamefont {Lee},\ and\ \citenamefont
  {Jalil}}]{rafi2022critical}%
  \BibitemOpen
  \bibfield  {author} {\bibinfo {author} {\bibfnamefont {S.~M.}\ \bibnamefont
  {Rafi-Ul-Islam}}, \bibinfo {author} {\bibfnamefont {Z.~B.}\ \bibnamefont
  {Siu}}, \bibinfo {author} {\bibfnamefont {H.}~\bibnamefont {Sahin}}, \bibinfo
  {author} {\bibfnamefont {C.~H.}\ \bibnamefont {Lee}},\ and\ \bibinfo {author}
  {\bibfnamefont {M.~B.~A.}\ \bibnamefont {Jalil}},\ }\bibfield  {title}
  {\bibinfo {title} {Critical hybridization of skin modes in coupled
  non-hermitian chains},\ }\href@noop {} {\bibfield  {journal} {\bibinfo
  {journal} {Phys. Rev. Res.}\ }\textbf {\bibinfo {volume} {4}},\ \bibinfo
  {pages} {013243} (\bibinfo {year} {2022}{\natexlab{a}})}\BibitemShut
  {NoStop}%
\bibitem [{\citenamefont {Yokomizo}\ and\ \citenamefont
  {Murakami}(2019)}]{yokomizo2019non}%
  \BibitemOpen
  \bibfield  {author} {\bibinfo {author} {\bibfnamefont {K.}~\bibnamefont
  {Yokomizo}}\ and\ \bibinfo {author} {\bibfnamefont {S.}~\bibnamefont
  {Murakami}},\ }\bibfield  {title} {\bibinfo {title} {Non-bloch band theory of
  non-hermitian systems},\ }\href@noop {} {\bibfield  {journal} {\bibinfo
  {journal} {Phys. Rev. lett.}\ }\textbf {\bibinfo {volume} {123}},\ \bibinfo
  {pages} {066404} (\bibinfo {year} {2019})}\BibitemShut {NoStop}%
\bibitem [{\citenamefont {El-Ganainy}\ \emph {et~al.}(2019)\citenamefont
  {El-Ganainy}, \citenamefont {Khajavikhan}, \citenamefont {Christodoulides},\
  and\ \citenamefont {Ozdemir}}]{el2019dawn}%
  \BibitemOpen
  \bibfield  {author} {\bibinfo {author} {\bibfnamefont {R.}~\bibnamefont
  {El-Ganainy}}, \bibinfo {author} {\bibfnamefont {M.}~\bibnamefont
  {Khajavikhan}}, \bibinfo {author} {\bibfnamefont {D.~N.}\ \bibnamefont
  {Christodoulides}},\ and\ \bibinfo {author} {\bibfnamefont {S.~K.}\
  \bibnamefont {Ozdemir}},\ }\bibfield  {title} {\bibinfo {title} {The dawn of
  non-hermitian optics},\ }\href@noop {} {\bibfield  {journal} {\bibinfo
  {journal} {Commun. Phys.}\ }\textbf {\bibinfo {volume} {2}},\ \bibinfo
  {pages} {37} (\bibinfo {year} {2019})}\BibitemShut {NoStop}%
\bibitem [{\citenamefont {Cao}\ \emph {et~al.}(2020)\citenamefont {Cao},
  \citenamefont {Lu}, \citenamefont {Meng}, \citenamefont {Sun}, \citenamefont
  {Shen},\ and\ \citenamefont {Xiao}}]{cao2020reservoir}%
  \BibitemOpen
  \bibfield  {author} {\bibinfo {author} {\bibfnamefont {W.}~\bibnamefont
  {Cao}}, \bibinfo {author} {\bibfnamefont {X.}~\bibnamefont {Lu}}, \bibinfo
  {author} {\bibfnamefont {X.}~\bibnamefont {Meng}}, \bibinfo {author}
  {\bibfnamefont {J.}~\bibnamefont {Sun}}, \bibinfo {author} {\bibfnamefont
  {H.}~\bibnamefont {Shen}},\ and\ \bibinfo {author} {\bibfnamefont
  {Y.}~\bibnamefont {Xiao}},\ }\bibfield  {title} {\bibinfo {title}
  {Reservoir-mediated quantum correlations in non-hermitian optical system},\
  }\href@noop {} {\bibfield  {journal} {\bibinfo  {journal} {Phys. Rev. Lett.}\
  }\textbf {\bibinfo {volume} {124}},\ \bibinfo {pages} {030401} (\bibinfo
  {year} {2020})}\BibitemShut {NoStop}%
\bibitem [{\citenamefont {Zhang}\ \emph {et~al.}(2018)\citenamefont {Zhang},
  \citenamefont {Ma}, \citenamefont {Sheng}, \citenamefont {Zhang},
  \citenamefont {Zhang},\ and\ \citenamefont {Xiao}}]{zhang2018non}%
  \BibitemOpen
  \bibfield  {author} {\bibinfo {author} {\bibfnamefont {Z.}~\bibnamefont
  {Zhang}}, \bibinfo {author} {\bibfnamefont {D.}~\bibnamefont {Ma}}, \bibinfo
  {author} {\bibfnamefont {J.}~\bibnamefont {Sheng}}, \bibinfo {author}
  {\bibfnamefont {Y.}~\bibnamefont {Zhang}}, \bibinfo {author} {\bibfnamefont
  {Y.}~\bibnamefont {Zhang}},\ and\ \bibinfo {author} {\bibfnamefont
  {M.}~\bibnamefont {Xiao}},\ }\bibfield  {title} {\bibinfo {title}
  {Non-hermitian optics in atomic systems},\ }\href@noop {} {\bibfield
  {journal} {\bibinfo  {journal} {J. Phys. B: Atom, Mol. Opt. Phys.}\ }\textbf
  {\bibinfo {volume} {51}},\ \bibinfo {pages} {072001} (\bibinfo {year}
  {2018})}\BibitemShut {NoStop}%
\bibitem [{\citenamefont {Feng}\ \emph {et~al.}(2017)\citenamefont {Feng},
  \citenamefont {El-Ganainy},\ and\ \citenamefont {Ge}}]{feng2017non}%
  \BibitemOpen
  \bibfield  {author} {\bibinfo {author} {\bibfnamefont {L.}~\bibnamefont
  {Feng}}, \bibinfo {author} {\bibfnamefont {R.}~\bibnamefont {El-Ganainy}},\
  and\ \bibinfo {author} {\bibfnamefont {L.}~\bibnamefont {Ge}},\ }\bibfield
  {title} {\bibinfo {title} {Non-hermitian photonics based on parity--time
  symmetry},\ }\href@noop {} {\bibfield  {journal} {\bibinfo  {journal} {Nat.
  Photon.}\ }\textbf {\bibinfo {volume} {11}},\ \bibinfo {pages} {752}
  (\bibinfo {year} {2017})}\BibitemShut {NoStop}%
\bibitem [{\citenamefont {Pan}\ \emph {et~al.}(2018)\citenamefont {Pan},
  \citenamefont {Zhao}, \citenamefont {Miao}, \citenamefont {Longhi},\ and\
  \citenamefont {Feng}}]{pan2018photonic}%
  \BibitemOpen
  \bibfield  {author} {\bibinfo {author} {\bibfnamefont {M.}~\bibnamefont
  {Pan}}, \bibinfo {author} {\bibfnamefont {H.}~\bibnamefont {Zhao}}, \bibinfo
  {author} {\bibfnamefont {P.}~\bibnamefont {Miao}}, \bibinfo {author}
  {\bibfnamefont {S.}~\bibnamefont {Longhi}},\ and\ \bibinfo {author}
  {\bibfnamefont {L.}~\bibnamefont {Feng}},\ }\bibfield  {title} {\bibinfo
  {title} {Photonic zero mode in a non-hermitian photonic lattice},\
  }\href@noop {} {\bibfield  {journal} {\bibinfo  {journal} {Nat. Commun.}\
  }\textbf {\bibinfo {volume} {9}},\ \bibinfo {pages} {1308} (\bibinfo {year}
  {2018})}\BibitemShut {NoStop}%
\bibitem [{\citenamefont {Longhi}\ \emph {et~al.}(2015)\citenamefont {Longhi},
  \citenamefont {Gatti},\ and\ \citenamefont {Valle}}]{longhi2015robust}%
  \BibitemOpen
  \bibfield  {author} {\bibinfo {author} {\bibfnamefont {S.}~\bibnamefont
  {Longhi}}, \bibinfo {author} {\bibfnamefont {D.}~\bibnamefont {Gatti}},\ and\
  \bibinfo {author} {\bibfnamefont {G.~D.}\ \bibnamefont {Valle}},\ }\bibfield
  {title} {\bibinfo {title} {Robust light transport in non-hermitian photonic
  lattices},\ }\href@noop {} {\bibfield  {journal} {\bibinfo  {journal} {Sci.
  Rep.}\ }\textbf {\bibinfo {volume} {5}},\ \bibinfo {pages} {1} (\bibinfo
  {year} {2015})}\BibitemShut {NoStop}%
\bibitem [{\citenamefont {Midya}\ \emph {et~al.}(2018)\citenamefont {Midya},
  \citenamefont {Zhao},\ and\ \citenamefont {Feng}}]{midya2018non}%
  \BibitemOpen
  \bibfield  {author} {\bibinfo {author} {\bibfnamefont {B.}~\bibnamefont
  {Midya}}, \bibinfo {author} {\bibfnamefont {H.}~\bibnamefont {Zhao}},\ and\
  \bibinfo {author} {\bibfnamefont {L.}~\bibnamefont {Feng}},\ }\bibfield
  {title} {\bibinfo {title} {Non-hermitian photonics promises exceptional
  topology of light},\ }\href@noop {} {\bibfield  {journal} {\bibinfo
  {journal} {Nat. Commun.}\ }\textbf {\bibinfo {volume} {9}},\ \bibinfo {pages}
  {2674} (\bibinfo {year} {2018})}\BibitemShut {NoStop}%
\bibitem [{\citenamefont {Ghatak}\ \emph {et~al.}(2020)\citenamefont {Ghatak},
  \citenamefont {Brandenbourger}, \citenamefont {Van~Wezel},\ and\
  \citenamefont {Coulais}}]{ghatak2020observation}%
  \BibitemOpen
  \bibfield  {author} {\bibinfo {author} {\bibfnamefont {A.}~\bibnamefont
  {Ghatak}}, \bibinfo {author} {\bibfnamefont {M.}~\bibnamefont
  {Brandenbourger}}, \bibinfo {author} {\bibfnamefont {J.}~\bibnamefont
  {Van~Wezel}},\ and\ \bibinfo {author} {\bibfnamefont {C.}~\bibnamefont
  {Coulais}},\ }\bibfield  {title} {\bibinfo {title} {Observation of
  non-hermitian topology and its bulk--edge correspondence in an active
  mechanical metamaterial},\ }\href@noop {} {\bibfield  {journal} {\bibinfo
  {journal} {Proc. Nat. Aca. Sci.}\ }\textbf {\bibinfo {volume} {117}},\
  \bibinfo {pages} {29561} (\bibinfo {year} {2020})}\BibitemShut {NoStop}%
\bibitem [{\citenamefont {Hou}\ \emph {et~al.}(2020)\citenamefont {Hou},
  \citenamefont {Li}, \citenamefont {Luo}, \citenamefont {Gu},\ and\
  \citenamefont {Zhang}}]{hou2020topological}%
  \BibitemOpen
  \bibfield  {author} {\bibinfo {author} {\bibfnamefont {J.}~\bibnamefont
  {Hou}}, \bibinfo {author} {\bibfnamefont {Z.}~\bibnamefont {Li}}, \bibinfo
  {author} {\bibfnamefont {X.-W.}\ \bibnamefont {Luo}}, \bibinfo {author}
  {\bibfnamefont {Q.}~\bibnamefont {Gu}},\ and\ \bibinfo {author}
  {\bibfnamefont {C.}~\bibnamefont {Zhang}},\ }\bibfield  {title} {\bibinfo
  {title} {Topological bands and triply degenerate points in non-hermitian
  hyperbolic metamaterials},\ }\href@noop {} {\bibfield  {journal} {\bibinfo
  {journal} {Phys. Rev. Lett.}\ }\textbf {\bibinfo {volume} {124}},\ \bibinfo
  {pages} {073603} (\bibinfo {year} {2020})}\BibitemShut {NoStop}%
\bibitem [{\citenamefont {Zhou}\ and\ \citenamefont
  {Zhang}(2020)}]{zhou2020non}%
  \BibitemOpen
  \bibfield  {author} {\bibinfo {author} {\bibfnamefont {D.}~\bibnamefont
  {Zhou}}\ and\ \bibinfo {author} {\bibfnamefont {J.}~\bibnamefont {Zhang}},\
  }\bibfield  {title} {\bibinfo {title} {Non-hermitian topological
  metamaterials with odd elasticity},\ }\href@noop {} {\bibfield  {journal}
  {\bibinfo  {journal} {Phys. Rev. Res.}\ }\textbf {\bibinfo {volume} {2}},\
  \bibinfo {pages} {023173} (\bibinfo {year} {2020})}\BibitemShut {NoStop}%
\bibitem [{\citenamefont {Rosenthal}\ \emph {et~al.}(2018)\citenamefont
  {Rosenthal}, \citenamefont {Ehrlich}, \citenamefont {Rudner}, \citenamefont
  {Higginbotham},\ and\ \citenamefont {Lehnert}}]{rosenthal2018topological}%
  \BibitemOpen
  \bibfield  {author} {\bibinfo {author} {\bibfnamefont {E.~I.}\ \bibnamefont
  {Rosenthal}}, \bibinfo {author} {\bibfnamefont {N.~K.}\ \bibnamefont
  {Ehrlich}}, \bibinfo {author} {\bibfnamefont {M.~S.}\ \bibnamefont {Rudner}},
  \bibinfo {author} {\bibfnamefont {A.~P.}\ \bibnamefont {Higginbotham}},\ and\
  \bibinfo {author} {\bibfnamefont {K.~W.}\ \bibnamefont {Lehnert}},\
  }\bibfield  {title} {\bibinfo {title} {Topological phase transition measured
  in a dissipative metamaterial},\ }\href@noop {} {\bibfield  {journal}
  {\bibinfo  {journal} {Phys. Rev. B}\ }\textbf {\bibinfo {volume} {97}},\
  \bibinfo {pages} {220301} (\bibinfo {year} {2018})}\BibitemShut {NoStop}%
\bibitem [{\citenamefont {Fring}\ and\ \citenamefont
  {Taira}(2020{\natexlab{a}})}]{fring2020pseudo}%
  \BibitemOpen
  \bibfield  {author} {\bibinfo {author} {\bibfnamefont {A.}~\bibnamefont
  {Fring}}\ and\ \bibinfo {author} {\bibfnamefont {T.}~\bibnamefont {Taira}},\
  }\bibfield  {title} {\bibinfo {title} {Pseudo-hermitian approach to
  goldstone’s theorem in non-abelian non-hermitian quantum field theories},\
  }\href@noop {} {\bibfield  {journal} {\bibinfo  {journal} {Phys. Rev. D}\
  }\textbf {\bibinfo {volume} {101}},\ \bibinfo {pages} {045014} (\bibinfo
  {year} {2020}{\natexlab{a}})}\BibitemShut {NoStop}%
\bibitem [{\citenamefont {Moiseyev}(2011)}]{moiseyev2011non}%
  \BibitemOpen
  \bibfield  {author} {\bibinfo {author} {\bibfnamefont {N.}~\bibnamefont
  {Moiseyev}},\ }\href@noop {} {\emph {\bibinfo {title} {Non-Hermitian quantum
  mechanics}}}\ (\bibinfo  {publisher} {Cambridge University Press},\ \bibinfo
  {year} {2011})\BibitemShut {NoStop}%
\bibitem [{\citenamefont {Fring}\ and\ \citenamefont
  {Taira}(2020{\natexlab{b}})}]{fring2020goldstone}%
  \BibitemOpen
  \bibfield  {author} {\bibinfo {author} {\bibfnamefont {A.}~\bibnamefont
  {Fring}}\ and\ \bibinfo {author} {\bibfnamefont {T.}~\bibnamefont {Taira}},\
  }\bibfield  {title} {\bibinfo {title} {Goldstone bosons in different
  pt-regimes of non-hermitian scalar quantum field theories},\ }\href@noop {}
  {\bibfield  {journal} {\bibinfo  {journal} {Nucl. Phys. B}\ }\textbf
  {\bibinfo {volume} {950}},\ \bibinfo {pages} {114834} (\bibinfo {year}
  {2020}{\natexlab{b}})}\BibitemShut {NoStop}%
\bibitem [{\citenamefont {Rafi-Ul-Islam}\ \emph
  {et~al.}(2021{\natexlab{a}})\citenamefont {Rafi-Ul-Islam}, \citenamefont
  {Siu},\ and\ \citenamefont {Jalil}}]{rafi2021non}%
  \BibitemOpen
  \bibfield  {author} {\bibinfo {author} {\bibfnamefont {S.~M.}\ \bibnamefont
  {Rafi-Ul-Islam}}, \bibinfo {author} {\bibfnamefont {Z.~B.}\ \bibnamefont
  {Siu}},\ and\ \bibinfo {author} {\bibfnamefont {M.~B.~A.}\ \bibnamefont
  {Jalil}},\ }\bibfield  {title} {\bibinfo {title} {Non-hermitian topological
  phases and exceptional lines in topolectrical circuits},\ }\href@noop {}
  {\bibfield  {journal} {\bibinfo  {journal} {New J. Phys.}\ }\textbf {\bibinfo
  {volume} {23}},\ \bibinfo {pages} {033014} (\bibinfo {year}
  {2021}{\natexlab{a}})}\BibitemShut {NoStop}%
\bibitem [{\citenamefont {Hofmann}\ \emph {et~al.}(2020)\citenamefont
  {Hofmann}, \citenamefont {Helbig}, \citenamefont {Schindler}, \citenamefont
  {Salgo}, \citenamefont {Brzezi{\'n}ska}, \citenamefont {Greiter},
  \citenamefont {Kiessling}, \citenamefont {Wolf}, \citenamefont {Vollhardt},
  \citenamefont {Kaba{\v{s}}i} \emph {et~al.}}]{hofmann2020reciprocal}%
  \BibitemOpen
  \bibfield  {author} {\bibinfo {author} {\bibfnamefont {T.}~\bibnamefont
  {Hofmann}}, \bibinfo {author} {\bibfnamefont {T.}~\bibnamefont {Helbig}},
  \bibinfo {author} {\bibfnamefont {F.}~\bibnamefont {Schindler}}, \bibinfo
  {author} {\bibfnamefont {N.}~\bibnamefont {Salgo}}, \bibinfo {author}
  {\bibfnamefont {M.}~\bibnamefont {Brzezi{\'n}ska}}, \bibinfo {author}
  {\bibfnamefont {M.}~\bibnamefont {Greiter}}, \bibinfo {author} {\bibfnamefont
  {T.}~\bibnamefont {Kiessling}}, \bibinfo {author} {\bibfnamefont
  {D.}~\bibnamefont {Wolf}}, \bibinfo {author} {\bibfnamefont {A.}~\bibnamefont
  {Vollhardt}}, \bibinfo {author} {\bibfnamefont {A.}~\bibnamefont
  {Kaba{\v{s}}i}}, \emph {et~al.},\ }\bibfield  {title} {\bibinfo {title}
  {Reciprocal skin effect and its realization in a topolectrical circuit},\
  }\href@noop {} {\bibfield  {journal} {\bibinfo  {journal} {Phys. Rev. Res.}\
  }\textbf {\bibinfo {volume} {2}},\ \bibinfo {pages} {023265} (\bibinfo {year}
  {2020})}\BibitemShut {NoStop}%
\bibitem [{\citenamefont {Zhang}\ \emph
  {et~al.}(2023{\natexlab{a}})\citenamefont {Zhang}, \citenamefont {Zhang},
  \citenamefont {Sahin}, \citenamefont {Siu}, \citenamefont {Rafi-Ul-Islam},
  \citenamefont {Kong}, \citenamefont {Shen}, \citenamefont {Jalil},
  \citenamefont {Thomale},\ and\ \citenamefont {Lee}}]{zhang2023anomalous}%
  \BibitemOpen
  \bibfield  {author} {\bibinfo {author} {\bibfnamefont {X.}~\bibnamefont
  {Zhang}}, \bibinfo {author} {\bibfnamefont {B.}~\bibnamefont {Zhang}},
  \bibinfo {author} {\bibfnamefont {H.}~\bibnamefont {Sahin}}, \bibinfo
  {author} {\bibfnamefont {Z.~B.}\ \bibnamefont {Siu}}, \bibinfo {author}
  {\bibfnamefont {S.~M.}\ \bibnamefont {Rafi-Ul-Islam}}, \bibinfo {author}
  {\bibfnamefont {J.~F.}\ \bibnamefont {Kong}}, \bibinfo {author}
  {\bibfnamefont {B.}~\bibnamefont {Shen}}, \bibinfo {author} {\bibfnamefont
  {M.~B.~A.}\ \bibnamefont {Jalil}}, \bibinfo {author} {\bibfnamefont
  {R.}~\bibnamefont {Thomale}},\ and\ \bibinfo {author} {\bibfnamefont {C.~H.}\
  \bibnamefont {Lee}},\ }\bibfield  {title} {\bibinfo {title} {Anomalous
  fractal scaling in two-dimensional electric networks},\ }\href@noop {}
  {\bibfield  {journal} {\bibinfo  {journal} {Commun. Phys.}\ }\textbf
  {\bibinfo {volume} {6}},\ \bibinfo {pages} {151} (\bibinfo {year}
  {2023}{\natexlab{a}})}\BibitemShut {NoStop}%
\bibitem [{\citenamefont {Sahin}\ \emph {et~al.}(2023)\citenamefont {Sahin},
  \citenamefont {Siu}, \citenamefont {Rafi-Ul-Islam}, \citenamefont {Kong},
  \citenamefont {Jalil},\ and\ \citenamefont {Lee}}]{sahin2023impedance}%
  \BibitemOpen
  \bibfield  {author} {\bibinfo {author} {\bibfnamefont {H.}~\bibnamefont
  {Sahin}}, \bibinfo {author} {\bibfnamefont {Z.~B.}\ \bibnamefont {Siu}},
  \bibinfo {author} {\bibfnamefont {S.~M.}\ \bibnamefont {Rafi-Ul-Islam}},
  \bibinfo {author} {\bibfnamefont {J.~F.}\ \bibnamefont {Kong}}, \bibinfo
  {author} {\bibfnamefont {M.~B.~A.}\ \bibnamefont {Jalil}},\ and\ \bibinfo
  {author} {\bibfnamefont {C.~H.}\ \bibnamefont {Lee}},\ }\bibfield  {title}
  {\bibinfo {title} {Impedance responses and size-dependent resonances in
  topolectrical circuits via the method of images},\ }\href@noop {} {\bibfield
  {journal} {\bibinfo  {journal} {Phys. Rev. B}\ }\textbf {\bibinfo {volume}
  {107}},\ \bibinfo {pages} {245114} (\bibinfo {year} {2023})}\BibitemShut
  {NoStop}%
\bibitem [{\citenamefont {Rafi-Ul-Islam}\ \emph
  {et~al.}(2020{\natexlab{a}})\citenamefont {Rafi-Ul-Islam}, \citenamefont
  {Bin~Siu},\ and\ \citenamefont {Jalil}}]{rafi2020topoelectrical}%
  \BibitemOpen
  \bibfield  {author} {\bibinfo {author} {\bibfnamefont {S.~M.}\ \bibnamefont
  {Rafi-Ul-Islam}}, \bibinfo {author} {\bibfnamefont {Z.}~\bibnamefont
  {Bin~Siu}},\ and\ \bibinfo {author} {\bibfnamefont {M.~B.~A.}\ \bibnamefont
  {Jalil}},\ }\bibfield  {title} {\bibinfo {title} {Topoelectrical circuit
  realization of a weyl semimetal heterojunction},\ }\href@noop {} {\bibfield
  {journal} {\bibinfo  {journal} {Commun. Phys.}\ }\textbf {\bibinfo {volume}
  {3}},\ \bibinfo {pages} {72} (\bibinfo {year}
  {2020}{\natexlab{a}})}\BibitemShut {NoStop}%
\bibitem [{\citenamefont {Rafi-Ul-Islam}\ \emph
  {et~al.}(2020{\natexlab{b}})\citenamefont {Rafi-Ul-Islam}, \citenamefont
  {Siu}, \citenamefont {Sun},\ and\ \citenamefont
  {Jalil}}]{rafi2020realization}%
  \BibitemOpen
  \bibfield  {author} {\bibinfo {author} {\bibfnamefont {S.~M.}\ \bibnamefont
  {Rafi-Ul-Islam}}, \bibinfo {author} {\bibfnamefont {Z.~B.}\ \bibnamefont
  {Siu}}, \bibinfo {author} {\bibfnamefont {C.}~\bibnamefont {Sun}},\ and\
  \bibinfo {author} {\bibfnamefont {M.~B.~A.}\ \bibnamefont {Jalil}},\
  }\bibfield  {title} {\bibinfo {title} {Realization of weyl semimetal phases
  in topoelectrical circuits},\ }\href@noop {} {\bibfield  {journal} {\bibinfo
  {journal} {New J. Phys.}\ }\textbf {\bibinfo {volume} {22}},\ \bibinfo
  {pages} {023025} (\bibinfo {year} {2020}{\natexlab{b}})}\BibitemShut
  {NoStop}%
\bibitem [{\citenamefont {Rafi-Ul-Islam}\ \emph
  {et~al.}(2020{\natexlab{c}})\citenamefont {Rafi-Ul-Islam}, \citenamefont
  {Siu},\ and\ \citenamefont {Jalil}}]{rafi2020anti}%
  \BibitemOpen
  \bibfield  {author} {\bibinfo {author} {\bibfnamefont {S.~M.}\ \bibnamefont
  {Rafi-Ul-Islam}}, \bibinfo {author} {\bibfnamefont {Z.~B.}\ \bibnamefont
  {Siu}},\ and\ \bibinfo {author} {\bibfnamefont {M.~B.~A.}\ \bibnamefont
  {Jalil}},\ }\bibfield  {title} {\bibinfo {title} {Anti-klein tunneling in
  topoelectrical weyl semimetal circuits},\ }\href@noop {} {\bibfield
  {journal} {\bibinfo  {journal} {Appl. Phys. Lett.}\ }\textbf {\bibinfo
  {volume} {116}} (\bibinfo {year} {2020}{\natexlab{c}})}\BibitemShut {NoStop}%
\bibitem [{\citenamefont {Rafi-Ul-Islam}\ \emph
  {et~al.}(2023{\natexlab{a}})\citenamefont {Rafi-Ul-Islam}, \citenamefont
  {Siu}, \citenamefont {Sahin},\ and\ \citenamefont {Jalil}}]{rafi2023valley}%
  \BibitemOpen
  \bibfield  {author} {\bibinfo {author} {\bibfnamefont {S.~M.}\ \bibnamefont
  {Rafi-Ul-Islam}}, \bibinfo {author} {\bibfnamefont {Z.~B.}\ \bibnamefont
  {Siu}}, \bibinfo {author} {\bibfnamefont {H.}~\bibnamefont {Sahin}},\ and\
  \bibinfo {author} {\bibfnamefont {M.~B.~A.}\ \bibnamefont {Jalil}},\
  }\bibfield  {title} {\bibinfo {title} {Valley hall effect and kink states in
  topolectrical circuits},\ }\href@noop {} {\bibfield  {journal} {\bibinfo
  {journal} {Phys. Rev. Res.}\ }\textbf {\bibinfo {volume} {5}},\ \bibinfo
  {pages} {013107} (\bibinfo {year} {2023}{\natexlab{a}})}\BibitemShut
  {NoStop}%
\bibitem [{\citenamefont {Rafi-Ul-Islam}\ \emph
  {et~al.}(2021{\natexlab{b}})\citenamefont {Rafi-Ul-Islam}, \citenamefont
  {Siu},\ and\ \citenamefont {Jalil}}]{rafi2021topological}%
  \BibitemOpen
  \bibfield  {author} {\bibinfo {author} {\bibfnamefont {S.~M.}\ \bibnamefont
  {Rafi-Ul-Islam}}, \bibinfo {author} {\bibfnamefont {Z.~B.}\ \bibnamefont
  {Siu}},\ and\ \bibinfo {author} {\bibfnamefont {M.~B.~A.}\ \bibnamefont
  {Jalil}},\ }\bibfield  {title} {\bibinfo {title} {Topological phases with
  higher winding numbers in nonreciprocal one-dimensional topolectrical
  circuits},\ }\href@noop {} {\bibfield  {journal} {\bibinfo  {journal} {Phys.
  Rev. B}\ }\textbf {\bibinfo {volume} {103}},\ \bibinfo {pages} {035420}
  (\bibinfo {year} {2021}{\natexlab{b}})}\BibitemShut {NoStop}%
\bibitem [{\citenamefont {Zou}\ \emph {et~al.}(2021)\citenamefont {Zou},
  \citenamefont {Chen}, \citenamefont {He}, \citenamefont {Bao}, \citenamefont
  {Lee}, \citenamefont {Sun},\ and\ \citenamefont
  {Zhang}}]{zou2021observation}%
  \BibitemOpen
  \bibfield  {author} {\bibinfo {author} {\bibfnamefont {D.}~\bibnamefont
  {Zou}}, \bibinfo {author} {\bibfnamefont {T.}~\bibnamefont {Chen}}, \bibinfo
  {author} {\bibfnamefont {W.}~\bibnamefont {He}}, \bibinfo {author}
  {\bibfnamefont {J.}~\bibnamefont {Bao}}, \bibinfo {author} {\bibfnamefont
  {C.~H.}\ \bibnamefont {Lee}}, \bibinfo {author} {\bibfnamefont
  {H.}~\bibnamefont {Sun}},\ and\ \bibinfo {author} {\bibfnamefont
  {X.}~\bibnamefont {Zhang}},\ }\bibfield  {title} {\bibinfo {title}
  {Observation of hybrid higher-order skin-topological effect in non-hermitian
  topolectrical circuits},\ }\href@noop {} {\bibfield  {journal} {\bibinfo
  {journal} {Nat. Commun.}\ }\textbf {\bibinfo {volume} {12}},\ \bibinfo
  {pages} {7201} (\bibinfo {year} {2021})}\BibitemShut {NoStop}%
\bibitem [{\citenamefont {Helbig}\ \emph {et~al.}(2020)\citenamefont {Helbig},
  \citenamefont {Hofmann}, \citenamefont {Imhof}, \citenamefont {Abdelghany},
  \citenamefont {Kiessling}, \citenamefont {Molenkamp}, \citenamefont {Lee},
  \citenamefont {Szameit}, \citenamefont {Greiter},\ and\ \citenamefont
  {Thomale}}]{helbig2020generalized}%
  \BibitemOpen
  \bibfield  {author} {\bibinfo {author} {\bibfnamefont {T.}~\bibnamefont
  {Helbig}}, \bibinfo {author} {\bibfnamefont {T.}~\bibnamefont {Hofmann}},
  \bibinfo {author} {\bibfnamefont {S.}~\bibnamefont {Imhof}}, \bibinfo
  {author} {\bibfnamefont {M.}~\bibnamefont {Abdelghany}}, \bibinfo {author}
  {\bibfnamefont {T.}~\bibnamefont {Kiessling}}, \bibinfo {author}
  {\bibfnamefont {L.}~\bibnamefont {Molenkamp}}, \bibinfo {author}
  {\bibfnamefont {C.}~\bibnamefont {Lee}}, \bibinfo {author} {\bibfnamefont
  {A.}~\bibnamefont {Szameit}}, \bibinfo {author} {\bibfnamefont
  {M.}~\bibnamefont {Greiter}},\ and\ \bibinfo {author} {\bibfnamefont
  {R.}~\bibnamefont {Thomale}},\ }\bibfield  {title} {\bibinfo {title}
  {Generalized bulk--boundary correspondence in non-hermitian topolectrical
  circuits},\ }\href@noop {} {\bibfield  {journal} {\bibinfo  {journal} {Nat.
  Phys.}\ }\textbf {\bibinfo {volume} {16}},\ \bibinfo {pages} {747} (\bibinfo
  {year} {2020})}\BibitemShut {NoStop}%
\bibitem [{\citenamefont {Rafi-Ul-Islam}\ \emph
  {et~al.}(2022{\natexlab{b}})\citenamefont {Rafi-Ul-Islam}, \citenamefont
  {Siu}, \citenamefont {Sahin}, \citenamefont {Lee},\ and\ \citenamefont
  {Jalil}}]{rafi2022system}%
  \BibitemOpen
  \bibfield  {author} {\bibinfo {author} {\bibfnamefont {S.~M.}\ \bibnamefont
  {Rafi-Ul-Islam}}, \bibinfo {author} {\bibfnamefont {Z.~B.}\ \bibnamefont
  {Siu}}, \bibinfo {author} {\bibfnamefont {H.}~\bibnamefont {Sahin}}, \bibinfo
  {author} {\bibfnamefont {C.~H.}\ \bibnamefont {Lee}},\ and\ \bibinfo {author}
  {\bibfnamefont {M.~B.~A.}\ \bibnamefont {Jalil}},\ }\bibfield  {title}
  {\bibinfo {title} {System size dependent topological zero modes in coupled
  topolectrical chains},\ }\href@noop {} {\bibfield  {journal} {\bibinfo
  {journal} {Phys. Rev. B}\ }\textbf {\bibinfo {volume} {106}},\ \bibinfo
  {pages} {075158} (\bibinfo {year} {2022}{\natexlab{b}})}\BibitemShut
  {NoStop}%
\bibitem [{\citenamefont {Lee}\ \emph {et~al.}(2018)\citenamefont {Lee},
  \citenamefont {Imhof}, \citenamefont {Berger}, \citenamefont {Bayer},
  \citenamefont {Brehm}, \citenamefont {Molenkamp}, \citenamefont {Kiessling},\
  and\ \citenamefont {Thomale}}]{lee2018topolectrical}%
  \BibitemOpen
  \bibfield  {author} {\bibinfo {author} {\bibfnamefont {C.~H.}\ \bibnamefont
  {Lee}}, \bibinfo {author} {\bibfnamefont {S.}~\bibnamefont {Imhof}}, \bibinfo
  {author} {\bibfnamefont {C.}~\bibnamefont {Berger}}, \bibinfo {author}
  {\bibfnamefont {F.}~\bibnamefont {Bayer}}, \bibinfo {author} {\bibfnamefont
  {J.}~\bibnamefont {Brehm}}, \bibinfo {author} {\bibfnamefont {L.~W.}\
  \bibnamefont {Molenkamp}}, \bibinfo {author} {\bibfnamefont {T.}~\bibnamefont
  {Kiessling}},\ and\ \bibinfo {author} {\bibfnamefont {R.}~\bibnamefont
  {Thomale}},\ }\bibfield  {title} {\bibinfo {title} {Topolectrical circuits},\
  }\href@noop {} {\bibfield  {journal} {\bibinfo  {journal} {Commun. Phys.}\
  }\textbf {\bibinfo {volume} {1}},\ \bibinfo {pages} {39} (\bibinfo {year}
  {2018})}\BibitemShut {NoStop}%
\bibitem [{\citenamefont {Rafi-Ul-Islam}\ \emph
  {et~al.}(2022{\natexlab{c}})\citenamefont {Rafi-Ul-Islam}, \citenamefont
  {Sahin}, \citenamefont {Siu},\ and\ \citenamefont
  {Jalil}}]{rafi2022interfacial}%
  \BibitemOpen
  \bibfield  {author} {\bibinfo {author} {\bibfnamefont {S.~M.}\ \bibnamefont
  {Rafi-Ul-Islam}}, \bibinfo {author} {\bibfnamefont {H.}~\bibnamefont
  {Sahin}}, \bibinfo {author} {\bibfnamefont {Z.~B.}\ \bibnamefont {Siu}},\
  and\ \bibinfo {author} {\bibfnamefont {M.~B.~A.}\ \bibnamefont {Jalil}},\
  }\bibfield  {title} {\bibinfo {title} {Interfacial skin modes at a
  non-hermitian heterojunction},\ }\href@noop {} {\bibfield  {journal}
  {\bibinfo  {journal} {Phys. Rev. Res.}\ }\textbf {\bibinfo {volume} {4}},\
  \bibinfo {pages} {043021} (\bibinfo {year} {2022}{\natexlab{c}})}\BibitemShut
  {NoStop}%
\bibitem [{\citenamefont {Rafi-Ul-Islam}\ \emph
  {et~al.}(2022{\natexlab{d}})\citenamefont {Rafi-Ul-Islam}, \citenamefont
  {Siu}, \citenamefont {Sahin},\ and\ \citenamefont {Jalil}}]{rafi2022type}%
  \BibitemOpen
  \bibfield  {author} {\bibinfo {author} {\bibfnamefont {S.~M.}\ \bibnamefont
  {Rafi-Ul-Islam}}, \bibinfo {author} {\bibfnamefont {Z.~B.}\ \bibnamefont
  {Siu}}, \bibinfo {author} {\bibfnamefont {H.}~\bibnamefont {Sahin}},\ and\
  \bibinfo {author} {\bibfnamefont {M.~B.~A.}\ \bibnamefont {Jalil}},\
  }\bibfield  {title} {\bibinfo {title} {Type-ii corner modes in topolectrical
  circuits},\ }\href@noop {} {\bibfield  {journal} {\bibinfo  {journal} {Phys.
  Rev. B}\ }\textbf {\bibinfo {volume} {106}},\ \bibinfo {pages} {245128}
  (\bibinfo {year} {2022}{\natexlab{d}})}\BibitemShut {NoStop}%
\bibitem [{\citenamefont {Martinez~Alvarez}\ \emph {et~al.}(2018)\citenamefont
  {Martinez~Alvarez}, \citenamefont {Barrios~Vargas}, \citenamefont
  {Berdakin},\ and\ \citenamefont {Foa~Torres}}]{martinez2018topological}%
  \BibitemOpen
  \bibfield  {author} {\bibinfo {author} {\bibfnamefont {V.}~\bibnamefont
  {Martinez~Alvarez}}, \bibinfo {author} {\bibfnamefont {J.}~\bibnamefont
  {Barrios~Vargas}}, \bibinfo {author} {\bibfnamefont {M.}~\bibnamefont
  {Berdakin}},\ and\ \bibinfo {author} {\bibfnamefont {L.}~\bibnamefont
  {Foa~Torres}},\ }\bibfield  {title} {\bibinfo {title} {Topological states of
  non-hermitian systems},\ }\href@noop {} {\bibfield  {journal} {\bibinfo
  {journal} {Eur. Phys. J. Spec. Top.}\ }\textbf {\bibinfo {volume} {227}},\
  \bibinfo {pages} {1295} (\bibinfo {year} {2018})}\BibitemShut {NoStop}%
\bibitem [{\citenamefont {Ghatak}\ and\ \citenamefont
  {Das}(2019)}]{ghatak2019new}%
  \BibitemOpen
  \bibfield  {author} {\bibinfo {author} {\bibfnamefont {A.}~\bibnamefont
  {Ghatak}}\ and\ \bibinfo {author} {\bibfnamefont {T.}~\bibnamefont {Das}},\
  }\bibfield  {title} {\bibinfo {title} {New topological invariants in
  non-hermitian systems},\ }\href@noop {} {\bibfield  {journal} {\bibinfo
  {journal} {J. Phys. Condens. Matt.}\ }\textbf {\bibinfo {volume} {31}},\
  \bibinfo {pages} {263001} (\bibinfo {year} {2019})}\BibitemShut {NoStop}%
\bibitem [{\citenamefont {Rafi-Ul-Islam}\ \emph
  {et~al.}(2020{\natexlab{d}})\citenamefont {Rafi-Ul-Islam}, \citenamefont
  {Siu}, \citenamefont {Sun},\ and\ \citenamefont {Jalil}}]{rafi2020strain}%
  \BibitemOpen
  \bibfield  {author} {\bibinfo {author} {\bibfnamefont {S.~M.}\ \bibnamefont
  {Rafi-Ul-Islam}}, \bibinfo {author} {\bibfnamefont {Z.~B.}\ \bibnamefont
  {Siu}}, \bibinfo {author} {\bibfnamefont {C.}~\bibnamefont {Sun}},\ and\
  \bibinfo {author} {\bibfnamefont {M.~B.~A.}\ \bibnamefont {Jalil}},\
  }\bibfield  {title} {\bibinfo {title} {Strain-controlled current switching in
  weyl semimetals},\ }\href@noop {} {\bibfield  {journal} {\bibinfo  {journal}
  {Phys. Rev. Appl.}\ }\textbf {\bibinfo {volume} {14}},\ \bibinfo {pages}
  {034007} (\bibinfo {year} {2020}{\natexlab{d}})}\BibitemShut {NoStop}%
\bibitem [{\citenamefont {Lee}\ \emph {et~al.}(2014)\citenamefont {Lee},
  \citenamefont {Reiter},\ and\ \citenamefont
  {Moiseyev}}]{lee2014entanglement}%
  \BibitemOpen
  \bibfield  {author} {\bibinfo {author} {\bibfnamefont {T.~E.}\ \bibnamefont
  {Lee}}, \bibinfo {author} {\bibfnamefont {F.}~\bibnamefont {Reiter}},\ and\
  \bibinfo {author} {\bibfnamefont {N.}~\bibnamefont {Moiseyev}},\ }\bibfield
  {title} {\bibinfo {title} {Entanglement and spin squeezing in non-hermitian
  phase transitions},\ }\href@noop {} {\bibfield  {journal} {\bibinfo
  {journal} {Phys. Rev. Lett.}\ }\textbf {\bibinfo {volume} {113}},\ \bibinfo
  {pages} {250401} (\bibinfo {year} {2014})}\BibitemShut {NoStop}%
\bibitem [{\citenamefont {Lee}\ and\ \citenamefont
  {Thomale}(2019)}]{lee2019anatomy}%
  \BibitemOpen
  \bibfield  {author} {\bibinfo {author} {\bibfnamefont {C.~H.}\ \bibnamefont
  {Lee}}\ and\ \bibinfo {author} {\bibfnamefont {R.}~\bibnamefont {Thomale}},\
  }\bibfield  {title} {\bibinfo {title} {Anatomy of skin modes and topology in
  non-hermitian systems},\ }\href@noop {} {\bibfield  {journal} {\bibinfo
  {journal} {Phys. Rev. B}\ }\textbf {\bibinfo {volume} {99}},\ \bibinfo
  {pages} {201103} (\bibinfo {year} {2019})}\BibitemShut {NoStop}%
\bibitem [{\citenamefont {Li}\ and\ \citenamefont {Lee}(2022)}]{li2022non}%
  \BibitemOpen
  \bibfield  {author} {\bibinfo {author} {\bibfnamefont {L.}~\bibnamefont
  {Li}}\ and\ \bibinfo {author} {\bibfnamefont {C.~H.}\ \bibnamefont {Lee}},\
  }\bibfield  {title} {\bibinfo {title} {Non-hermitian pseudo-gaps},\
  }\href@noop {} {\bibfield  {journal} {\bibinfo  {journal} {Sci. Bull.}\
  }\textbf {\bibinfo {volume} {67}},\ \bibinfo {pages} {685} (\bibinfo {year}
  {2022})}\BibitemShut {NoStop}%
\bibitem [{\citenamefont {Rafi-Ul-Islam}\ \emph
  {et~al.}(2022{\natexlab{e}})\citenamefont {Rafi-Ul-Islam}, \citenamefont
  {Siu}, \citenamefont {Sahin}, \citenamefont {Lee},\ and\ \citenamefont
  {Jalil}}]{rafi2022unconventional}%
  \BibitemOpen
  \bibfield  {author} {\bibinfo {author} {\bibfnamefont {S.~M.}\ \bibnamefont
  {Rafi-Ul-Islam}}, \bibinfo {author} {\bibfnamefont {Z.~B.}\ \bibnamefont
  {Siu}}, \bibinfo {author} {\bibfnamefont {H.}~\bibnamefont {Sahin}}, \bibinfo
  {author} {\bibfnamefont {C.~H.}\ \bibnamefont {Lee}},\ and\ \bibinfo {author}
  {\bibfnamefont {M.~B.~A.}\ \bibnamefont {Jalil}},\ }\bibfield  {title}
  {\bibinfo {title} {Unconventional skin modes in generalized topolectrical
  circuits with multiple asymmetric couplings},\ }\href@noop {} {\bibfield
  {journal} {\bibinfo  {journal} {Phys. Rev. Res.}\ }\textbf {\bibinfo {volume}
  {4}},\ \bibinfo {pages} {043108} (\bibinfo {year}
  {2022}{\natexlab{e}})}\BibitemShut {NoStop}%
\bibitem [{\citenamefont {Kawabata}\ \emph
  {et~al.}(2020{\natexlab{a}})\citenamefont {Kawabata}, \citenamefont {Okuma},\
  and\ \citenamefont {Sato}}]{kawabata2020non}%
  \BibitemOpen
  \bibfield  {author} {\bibinfo {author} {\bibfnamefont {K.}~\bibnamefont
  {Kawabata}}, \bibinfo {author} {\bibfnamefont {N.}~\bibnamefont {Okuma}},\
  and\ \bibinfo {author} {\bibfnamefont {M.}~\bibnamefont {Sato}},\ }\bibfield
  {title} {\bibinfo {title} {Non-bloch band theory of non-hermitian
  hamiltonians in the symplectic class},\ }\href@noop {} {\bibfield  {journal}
  {\bibinfo  {journal} {Phys. Rev. B}\ }\textbf {\bibinfo {volume} {101}},\
  \bibinfo {pages} {195147} (\bibinfo {year} {2020}{\natexlab{a}})}\BibitemShut
  {NoStop}%
\bibitem [{\citenamefont {Zhang}\ \emph {et~al.}(2020)\citenamefont {Zhang},
  \citenamefont {Yang},\ and\ \citenamefont {Fang}}]{zhang2020correspondence}%
  \BibitemOpen
  \bibfield  {author} {\bibinfo {author} {\bibfnamefont {K.}~\bibnamefont
  {Zhang}}, \bibinfo {author} {\bibfnamefont {Z.}~\bibnamefont {Yang}},\ and\
  \bibinfo {author} {\bibfnamefont {C.}~\bibnamefont {Fang}},\ }\bibfield
  {title} {\bibinfo {title} {Correspondence between winding numbers and skin
  modes in non-hermitian systems},\ }\href@noop {} {\bibfield  {journal}
  {\bibinfo  {journal} {Phys. Rev. Lett.}\ }\textbf {\bibinfo {volume} {125}},\
  \bibinfo {pages} {126402} (\bibinfo {year} {2020})}\BibitemShut {NoStop}%
\bibitem [{\citenamefont {Liu}\ \emph {et~al.}(2019)\citenamefont {Liu},
  \citenamefont {Zhang}, \citenamefont {Ai}, \citenamefont {Gong},
  \citenamefont {Kawabata}, \citenamefont {Ueda},\ and\ \citenamefont
  {Nori}}]{liu2019second}%
  \BibitemOpen
  \bibfield  {author} {\bibinfo {author} {\bibfnamefont {T.}~\bibnamefont
  {Liu}}, \bibinfo {author} {\bibfnamefont {Y.-R.}\ \bibnamefont {Zhang}},
  \bibinfo {author} {\bibfnamefont {Q.}~\bibnamefont {Ai}}, \bibinfo {author}
  {\bibfnamefont {Z.}~\bibnamefont {Gong}}, \bibinfo {author} {\bibfnamefont
  {K.}~\bibnamefont {Kawabata}}, \bibinfo {author} {\bibfnamefont
  {M.}~\bibnamefont {Ueda}},\ and\ \bibinfo {author} {\bibfnamefont
  {F.}~\bibnamefont {Nori}},\ }\bibfield  {title} {\bibinfo {title}
  {Second-order topological phases in non-hermitian systems},\ }\href@noop {}
  {\bibfield  {journal} {\bibinfo  {journal} {Phys. Rev. Lett.}\ }\textbf
  {\bibinfo {volume} {122}},\ \bibinfo {pages} {076801} (\bibinfo {year}
  {2019})}\BibitemShut {NoStop}%
\bibitem [{\citenamefont {Hamazaki}\ \emph {et~al.}(2019)\citenamefont
  {Hamazaki}, \citenamefont {Kawabata},\ and\ \citenamefont
  {Ueda}}]{hamazaki2019non}%
  \BibitemOpen
  \bibfield  {author} {\bibinfo {author} {\bibfnamefont {R.}~\bibnamefont
  {Hamazaki}}, \bibinfo {author} {\bibfnamefont {K.}~\bibnamefont {Kawabata}},\
  and\ \bibinfo {author} {\bibfnamefont {M.}~\bibnamefont {Ueda}},\ }\bibfield
  {title} {\bibinfo {title} {Non-hermitian many-body localization},\
  }\href@noop {} {\bibfield  {journal} {\bibinfo  {journal} {Phys. Rev. Lett.}\
  }\textbf {\bibinfo {volume} {123}},\ \bibinfo {pages} {090603} (\bibinfo
  {year} {2019})}\BibitemShut {NoStop}%
\bibitem [{\citenamefont {Hu}\ and\ \citenamefont {Zhao}(2021)}]{hu2021knots}%
  \BibitemOpen
  \bibfield  {author} {\bibinfo {author} {\bibfnamefont {H.}~\bibnamefont
  {Hu}}\ and\ \bibinfo {author} {\bibfnamefont {E.}~\bibnamefont {Zhao}},\
  }\bibfield  {title} {\bibinfo {title} {Knots and non-hermitian bloch bands},\
  }\href@noop {} {\bibfield  {journal} {\bibinfo  {journal} {Phys. Rev. lett.}\
  }\textbf {\bibinfo {volume} {126}},\ \bibinfo {pages} {010401} (\bibinfo
  {year} {2021})}\BibitemShut {NoStop}%
\bibitem [{\citenamefont {Wang}\ \emph
  {et~al.}(2021{\natexlab{a}})\citenamefont {Wang}, \citenamefont {Dutt},
  \citenamefont {Wojcik},\ and\ \citenamefont {Fan}}]{wang2021topological}%
  \BibitemOpen
  \bibfield  {author} {\bibinfo {author} {\bibfnamefont {K.}~\bibnamefont
  {Wang}}, \bibinfo {author} {\bibfnamefont {A.}~\bibnamefont {Dutt}}, \bibinfo
  {author} {\bibfnamefont {C.~C.}\ \bibnamefont {Wojcik}},\ and\ \bibinfo
  {author} {\bibfnamefont {S.}~\bibnamefont {Fan}},\ }\bibfield  {title}
  {\bibinfo {title} {Topological complex-energy braiding of non-hermitian
  bands},\ }\href@noop {} {\bibfield  {journal} {\bibinfo  {journal} {Nature}\
  }\textbf {\bibinfo {volume} {598}},\ \bibinfo {pages} {59} (\bibinfo {year}
  {2021}{\natexlab{a}})}\BibitemShut {NoStop}%
\bibitem [{\citenamefont {Lee}\ \emph {et~al.}(2020)\citenamefont {Lee},
  \citenamefont {Sutrisno}, \citenamefont {Hofmann}, \citenamefont {Helbig},
  \citenamefont {Liu}, \citenamefont {Ang}, \citenamefont {Ang}, \citenamefont
  {Zhang}, \citenamefont {Greiter},\ and\ \citenamefont
  {Thomale}}]{lee2020imaging}%
  \BibitemOpen
  \bibfield  {author} {\bibinfo {author} {\bibfnamefont {C.~H.}\ \bibnamefont
  {Lee}}, \bibinfo {author} {\bibfnamefont {A.}~\bibnamefont {Sutrisno}},
  \bibinfo {author} {\bibfnamefont {T.}~\bibnamefont {Hofmann}}, \bibinfo
  {author} {\bibfnamefont {T.}~\bibnamefont {Helbig}}, \bibinfo {author}
  {\bibfnamefont {Y.}~\bibnamefont {Liu}}, \bibinfo {author} {\bibfnamefont
  {Y.~S.}\ \bibnamefont {Ang}}, \bibinfo {author} {\bibfnamefont {L.~K.}\
  \bibnamefont {Ang}}, \bibinfo {author} {\bibfnamefont {X.}~\bibnamefont
  {Zhang}}, \bibinfo {author} {\bibfnamefont {M.}~\bibnamefont {Greiter}},\
  and\ \bibinfo {author} {\bibfnamefont {R.}~\bibnamefont {Thomale}},\
  }\bibfield  {title} {\bibinfo {title} {Imaging nodal knots in momentum space
  through topolectrical circuits},\ }\href@noop {} {\bibfield  {journal}
  {\bibinfo  {journal} {Nat. Commun.}\ }\textbf {\bibinfo {volume} {11}},\
  \bibinfo {pages} {4385} (\bibinfo {year} {2020})}\BibitemShut {NoStop}%
\bibitem [{\citenamefont {Li}\ and\ \citenamefont
  {Mong}(2021)}]{li2021homotopical}%
  \BibitemOpen
  \bibfield  {author} {\bibinfo {author} {\bibfnamefont {Z.}~\bibnamefont
  {Li}}\ and\ \bibinfo {author} {\bibfnamefont {R.~S.}\ \bibnamefont {Mong}},\
  }\bibfield  {title} {\bibinfo {title} {Homotopical characterization of
  non-hermitian band structures},\ }\href@noop {} {\bibfield  {journal}
  {\bibinfo  {journal} {Phys. Rev. B}\ }\textbf {\bibinfo {volume} {103}},\
  \bibinfo {pages} {155129} (\bibinfo {year} {2021})}\BibitemShut {NoStop}%
\bibitem [{\citenamefont {Ezawa}(2019)}]{ezawa2019braiding}%
  \BibitemOpen
  \bibfield  {author} {\bibinfo {author} {\bibfnamefont {M.}~\bibnamefont
  {Ezawa}},\ }\bibfield  {title} {\bibinfo {title} {Braiding of majorana-like
  corner states in electric circuits and its non-hermitian generalization},\
  }\href@noop {} {\bibfield  {journal} {\bibinfo  {journal} {Phys. Rev. B}\
  }\textbf {\bibinfo {volume} {100}},\ \bibinfo {pages} {045407} (\bibinfo
  {year} {2019})}\BibitemShut {NoStop}%
\bibitem [{\citenamefont {Hu}\ \emph {et~al.}(2022)\citenamefont {Hu},
  \citenamefont {Sun},\ and\ \citenamefont {Chen}}]{hu2022knot}%
  \BibitemOpen
  \bibfield  {author} {\bibinfo {author} {\bibfnamefont {H.}~\bibnamefont
  {Hu}}, \bibinfo {author} {\bibfnamefont {S.}~\bibnamefont {Sun}},\ and\
  \bibinfo {author} {\bibfnamefont {S.}~\bibnamefont {Chen}},\ }\bibfield
  {title} {\bibinfo {title} {Knot topology of exceptional point and
  non-hermitian no-go theorem},\ }\href@noop {} {\bibfield  {journal} {\bibinfo
   {journal} {Phys. Rev. Res.}\ }\textbf {\bibinfo {volume} {4}},\ \bibinfo
  {pages} {L022064} (\bibinfo {year} {2022})}\BibitemShut {NoStop}%
\bibitem [{\citenamefont {Zhang}\ \emph
  {et~al.}(2023{\natexlab{b}})\citenamefont {Zhang}, \citenamefont {Li},
  \citenamefont {Sun}, \citenamefont {Liu}, \citenamefont {Zhao}, \citenamefont
  {Feng}, \citenamefont {Fan},\ and\ \citenamefont
  {Qiu}}]{zhang2023observation}%
  \BibitemOpen
  \bibfield  {author} {\bibinfo {author} {\bibfnamefont {Q.}~\bibnamefont
  {Zhang}}, \bibinfo {author} {\bibfnamefont {Y.}~\bibnamefont {Li}}, \bibinfo
  {author} {\bibfnamefont {H.}~\bibnamefont {Sun}}, \bibinfo {author}
  {\bibfnamefont {X.}~\bibnamefont {Liu}}, \bibinfo {author} {\bibfnamefont
  {L.}~\bibnamefont {Zhao}}, \bibinfo {author} {\bibfnamefont {X.}~\bibnamefont
  {Feng}}, \bibinfo {author} {\bibfnamefont {X.}~\bibnamefont {Fan}},\ and\
  \bibinfo {author} {\bibfnamefont {C.}~\bibnamefont {Qiu}},\ }\bibfield
  {title} {\bibinfo {title} {Observation of acoustic non-hermitian bloch braids
  and associated topological phase transitions},\ }\href@noop {} {\bibfield
  {journal} {\bibinfo  {journal} {Phys. Rev. Lett.}\ }\textbf {\bibinfo
  {volume} {130}},\ \bibinfo {pages} {017201} (\bibinfo {year}
  {2023}{\natexlab{b}})}\BibitemShut {NoStop}%
\bibitem [{\citenamefont {Li}\ \emph {et~al.}(2022)\citenamefont {Li},
  \citenamefont {Chen},\ and\ \citenamefont {Yang}}]{li2022topological}%
  \BibitemOpen
  \bibfield  {author} {\bibinfo {author} {\bibfnamefont {Y.}~\bibnamefont
  {Li}}, \bibinfo {author} {\bibfnamefont {Y.}~\bibnamefont {Chen}},\ and\
  \bibinfo {author} {\bibfnamefont {X.}~\bibnamefont {Yang}},\ }\bibfield
  {title} {\bibinfo {title} {Topological energy braiding of the non-bloch
  bands},\ }\href@noop {} {\bibfield  {journal} {\bibinfo  {journal} {ArXiv
  preprint arXiv:2204.12857}\ } (\bibinfo {year} {2022})}\BibitemShut {NoStop}%
\bibitem [{\citenamefont {Okuma}\ \emph
  {et~al.}(2020{\natexlab{a}})\citenamefont {Okuma}, \citenamefont {Kawabata},
  \citenamefont {Shiozaki},\ and\ \citenamefont {Sato}}]{okuma2020topological}%
  \BibitemOpen
  \bibfield  {author} {\bibinfo {author} {\bibfnamefont {N.}~\bibnamefont
  {Okuma}}, \bibinfo {author} {\bibfnamefont {K.}~\bibnamefont {Kawabata}},
  \bibinfo {author} {\bibfnamefont {K.}~\bibnamefont {Shiozaki}},\ and\
  \bibinfo {author} {\bibfnamefont {M.}~\bibnamefont {Sato}},\ }\bibfield
  {title} {\bibinfo {title} {Topological origin of non-hermitian skin
  effects},\ }\href@noop {} {\bibfield  {journal} {\bibinfo  {journal} {Phys.
  Rev. Lett.}\ }\textbf {\bibinfo {volume} {124}},\ \bibinfo {pages} {086801}
  (\bibinfo {year} {2020}{\natexlab{a}})}\BibitemShut {NoStop}%
\bibitem [{\citenamefont {Siu}\ \emph {et~al.}(2023)\citenamefont {Siu},
  \citenamefont {Rafi-Ul-Islam},\ and\ \citenamefont
  {Jalil}}]{siu2023terminal}%
  \BibitemOpen
  \bibfield  {author} {\bibinfo {author} {\bibfnamefont {Z.~B.}\ \bibnamefont
  {Siu}}, \bibinfo {author} {\bibfnamefont {S.~M.}\ \bibnamefont
  {Rafi-Ul-Islam}},\ and\ \bibinfo {author} {\bibfnamefont {M.~B.~A.}\
  \bibnamefont {Jalil}},\ }\bibfield  {title} {\bibinfo {title}
  {Terminal-coupling induced critical eigenspectrum transition in closed
  non-hermitian loops},\ }\href@noop {} {\bibfield  {journal} {\bibinfo
  {journal} {Sci. Rep.}\ }\textbf {\bibinfo {volume} {13}},\ \bibinfo {pages}
  {22770} (\bibinfo {year} {2023})}\BibitemShut {NoStop}%
\bibitem [{\citenamefont {Rafi-Ul-Islam}\ \emph
  {et~al.}(2024{\natexlab{a}})\citenamefont {Rafi-Ul-Islam}, \citenamefont
  {Siu}, \citenamefont {Razo},\ and\ \citenamefont
  {Jalil}}]{rafi2024saturation}%
  \BibitemOpen
  \bibfield  {author} {\bibinfo {author} {\bibfnamefont {S.~M.}\ \bibnamefont
  {Rafi-Ul-Islam}}, \bibinfo {author} {\bibfnamefont {Z.~B.}\ \bibnamefont
  {Siu}}, \bibinfo {author} {\bibfnamefont {M.~S.~H.}\ \bibnamefont {Razo}},\
  and\ \bibinfo {author} {\bibfnamefont {M.~B.~A.}\ \bibnamefont {Jalil}},\
  }\bibfield  {title} {\bibinfo {title} {Saturation dynamics in non-hermitian
  topological sensing systems},\ }\href@noop {} {\bibfield  {journal} {\bibinfo
   {journal} {ArXiv Preprint ArXiv:2406.19629}\ } (\bibinfo {year}
  {2024}{\natexlab{a}})}\BibitemShut {NoStop}%
\bibitem [{\citenamefont {Li}\ \emph {et~al.}(2020)\citenamefont {Li},
  \citenamefont {Lee}, \citenamefont {Mu},\ and\ \citenamefont
  {Gong}}]{li2020critical}%
  \BibitemOpen
  \bibfield  {author} {\bibinfo {author} {\bibfnamefont {L.}~\bibnamefont
  {Li}}, \bibinfo {author} {\bibfnamefont {C.~H.}\ \bibnamefont {Lee}},
  \bibinfo {author} {\bibfnamefont {S.}~\bibnamefont {Mu}},\ and\ \bibinfo
  {author} {\bibfnamefont {J.}~\bibnamefont {Gong}},\ }\bibfield  {title}
  {\bibinfo {title} {Critical non-hermitian skin effect},\ }\href@noop {}
  {\bibfield  {journal} {\bibinfo  {journal} {Nat. Commun.}\ }\textbf {\bibinfo
  {volume} {11}},\ \bibinfo {pages} {5491} (\bibinfo {year}
  {2020})}\BibitemShut {NoStop}%
\bibitem [{\citenamefont {Rafi-Ul-Islam}\ \emph
  {et~al.}(2024{\natexlab{b}})\citenamefont {Rafi-Ul-Islam}, \citenamefont
  {Siu}, \citenamefont {Sahin}, \citenamefont {Razo},\ and\ \citenamefont
  {Jalil}}]{rafi2024twisted}%
  \BibitemOpen
  \bibfield  {author} {\bibinfo {author} {\bibfnamefont {S.~M.}\ \bibnamefont
  {Rafi-Ul-Islam}}, \bibinfo {author} {\bibfnamefont {Z.~B.}\ \bibnamefont
  {Siu}}, \bibinfo {author} {\bibfnamefont {H.}~\bibnamefont {Sahin}}, \bibinfo
  {author} {\bibfnamefont {M.~S.~H.}\ \bibnamefont {Razo}},\ and\ \bibinfo
  {author} {\bibfnamefont {M.~B.~A.}\ \bibnamefont {Jalil}},\ }\bibfield
  {title} {\bibinfo {title} {Twisted topology of non-hermitian systems induced
  by long-range coupling},\ }\href@noop {} {\bibfield  {journal} {\bibinfo
  {journal} {Phys. Rev. B}\ }\textbf {\bibinfo {volume} {109}},\ \bibinfo
  {pages} {045410} (\bibinfo {year} {2024}{\natexlab{b}})}\BibitemShut
  {NoStop}%
\bibitem [{\citenamefont {Longhi}(2019)}]{longhi2019probing}%
  \BibitemOpen
  \bibfield  {author} {\bibinfo {author} {\bibfnamefont {S.}~\bibnamefont
  {Longhi}},\ }\bibfield  {title} {\bibinfo {title} {Probing non-hermitian skin
  effect and non-bloch phase transitions},\ }\href@noop {} {\bibfield
  {journal} {\bibinfo  {journal} {Phys. Rev. Res.}\ }\textbf {\bibinfo {volume}
  {1}},\ \bibinfo {pages} {023013} (\bibinfo {year} {2019})}\BibitemShut
  {NoStop}%
\bibitem [{\citenamefont {Song}\ \emph {et~al.}(2019)\citenamefont {Song},
  \citenamefont {Yao},\ and\ \citenamefont {Wang}}]{song2019non}%
  \BibitemOpen
  \bibfield  {author} {\bibinfo {author} {\bibfnamefont {F.}~\bibnamefont
  {Song}}, \bibinfo {author} {\bibfnamefont {S.}~\bibnamefont {Yao}},\ and\
  \bibinfo {author} {\bibfnamefont {Z.}~\bibnamefont {Wang}},\ }\bibfield
  {title} {\bibinfo {title} {Non-hermitian skin effect and chiral damping in
  open quantum systems},\ }\href@noop {} {\bibfield  {journal} {\bibinfo
  {journal} {Phys. Rev. Lett.}\ }\textbf {\bibinfo {volume} {123}},\ \bibinfo
  {pages} {170401} (\bibinfo {year} {2019})}\BibitemShut {NoStop}%
\bibitem [{\citenamefont {Kawabata}\ \emph
  {et~al.}(2020{\natexlab{b}})\citenamefont {Kawabata}, \citenamefont {Sato},\
  and\ \citenamefont {Shiozaki}}]{kawabata2020higher}%
  \BibitemOpen
  \bibfield  {author} {\bibinfo {author} {\bibfnamefont {K.}~\bibnamefont
  {Kawabata}}, \bibinfo {author} {\bibfnamefont {M.}~\bibnamefont {Sato}},\
  and\ \bibinfo {author} {\bibfnamefont {K.}~\bibnamefont {Shiozaki}},\
  }\bibfield  {title} {\bibinfo {title} {Higher-order non-hermitian skin
  effect},\ }\href@noop {} {\bibfield  {journal} {\bibinfo  {journal} {Phys.
  Rev. B}\ }\textbf {\bibinfo {volume} {102}},\ \bibinfo {pages} {205118}
  (\bibinfo {year} {2020}{\natexlab{b}})}\BibitemShut {NoStop}%
\bibitem [{\citenamefont {Zhang}\ \emph
  {et~al.}(2021{\natexlab{a}})\citenamefont {Zhang}, \citenamefont {Tian},
  \citenamefont {Jiang}, \citenamefont {Lu},\ and\ \citenamefont
  {Chen}}]{zhang2021observation}%
  \BibitemOpen
  \bibfield  {author} {\bibinfo {author} {\bibfnamefont {X.}~\bibnamefont
  {Zhang}}, \bibinfo {author} {\bibfnamefont {Y.}~\bibnamefont {Tian}},
  \bibinfo {author} {\bibfnamefont {J.-H.}\ \bibnamefont {Jiang}}, \bibinfo
  {author} {\bibfnamefont {M.-H.}\ \bibnamefont {Lu}},\ and\ \bibinfo {author}
  {\bibfnamefont {Y.-F.}\ \bibnamefont {Chen}},\ }\bibfield  {title} {\bibinfo
  {title} {Observation of higher-order non-hermitian skin effect},\ }\href@noop
  {} {\bibfield  {journal} {\bibinfo  {journal} {Nat. Commun.}\ }\textbf
  {\bibinfo {volume} {12}},\ \bibinfo {pages} {5377} (\bibinfo {year}
  {2021}{\natexlab{a}})}\BibitemShut {NoStop}%
\bibitem [{\citenamefont {Zhang}\ \emph {et~al.}(2022)\citenamefont {Zhang},
  \citenamefont {Yang},\ and\ \citenamefont {Fang}}]{zhang2022universal}%
  \BibitemOpen
  \bibfield  {author} {\bibinfo {author} {\bibfnamefont {K.}~\bibnamefont
  {Zhang}}, \bibinfo {author} {\bibfnamefont {Z.}~\bibnamefont {Yang}},\ and\
  \bibinfo {author} {\bibfnamefont {C.}~\bibnamefont {Fang}},\ }\bibfield
  {title} {\bibinfo {title} {Universal non-hermitian skin effect in two and
  higher dimensions},\ }\href@noop {} {\bibfield  {journal} {\bibinfo
  {journal} {Nat. Commun.}\ }\textbf {\bibinfo {volume} {13}},\ \bibinfo
  {pages} {2496} (\bibinfo {year} {2022})}\BibitemShut {NoStop}%
\bibitem [{\citenamefont {Kawabata}\ \emph {et~al.}(2019)\citenamefont
  {Kawabata}, \citenamefont {Bessho},\ and\ \citenamefont
  {Sato}}]{kawabata2019classification}%
  \BibitemOpen
  \bibfield  {author} {\bibinfo {author} {\bibfnamefont {K.}~\bibnamefont
  {Kawabata}}, \bibinfo {author} {\bibfnamefont {T.}~\bibnamefont {Bessho}},\
  and\ \bibinfo {author} {\bibfnamefont {M.}~\bibnamefont {Sato}},\ }\bibfield
  {title} {\bibinfo {title} {Classification of exceptional points and
  non-hermitian topological semimetals},\ }\href@noop {} {\bibfield  {journal}
  {\bibinfo  {journal} {Phys. Rev. Lett.}\ }\textbf {\bibinfo {volume} {123}},\
  \bibinfo {pages} {066405} (\bibinfo {year} {2019})}\BibitemShut {NoStop}%
\bibitem [{\citenamefont {Minganti}\ \emph {et~al.}(2019)\citenamefont
  {Minganti}, \citenamefont {Miranowicz}, \citenamefont {Chhajlany},\ and\
  \citenamefont {Nori}}]{minganti2019quantum}%
  \BibitemOpen
  \bibfield  {author} {\bibinfo {author} {\bibfnamefont {F.}~\bibnamefont
  {Minganti}}, \bibinfo {author} {\bibfnamefont {A.}~\bibnamefont
  {Miranowicz}}, \bibinfo {author} {\bibfnamefont {R.~W.}\ \bibnamefont
  {Chhajlany}},\ and\ \bibinfo {author} {\bibfnamefont {F.}~\bibnamefont
  {Nori}},\ }\bibfield  {title} {\bibinfo {title} {Quantum exceptional points
  of non-hermitian hamiltonians and liouvillians: The effects of quantum
  jumps},\ }\href@noop {} {\bibfield  {journal} {\bibinfo  {journal} {Phys.
  Rev. A}\ }\textbf {\bibinfo {volume} {100}},\ \bibinfo {pages} {062131}
  (\bibinfo {year} {2019})}\BibitemShut {NoStop}%
\bibitem [{\citenamefont {Hu}\ \emph {et~al.}(2017)\citenamefont {Hu},
  \citenamefont {Wang}, \citenamefont {Shum},\ and\ \citenamefont
  {Chong}}]{hu2017exceptional}%
  \BibitemOpen
  \bibfield  {author} {\bibinfo {author} {\bibfnamefont {W.}~\bibnamefont
  {Hu}}, \bibinfo {author} {\bibfnamefont {H.}~\bibnamefont {Wang}}, \bibinfo
  {author} {\bibfnamefont {P.~P.}\ \bibnamefont {Shum}},\ and\ \bibinfo
  {author} {\bibfnamefont {Y.~D.}\ \bibnamefont {Chong}},\ }\bibfield  {title}
  {\bibinfo {title} {Exceptional points in a non-hermitian topological pump},\
  }\href@noop {} {\bibfield  {journal} {\bibinfo  {journal} {Phys. Rev. B}\
  }\textbf {\bibinfo {volume} {95}},\ \bibinfo {pages} {184306} (\bibinfo
  {year} {2017})}\BibitemShut {NoStop}%
\bibitem [{\citenamefont {Parto}\ \emph {et~al.}(2021)\citenamefont {Parto},
  \citenamefont {Liu}, \citenamefont {Bahari}, \citenamefont {Khajavikhan},\
  and\ \citenamefont {Christodoulides}}]{parto2021non}%
  \BibitemOpen
  \bibfield  {author} {\bibinfo {author} {\bibfnamefont {M.}~\bibnamefont
  {Parto}}, \bibinfo {author} {\bibfnamefont {Y.~G.}\ \bibnamefont {Liu}},
  \bibinfo {author} {\bibfnamefont {B.}~\bibnamefont {Bahari}}, \bibinfo
  {author} {\bibfnamefont {M.}~\bibnamefont {Khajavikhan}},\ and\ \bibinfo
  {author} {\bibfnamefont {D.~N.}\ \bibnamefont {Christodoulides}},\ }\bibfield
   {title} {\bibinfo {title} {Non-hermitian and topological photonics: optics
  at an exceptional point},\ }\href@noop {} {\bibfield  {journal} {\bibinfo
  {journal} {Nanophotonics}\ }\textbf {\bibinfo {volume} {10}},\ \bibinfo
  {pages} {403} (\bibinfo {year} {2021})}\BibitemShut {NoStop}%
\bibitem [{\citenamefont {Rafi-Ul-Islam}\ \emph
  {et~al.}(2022{\natexlab{f}})\citenamefont {Rafi-Ul-Islam}, \citenamefont
  {Siu}, \citenamefont {Sahin},\ and\ \citenamefont {Jalil}}]{rafi2022valley}%
  \BibitemOpen
  \bibfield  {author} {\bibinfo {author} {\bibfnamefont {S.~M.}\ \bibnamefont
  {Rafi-Ul-Islam}}, \bibinfo {author} {\bibfnamefont {Z.~B.}\ \bibnamefont
  {Siu}}, \bibinfo {author} {\bibfnamefont {H.}~\bibnamefont {Sahin}},\ and\
  \bibinfo {author} {\bibfnamefont {M.~B.~A.}\ \bibnamefont {Jalil}},\
  }\bibfield  {title} {\bibinfo {title} {Valley and spin quantum hall
  conductance of silicene coupled to a ferroelectric layer},\ }\href@noop {}
  {\bibfield  {journal} {\bibinfo  {journal} {Front. Phys.}\ }\textbf {\bibinfo
  {volume} {10}},\ \bibinfo {pages} {1021192} (\bibinfo {year}
  {2022}{\natexlab{f}})}\BibitemShut {NoStop}%
\bibitem [{\citenamefont {Rafi-Ul-Islam}\ \emph
  {et~al.}(2023{\natexlab{b}})\citenamefont {Rafi-Ul-Islam}, \citenamefont
  {Siu}, \citenamefont {Sahin},\ and\ \citenamefont
  {Jalil}}]{rafi2023conductance}%
  \BibitemOpen
  \bibfield  {author} {\bibinfo {author} {\bibfnamefont {S.~M.}\ \bibnamefont
  {Rafi-Ul-Islam}}, \bibinfo {author} {\bibfnamefont {Z.~B.}\ \bibnamefont
  {Siu}}, \bibinfo {author} {\bibfnamefont {H.}~\bibnamefont {Sahin}},\ and\
  \bibinfo {author} {\bibfnamefont {M.~B.~A.}\ \bibnamefont {Jalil}},\
  }\bibfield  {title} {\bibinfo {title} {Conductance modulation and spin/valley
  polarized transmission in silicene coupled with ferroelectric layer},\
  }\href@noop {} {\bibfield  {journal} {\bibinfo  {journal} {J. Magn. Magn.
  Mater.}\ }\textbf {\bibinfo {volume} {571}},\ \bibinfo {pages} {170559}
  (\bibinfo {year} {2023}{\natexlab{b}})}\BibitemShut {NoStop}%
\bibitem [{\citenamefont {Rafi-Ul-Islam}\ \emph
  {et~al.}(2024{\natexlab{c}})\citenamefont {Rafi-Ul-Islam}, \citenamefont
  {Siu}, \citenamefont {Sahin},\ and\ \citenamefont {Jalil}}]{rafi2024chiral}%
  \BibitemOpen
  \bibfield  {author} {\bibinfo {author} {\bibfnamefont {S.~M.}\ \bibnamefont
  {Rafi-Ul-Islam}}, \bibinfo {author} {\bibfnamefont {Z.~B.}\ \bibnamefont
  {Siu}}, \bibinfo {author} {\bibfnamefont {H.}~\bibnamefont {Sahin}},\ and\
  \bibinfo {author} {\bibfnamefont {M.~B.~A.}\ \bibnamefont {Jalil}},\
  }\bibfield  {title} {\bibinfo {title} {Chiral surface and hinge states in
  higher-order weyl semimetallic circuits},\ }\href@noop {} {\bibfield
  {journal} {\bibinfo  {journal} {Phys. Rev. B}\ }\textbf {\bibinfo {volume}
  {109}},\ \bibinfo {pages} {085430} (\bibinfo {year}
  {2024}{\natexlab{c}})}\BibitemShut {NoStop}%
\bibitem [{\citenamefont {Sun}\ \emph {et~al.}(2019)\citenamefont {Sun},
  \citenamefont {Deng}, \citenamefont {Rafi-Ul-Islam}, \citenamefont {Liang},
  \citenamefont {Yang},\ and\ \citenamefont {Jalil}}]{sun2019field}%
  \BibitemOpen
  \bibfield  {author} {\bibinfo {author} {\bibfnamefont {C.}~\bibnamefont
  {Sun}}, \bibinfo {author} {\bibfnamefont {J.}~\bibnamefont {Deng}}, \bibinfo
  {author} {\bibfnamefont {S.~M.}\ \bibnamefont {Rafi-Ul-Islam}}, \bibinfo
  {author} {\bibfnamefont {G.}~\bibnamefont {Liang}}, \bibinfo {author}
  {\bibfnamefont {H.}~\bibnamefont {Yang}},\ and\ \bibinfo {author}
  {\bibfnamefont {M.~B.~A.}\ \bibnamefont {Jalil}},\ }\bibfield  {title}
  {\bibinfo {title} {Field-free switching of perpendicular magnetization
  through spin hall and anomalous hall effects in
  ferromagnet--heavy-metal--ferromagnet structures},\ }\href@noop {} {\bibfield
   {journal} {\bibinfo  {journal} {Phys. Rev. Appl.}\ }\textbf {\bibinfo
  {volume} {12}},\ \bibinfo {pages} {034022} (\bibinfo {year}
  {2019})}\BibitemShut {NoStop}%
\bibitem [{\citenamefont {Sun}\ \emph {et~al.}(2020)\citenamefont {Sun},
  \citenamefont {Rafi-Ul-Islam}, \citenamefont {Yang},\ and\ \citenamefont
  {Jalil}}]{sun2020spin}%
  \BibitemOpen
  \bibfield  {author} {\bibinfo {author} {\bibfnamefont {C.}~\bibnamefont
  {Sun}}, \bibinfo {author} {\bibfnamefont {S.~M.}\ \bibnamefont
  {Rafi-Ul-Islam}}, \bibinfo {author} {\bibfnamefont {H.}~\bibnamefont
  {Yang}},\ and\ \bibinfo {author} {\bibfnamefont {M.~B.~A.}\ \bibnamefont
  {Jalil}},\ }\bibfield  {title} {\bibinfo {title} {Spin nernst and anomalous
  nernst effects and their signature outputs in ferromagnet/nonmagnet
  heterostructures},\ }\href@noop {} {\bibfield  {journal} {\bibinfo  {journal}
  {Physical Review B}\ }\textbf {\bibinfo {volume} {102}},\ \bibinfo {pages}
  {214419} (\bibinfo {year} {2020})}\BibitemShut {NoStop}%
\bibitem [{\citenamefont {Malciu}\ \emph {et~al.}(2018)\citenamefont {Malciu},
  \citenamefont {Mazza},\ and\ \citenamefont {Mora}}]{malciu2018braiding}%
  \BibitemOpen
  \bibfield  {author} {\bibinfo {author} {\bibfnamefont {C.}~\bibnamefont
  {Malciu}}, \bibinfo {author} {\bibfnamefont {L.}~\bibnamefont {Mazza}},\ and\
  \bibinfo {author} {\bibfnamefont {C.}~\bibnamefont {Mora}},\ }\bibfield
  {title} {\bibinfo {title} {Braiding majorana zero modes using quantum dots},\
  }\href@noop {} {\bibfield  {journal} {\bibinfo  {journal} {Phys. Rev. B}\
  }\textbf {\bibinfo {volume} {98}},\ \bibinfo {pages} {165426} (\bibinfo
  {year} {2018})}\BibitemShut {NoStop}%
\bibitem [{\citenamefont {Homeier}\ \emph {et~al.}(2021)\citenamefont
  {Homeier}, \citenamefont {Schweizer}, \citenamefont {Aidelsburger},
  \citenamefont {Fedorov},\ and\ \citenamefont {Grusdt}}]{homeier2021z}%
  \BibitemOpen
  \bibfield  {author} {\bibinfo {author} {\bibfnamefont {L.}~\bibnamefont
  {Homeier}}, \bibinfo {author} {\bibfnamefont {C.}~\bibnamefont {Schweizer}},
  \bibinfo {author} {\bibfnamefont {M.}~\bibnamefont {Aidelsburger}}, \bibinfo
  {author} {\bibfnamefont {A.}~\bibnamefont {Fedorov}},\ and\ \bibinfo {author}
  {\bibfnamefont {F.}~\bibnamefont {Grusdt}},\ }\bibfield  {title} {\bibinfo
  {title} {Z 2 lattice gauge theories and kitaev's toric code: A scheme for
  analog quantum simulation},\ }\href@noop {} {\bibfield  {journal} {\bibinfo
  {journal} {Phys. Rev. B}\ }\textbf {\bibinfo {volume} {104}},\ \bibinfo
  {pages} {085138} (\bibinfo {year} {2021})}\BibitemShut {NoStop}%
\bibitem [{\citenamefont {Sekania}\ \emph {et~al.}(2017)\citenamefont
  {Sekania}, \citenamefont {Plugge}, \citenamefont {Greiter}, \citenamefont
  {Thomale},\ and\ \citenamefont {Schmitteckert}}]{sekania2017braiding}%
  \BibitemOpen
  \bibfield  {author} {\bibinfo {author} {\bibfnamefont {M.}~\bibnamefont
  {Sekania}}, \bibinfo {author} {\bibfnamefont {S.}~\bibnamefont {Plugge}},
  \bibinfo {author} {\bibfnamefont {M.}~\bibnamefont {Greiter}}, \bibinfo
  {author} {\bibfnamefont {R.}~\bibnamefont {Thomale}},\ and\ \bibinfo {author}
  {\bibfnamefont {P.}~\bibnamefont {Schmitteckert}},\ }\bibfield  {title}
  {\bibinfo {title} {Braiding errors in interacting majorana quantum wires},\
  }\href@noop {} {\bibfield  {journal} {\bibinfo  {journal} {Phys. Rev. B}\
  }\textbf {\bibinfo {volume} {96}},\ \bibinfo {pages} {094307} (\bibinfo
  {year} {2017})}\BibitemShut {NoStop}%
\bibitem [{\citenamefont {Ezawa}(2020)}]{ezawa2020non}%
  \BibitemOpen
  \bibfield  {author} {\bibinfo {author} {\bibfnamefont {M.}~\bibnamefont
  {Ezawa}},\ }\bibfield  {title} {\bibinfo {title} {Non-abelian braiding of
  majorana-like edge states and topological quantum computations in electric
  circuits},\ }\href@noop {} {\bibfield  {journal} {\bibinfo  {journal} {Phys.
  Rev. B}\ }\textbf {\bibinfo {volume} {102}},\ \bibinfo {pages} {075424}
  (\bibinfo {year} {2020})}\BibitemShut {NoStop}%
\bibitem [{\citenamefont {Backens}\ \emph {et~al.}(2017)\citenamefont
  {Backens}, \citenamefont {Shnirman}, \citenamefont {Makhlin}, \citenamefont
  {Gefen}, \citenamefont {Mooij},\ and\ \citenamefont
  {Sch{\"o}n}}]{backens2017emulating}%
  \BibitemOpen
  \bibfield  {author} {\bibinfo {author} {\bibfnamefont {S.}~\bibnamefont
  {Backens}}, \bibinfo {author} {\bibfnamefont {A.}~\bibnamefont {Shnirman}},
  \bibinfo {author} {\bibfnamefont {Y.}~\bibnamefont {Makhlin}}, \bibinfo
  {author} {\bibfnamefont {Y.}~\bibnamefont {Gefen}}, \bibinfo {author}
  {\bibfnamefont {J.~E.}\ \bibnamefont {Mooij}},\ and\ \bibinfo {author}
  {\bibfnamefont {G.}~\bibnamefont {Sch{\"o}n}},\ }\bibfield  {title} {\bibinfo
  {title} {Emulating majorana fermions and their braiding by ising spin
  chains},\ }\href@noop {} {\bibfield  {journal} {\bibinfo  {journal} {Phys.
  Rev. B}\ }\textbf {\bibinfo {volume} {96}},\ \bibinfo {pages} {195402}
  (\bibinfo {year} {2017})}\BibitemShut {NoStop}%
\bibitem [{\citenamefont {Pahomi}\ \emph {et~al.}(2020)\citenamefont {Pahomi},
  \citenamefont {Sigrist},\ and\ \citenamefont
  {Soluyanov}}]{pahomi2020braiding}%
  \BibitemOpen
  \bibfield  {author} {\bibinfo {author} {\bibfnamefont {T.~E.}\ \bibnamefont
  {Pahomi}}, \bibinfo {author} {\bibfnamefont {M.}~\bibnamefont {Sigrist}},\
  and\ \bibinfo {author} {\bibfnamefont {A.~A.}\ \bibnamefont {Soluyanov}},\
  }\bibfield  {title} {\bibinfo {title} {Braiding majorana corner modes in a
  second-order topological superconductor},\ }\href@noop {} {\bibfield
  {journal} {\bibinfo  {journal} {Phys. Rev. Res.}\ }\textbf {\bibinfo {volume}
  {2}},\ \bibinfo {pages} {032068} (\bibinfo {year} {2020})}\BibitemShut
  {NoStop}%
\bibitem [{\citenamefont {Levin}\ and\ \citenamefont
  {Gu}(2012)}]{levin2012braiding}%
  \BibitemOpen
  \bibfield  {author} {\bibinfo {author} {\bibfnamefont {M.}~\bibnamefont
  {Levin}}\ and\ \bibinfo {author} {\bibfnamefont {Z.-C.}\ \bibnamefont {Gu}},\
  }\bibfield  {title} {\bibinfo {title} {Braiding statistics approach to
  symmetry-protected topological phases},\ }\href@noop {} {\bibfield  {journal}
  {\bibinfo  {journal} {Phys. Rev. B}\ }\textbf {\bibinfo {volume} {86}},\
  \bibinfo {pages} {115109} (\bibinfo {year} {2012})}\BibitemShut {NoStop}%
\bibitem [{\citenamefont {Noh}\ \emph {et~al.}(2020)\citenamefont {Noh},
  \citenamefont {Schuster}, \citenamefont {Iadecola}, \citenamefont {Huang},
  \citenamefont {Wang}, \citenamefont {Chen}, \citenamefont {Chamon},\ and\
  \citenamefont {Rechtsman}}]{noh2020braiding}%
  \BibitemOpen
  \bibfield  {author} {\bibinfo {author} {\bibfnamefont {J.}~\bibnamefont
  {Noh}}, \bibinfo {author} {\bibfnamefont {T.}~\bibnamefont {Schuster}},
  \bibinfo {author} {\bibfnamefont {T.}~\bibnamefont {Iadecola}}, \bibinfo
  {author} {\bibfnamefont {S.}~\bibnamefont {Huang}}, \bibinfo {author}
  {\bibfnamefont {M.}~\bibnamefont {Wang}}, \bibinfo {author} {\bibfnamefont
  {K.~P.}\ \bibnamefont {Chen}}, \bibinfo {author} {\bibfnamefont
  {C.}~\bibnamefont {Chamon}},\ and\ \bibinfo {author} {\bibfnamefont {M.~C.}\
  \bibnamefont {Rechtsman}},\ }\bibfield  {title} {\bibinfo {title} {Braiding
  photonic topological zero modes},\ }\href@noop {} {\bibfield  {journal}
  {\bibinfo  {journal} {Nat. Phys.}\ }\textbf {\bibinfo {volume} {16}},\
  \bibinfo {pages} {989} (\bibinfo {year} {2020})}\BibitemShut {NoStop}%
\bibitem [{\citenamefont {Iadecola}\ \emph {et~al.}(2016)\citenamefont
  {Iadecola}, \citenamefont {Schuster},\ and\ \citenamefont
  {Chamon}}]{iadecola2016non}%
  \BibitemOpen
  \bibfield  {author} {\bibinfo {author} {\bibfnamefont {T.}~\bibnamefont
  {Iadecola}}, \bibinfo {author} {\bibfnamefont {T.}~\bibnamefont {Schuster}},\
  and\ \bibinfo {author} {\bibfnamefont {C.}~\bibnamefont {Chamon}},\
  }\bibfield  {title} {\bibinfo {title} {Non-abelian braiding of light},\
  }\href@noop {} {\bibfield  {journal} {\bibinfo  {journal} {Phys. Rev. Lett.}\
  }\textbf {\bibinfo {volume} {117}},\ \bibinfo {pages} {073901} (\bibinfo
  {year} {2016})}\BibitemShut {NoStop}%
\bibitem [{\citenamefont {Barlas}\ and\ \citenamefont
  {Prodan}(2020)}]{barlas2020topological}%
  \BibitemOpen
  \bibfield  {author} {\bibinfo {author} {\bibfnamefont {Y.}~\bibnamefont
  {Barlas}}\ and\ \bibinfo {author} {\bibfnamefont {E.}~\bibnamefont
  {Prodan}},\ }\bibfield  {title} {\bibinfo {title} {Topological braiding of
  non-abelian midgap defects in classical metamaterials},\ }\href@noop {}
  {\bibfield  {journal} {\bibinfo  {journal} {Phys. Rev. Lett.}\ }\textbf
  {\bibinfo {volume} {124}},\ \bibinfo {pages} {146801} (\bibinfo {year}
  {2020})}\BibitemShut {NoStop}%
\bibitem [{\citenamefont {Nayak}\ \emph {et~al.}(2008)\citenamefont {Nayak},
  \citenamefont {Simon}, \citenamefont {Stern}, \citenamefont {Freedman},\ and\
  \citenamefont {Sarma}}]{nayak2008non}%
  \BibitemOpen
  \bibfield  {author} {\bibinfo {author} {\bibfnamefont {C.}~\bibnamefont
  {Nayak}}, \bibinfo {author} {\bibfnamefont {S.~H.}\ \bibnamefont {Simon}},
  \bibinfo {author} {\bibfnamefont {A.}~\bibnamefont {Stern}}, \bibinfo
  {author} {\bibfnamefont {M.}~\bibnamefont {Freedman}},\ and\ \bibinfo
  {author} {\bibfnamefont {S.~D.}\ \bibnamefont {Sarma}},\ }\bibfield  {title}
  {\bibinfo {title} {Non-abelian anyons and topological quantum computation},\
  }\href@noop {} {\bibfield  {journal} {\bibinfo  {journal} {Rev. Mod. Phys.}\
  }\textbf {\bibinfo {volume} {80}},\ \bibinfo {pages} {1083} (\bibinfo {year}
  {2008})}\BibitemShut {NoStop}%
\bibitem [{\citenamefont {Wang}\ \emph
  {et~al.}(2021{\natexlab{b}})\citenamefont {Wang}, \citenamefont {Dutt},
  \citenamefont {Yang}, \citenamefont {Wojcik}, \citenamefont
  {Vu{\v{c}}kovi{\'c}},\ and\ \citenamefont {Fan}}]{wang2021generating}%
  \BibitemOpen
  \bibfield  {author} {\bibinfo {author} {\bibfnamefont {K.}~\bibnamefont
  {Wang}}, \bibinfo {author} {\bibfnamefont {A.}~\bibnamefont {Dutt}}, \bibinfo
  {author} {\bibfnamefont {K.~Y.}\ \bibnamefont {Yang}}, \bibinfo {author}
  {\bibfnamefont {C.~C.}\ \bibnamefont {Wojcik}}, \bibinfo {author}
  {\bibfnamefont {J.}~\bibnamefont {Vu{\v{c}}kovi{\'c}}},\ and\ \bibinfo
  {author} {\bibfnamefont {S.}~\bibnamefont {Fan}},\ }\bibfield  {title}
  {\bibinfo {title} {Generating arbitrary topological windings of a
  non-hermitian band},\ }\href@noop {} {\bibfield  {journal} {\bibinfo
  {journal} {Science}\ }\textbf {\bibinfo {volume} {371}},\ \bibinfo {pages}
  {1240} (\bibinfo {year} {2021}{\natexlab{b}})}\BibitemShut {NoStop}%
\bibitem [{\citenamefont {Patil}\ \emph {et~al.}(2022)\citenamefont {Patil},
  \citenamefont {H{\"o}ller}, \citenamefont {Henry}, \citenamefont {Guria},
  \citenamefont {Zhang}, \citenamefont {Jiang}, \citenamefont {Kralj},
  \citenamefont {Read},\ and\ \citenamefont {Harris}}]{patil2022measuring}%
  \BibitemOpen
  \bibfield  {author} {\bibinfo {author} {\bibfnamefont {Y.~S.}\ \bibnamefont
  {Patil}}, \bibinfo {author} {\bibfnamefont {J.}~\bibnamefont {H{\"o}ller}},
  \bibinfo {author} {\bibfnamefont {P.~A.}\ \bibnamefont {Henry}}, \bibinfo
  {author} {\bibfnamefont {C.}~\bibnamefont {Guria}}, \bibinfo {author}
  {\bibfnamefont {Y.}~\bibnamefont {Zhang}}, \bibinfo {author} {\bibfnamefont
  {L.}~\bibnamefont {Jiang}}, \bibinfo {author} {\bibfnamefont
  {N.}~\bibnamefont {Kralj}}, \bibinfo {author} {\bibfnamefont
  {N.}~\bibnamefont {Read}},\ and\ \bibinfo {author} {\bibfnamefont {J.~G.}\
  \bibnamefont {Harris}},\ }\bibfield  {title} {\bibinfo {title} {Measuring the
  knot of non-hermitian degeneracies and non-commuting braids},\ }\href@noop {}
  {\bibfield  {journal} {\bibinfo  {journal} {Nature}\ }\textbf {\bibinfo
  {volume} {607}},\ \bibinfo {pages} {271} (\bibinfo {year}
  {2022})}\BibitemShut {NoStop}%
\bibitem [{\citenamefont {Zhang}\ \emph
  {et~al.}(2021{\natexlab{b}})\citenamefont {Zhang}, \citenamefont {Yang},
  \citenamefont {Ge}, \citenamefont {Guan}, \citenamefont {Chen}, \citenamefont
  {Yan}, \citenamefont {Chen}, \citenamefont {Xi}, \citenamefont {Li},
  \citenamefont {Jia} \emph {et~al.}}]{zhang2021acoustic}%
  \BibitemOpen
  \bibfield  {author} {\bibinfo {author} {\bibfnamefont {L.}~\bibnamefont
  {Zhang}}, \bibinfo {author} {\bibfnamefont {Y.}~\bibnamefont {Yang}},
  \bibinfo {author} {\bibfnamefont {Y.}~\bibnamefont {Ge}}, \bibinfo {author}
  {\bibfnamefont {Y.-J.}\ \bibnamefont {Guan}}, \bibinfo {author}
  {\bibfnamefont {Q.}~\bibnamefont {Chen}}, \bibinfo {author} {\bibfnamefont
  {Q.}~\bibnamefont {Yan}}, \bibinfo {author} {\bibfnamefont {F.}~\bibnamefont
  {Chen}}, \bibinfo {author} {\bibfnamefont {R.}~\bibnamefont {Xi}}, \bibinfo
  {author} {\bibfnamefont {Y.}~\bibnamefont {Li}}, \bibinfo {author}
  {\bibfnamefont {D.}~\bibnamefont {Jia}}, \emph {et~al.},\ }\bibfield  {title}
  {\bibinfo {title} {Acoustic non-hermitian skin effect from twisted winding
  topology},\ }\href@noop {} {\bibfield  {journal} {\bibinfo  {journal} {Nat.
  Commun.}\ }\textbf {\bibinfo {volume} {12}},\ \bibinfo {pages} {6297}
  (\bibinfo {year} {2021}{\natexlab{b}})}\BibitemShut {NoStop}%
\bibitem [{\citenamefont {Zhang}\ and\ \citenamefont
  {Song}(2019)}]{zhang2019partial}%
  \BibitemOpen
  \bibfield  {author} {\bibinfo {author} {\bibfnamefont {X.}~\bibnamefont
  {Zhang}}\ and\ \bibinfo {author} {\bibfnamefont {Z.}~\bibnamefont {Song}},\
  }\bibfield  {title} {\bibinfo {title} {Partial topological zak phase and
  dynamical confinement in a non-hermitian bipartite system},\ }\href@noop {}
  {\bibfield  {journal} {\bibinfo  {journal} {Phys. Rev. A}\ }\textbf {\bibinfo
  {volume} {99}},\ \bibinfo {pages} {012113} (\bibinfo {year}
  {2019})}\BibitemShut {NoStop}%
\bibitem [{\citenamefont {Rehren}\ and\ \citenamefont
  {Schroer}(1989)}]{rehren1989einstein}%
  \BibitemOpen
  \bibfield  {author} {\bibinfo {author} {\bibfnamefont {K.-H.}\ \bibnamefont
  {Rehren}}\ and\ \bibinfo {author} {\bibfnamefont {B.}~\bibnamefont
  {Schroer}},\ }\bibfield  {title} {\bibinfo {title} {Einstein causality and
  artin braids},\ }\href@noop {} {\bibfield  {journal} {\bibinfo  {journal}
  {Nucl. Phys. B}\ }\textbf {\bibinfo {volume} {312}},\ \bibinfo {pages} {715}
  (\bibinfo {year} {1989})}\BibitemShut {NoStop}%
\bibitem [{\citenamefont {Okuma}\ \emph
  {et~al.}(2020{\natexlab{b}})\citenamefont {Okuma}, \citenamefont {Kawabata},
  \citenamefont {Shiozaki},\ and\ \citenamefont {Sato}}]{PRL124_086801}%
  \BibitemOpen
  \bibfield  {author} {\bibinfo {author} {\bibfnamefont {N.}~\bibnamefont
  {Okuma}}, \bibinfo {author} {\bibfnamefont {K.}~\bibnamefont {Kawabata}},
  \bibinfo {author} {\bibfnamefont {K.}~\bibnamefont {Shiozaki}},\ and\
  \bibinfo {author} {\bibfnamefont {M.}~\bibnamefont {Sato}},\ }\bibfield
  {title} {\bibinfo {title} {Topological origin of non-hermitian skin
  effects},\ }\href {https://doi.org/10.1103/PhysRevLett.124.086801} {\bibfield
   {journal} {\bibinfo  {journal} {Phys. Rev. Lett.}\ }\textbf {\bibinfo
  {volume} {124}},\ \bibinfo {pages} {086801} (\bibinfo {year}
  {2020}{\natexlab{b}})}\BibitemShut {NoStop}%
\bibitem [{\citenamefont {Gong}\ \emph
  {et~al.}(2018{\natexlab{b}})\citenamefont {Gong}, \citenamefont {Ashida},
  \citenamefont {Kawabata}, \citenamefont {Takasan}, \citenamefont
  {Higashikawa},\ and\ \citenamefont {Ueda}}]{PRX8_031079}%
  \BibitemOpen
  \bibfield  {author} {\bibinfo {author} {\bibfnamefont {Z.}~\bibnamefont
  {Gong}}, \bibinfo {author} {\bibfnamefont {Y.}~\bibnamefont {Ashida}},
  \bibinfo {author} {\bibfnamefont {K.}~\bibnamefont {Kawabata}}, \bibinfo
  {author} {\bibfnamefont {K.}~\bibnamefont {Takasan}}, \bibinfo {author}
  {\bibfnamefont {S.}~\bibnamefont {Higashikawa}},\ and\ \bibinfo {author}
  {\bibfnamefont {M.}~\bibnamefont {Ueda}},\ }\bibfield  {title} {\bibinfo
  {title} {Topological phases of non-hermitian systems},\ }\href
  {https://doi.org/10.1103/PhysRevX.8.031079} {\bibfield  {journal} {\bibinfo
  {journal} {Phys. Rev. X}\ }\textbf {\bibinfo {volume} {8}},\ \bibinfo {pages}
  {031079} (\bibinfo {year} {2018}{\natexlab{b}})}\BibitemShut {NoStop}%
\bibitem [{\citenamefont {Wang}\ \emph {et~al.}(2022)\citenamefont {Wang},
  \citenamefont {Dutt}, \citenamefont {Wojcik},\ and\ \citenamefont
  {Fan}}]{wang2022realization}%
  \BibitemOpen
  \bibfield  {author} {\bibinfo {author} {\bibfnamefont {K.}~\bibnamefont
  {Wang}}, \bibinfo {author} {\bibfnamefont {A.}~\bibnamefont {Dutt}}, \bibinfo
  {author} {\bibfnamefont {C.~C.}\ \bibnamefont {Wojcik}},\ and\ \bibinfo
  {author} {\bibfnamefont {S.}~\bibnamefont {Fan}},\ }\bibfield  {title}
  {\bibinfo {title} {Realization of topological complex-energy braids and knots
  of non-hermitian bands},\ }in\ \href@noop {} {\emph {\bibinfo {booktitle}
  {CLEO: QELS\_Fundamental Science}}}\ (\bibinfo {organization} {Optica
  Publishing Group},\ \bibinfo {year} {2022})\ pp.\ \bibinfo {pages}
  {JW4A--1}\BibitemShut {NoStop}%
\end{thebibliography}
%

\appendix

\section{\label{sec:Appx_Laplacian} Circuit Laplacian Formation}

In this section, we undertake a detailed derivation of the circuit Laplacian specifically for the case where $m=2$ and $n=1$, considering a circuit with 10 unit cells, corresponding to 20 nodes. Our objective is to apply Kirchhoff's Current Law (KCL) as expressed in Eq. \ref{eqn:KCL} to every node within the circuit illustrated in Fig. \ref{fig_circuitsetup}.

\begin{table}[H]
	\begin{tabular}{cc}
		Node 1: & $i\omega\left(C_{0}+C_{n}-\frac{1}{\omega^{2}L_{a}} +\frac{1}{i\omega R_{0}}\right)V_{1}-i\omega C_{0}V_{2}-i\omega C_{m}V_{18}=I_{1}$\tabularnewline
		Node 2: & $-i\omega C_{0}V_{1}+i\omega\left(C_{0}+C_{m}-\frac{1}{\omega^{2}L_{b}} +\frac{1}{i\omega R_{0}}\right)V_{2}-i\omega C_{n}V_{3}=I_{2}$\tabularnewline
		Node 3: & $i\omega\left(C_{0}+C_{n}-\frac{1}{\omega^{2}L_{a}} +\frac{1}{i\omega R_{0}}\right)V_{3}-i\omega C_{0}V_{4}-i\omega C_{m}V_{20}=I_{3}$\tabularnewline
		Node 4: & $-i\omega C_{0}V_{3}+i\omega\left(C_{0}+C_{m}-\frac{1}{\omega^{2}L_{b}} +\frac{1}{i\omega R_{0}}\right)V_{4}-i\omega C_{n}V_{5}=I_{4}$\tabularnewline
		$\vdots$ & $\vdots$\tabularnewline
		Node 19: & $-i\omega C_{n}V_{16}+i\omega\left(C_{0}+C_{n}-\frac{1}{\omega^{2}L_{a}}+\frac{1}{i\omega R_{0}}\right)V_{19}-i\omega C_{0}V_{20}=I_{19}$\tabularnewline
		Node 20: & $-i\omega C_{n}V_{1}-i\omega C_{0}V_{19}+i\omega\left(C_{0}+C_{m}-\frac{1}{\omega^{2}L_{b}}+\frac{1}{i\omega R_{0}}\right)V_{20}=I_{20}$\tabularnewline
	\end{tabular}
\end{table}

In matrix format, we can write
\begin{widetext}
	\begin{equation}
		i\omega\begin{pmatrix} d_a & -C_{0} & 0 & 0 & 0 & \cdots & {\color{red}-C_{m}} & 0 & 0\\
			-C_{0} & d_b & -C_{n} & 0 & 0 & \cdots & 0 & 0 & 0\\
			0 & 0 & d_a & -C_{0} & 0 & \cdots & 0 & 0 & {\color{red}-C_{m}}\\
			0 & 0 & -C_{0} & d_b & -C_{n} & \cdots & 0 & 0 & 0\\
			\cdots & \cdots & \cdots & \cdots & \cdots & \cdots & \ddots & \cdots & \cdots\\
			0 & 0 & 0 & 0 & -C_{n} & \cdots & 0 & d_a & -C_{0}\\
			{\color{red}-C_{n}} & 0 & 0 & 0 & 0 & \cdots & 0 & -C_{0} & d_b
		\end{pmatrix}\begin{pmatrix}V_{1}\\
			V_{2}\\
			V_{3}\\
			V_{4}\\
			V_{5}\\
			\vdots\\
			V_{16}\\
			V_{17}\\
			V_{18}\\
			V_{19}\\
			V_{20}
		\end{pmatrix}=\begin{pmatrix}I_{1}\\
			I_{2}\\
			I_{3}\\
			I_{4}\\
			\vdots\\
			I_{16}\\
			I_{17}\\
			I_{18}\\
			I_{19}\\
			I_{20}
		\end{pmatrix}\label{eq:Laplacian}
	\end{equation}
where $d_a=C_{0}+C_{n}-\frac{1}{\omega^{2}L_{a}} +\frac{1}{i\omega R_{0}}$ and $d_b=C_{0}+C_{m}-\frac{1}{\omega^{2}L_{b}}+\frac{1}{i\omega R_{0}}$ are the diagonal admittances. 
	The real-space representation of the model in Eq. \ref{H1} mirrors \ref{eq:Laplacian}, with the exception of the diagonal terms. By exciting the circuit at the resonant frequency $\omega_{r}=\frac{1}{\sqrt{L_{a}(C_{AB,0}+C_{AB,n})}}=\frac{1}{\sqrt{L_{b}(C_{AB,0}+C_{AB,-m})}}$, we can effectively eliminate these diagonal terms.

To realize the OBC system, the capacitors coupling the two ends must be removed, meaning the terms highlighted in red in Eq. \ref{eq:Laplacian} will be replaced by zero. Subsequently, when the Laplacian matrix is multiplied by the complex frequency $i\omega$, the real spectrum of the Laplacian corresponds to the imaginary spectrum of Eq. \ref{H1}, and vice versa.
\end{widetext}

\section{\label{sec:Appx_INIC}Unidirectional Capacitor}

\begin{figure}[htp]
	\centering
	\includegraphics[width=0.45\textwidth]{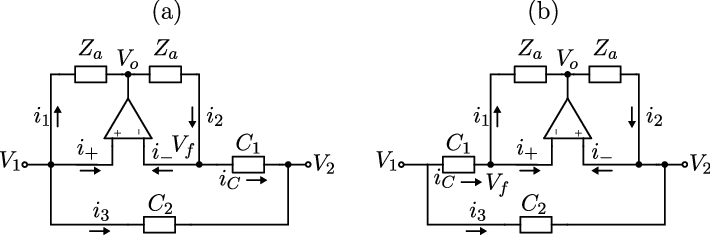} \caption{\label{fig:Unidirectional-Capacitor-using}Implementation of asymmetric capacitive coupling between two nodes using an INIC when the coupling direction is reversed. (a) Circuit setup for achieving unequal capacitance of $C_1+C_2$ for coupling from node 1 to 2, while the capacitance from node 2 to 1 is $C_2-C_1$. For $C_1=C_2=C/2$, unidirectional coupling capacitance of $C$ from 1 to 2 is achieved, while the capacitance from node 2 to 1 is approximately zero ($C_2-C_1 \approx 0$). (b) The same setup with the sign of $C_1$ reversed by reversing the polarity of the applied voltage to the OPamps. Hence, for realizing unidirectional coupling by setting $C_1=C_2=C$, coupling from node 1 to node 2 experiences almost zero capacitance ($C_2-C_1 \approx 0$), while coupling from node 2 to node 1 involves a coupling of $C$.}
\end{figure}

Achieving unidirectional coupling in an electrical circuit involves the utilization of a unidirectional capacitor, which can be effectively implemented through a negative impedance converter with current inversion (INIC). The circuit configuration is depicted in Fig. \ref{fig:Unidirectional-Capacitor-using}. Employing this INIC setup enables the realization of a unidirectional capacitor.

To establish a unidirectional capacitance, denoted as $C$, between node 1 and node 2, it is imperative to ensure that the capacitance from node 1 to 2 is $C$, while the capacitance from node 2 to 1 is effectively zero. The INIC setup, as illustrated in Fig. \ref{fig:Unidirectional-Capacitor-using}, facilitates the creation of this unidirectional capacitance by setting $C_1=C_2=C/2$ in Fig. \ref{fig:Unidirectional-Capacitor-using}a. For the purpose of stabilizing the Op-Amp output, a capacitor $C_{a}$ is added in parallel with a resistor $R_{a}$. Together, these components form an impedance denoted as $Z_{a}$. The unidirectionality of capacitive reactance, denoted as $Z_{c}$, is achieved through this setup.

\subsection{Zero-Capacitance} 

In this subsection, we delve into the intricacies of achieving nearly unidirectional capacitive coupling by approaching zero capacitance in one direction. In Fig. \ref{fig:Unidirectional-Capacitor-using}(a), we make the assumption of the Op-Amp's linear operation, and due to its very high open-loop gain, we can consider the voltages at its inverting and non-inverting terminals to be equal, i.e., $V_{+}=V_{-}$. This implies $V_{1}=V_{f}$. Additionally, owing to the Op-Amp's extremely high input impedance, the input currents at both inverting and non-inverting terminals are approximately zero, i.e., $i_{+}=i_{-}\approx0$.

Now, the current from node 1 to the output terminal of the Op-Amp is given by:

\begin{equation}
	i_{1}=\frac{V_{1}-V_{o}}{Z_{a}}\label{eq:i1}
\end{equation}

Similarly, the current from the output of the Op-Amp to the inverting terminal will be $i_{2}=\frac{V_{o}-V_{f}}{Z_{a}}=\frac{V_{o}-V_{1}}{Z_{a}}$ as $V_{1}=V_{f}$. Since the current going to the Op-Amp through the inverting terminal, $i_{-}$, is zero, the same $i_{2}$ will flow through $C_{1}$, i.e., $i_{2}=\frac{V_{1}-V_{2}}{Z_{C}}$.

\begin{eqnarray}
	\frac{V_{o}-V_{1}}{Z_{a}} & = & \frac{V_{1}-V_{2}}{Z_{C_{1}}}\nonumber \\
	V_{o} & = & V_{1}+\frac{Z_{a}}{Z_{C_{1}}}(V_{1}-V_{2})\label{eq:Vo}
\end{eqnarray}

where $Z_{C_{1}}=\frac{1}{j\omega C_{1}}$ is the capacitive reactance of capacitor $C_{1}$ at frequency $\omega$.

From \eqref{eq:i1}:
\begin{eqnarray*}
	i_{1} & = & \frac{V_{1}-V_{o}}{Z_{a}}\\
	Z_{a}i_{1} & = & V_{1}-V_{1}-\frac{Z_{a}}{Z_{C_{1}}}(V_{1}-V_{2})\\
	\frac{V_{1}-V_{2}}{i_{1}} & = & -Z_{C_{1}}
\end{eqnarray*}

This results in an equivalent negative capacitor. If another capacitor $C_{2}$ with impedance $Z_{C_{2}}$ is added in parallel, the total equivalent capacitance will be approximately $C_{2}-C_{1}$. Choosing $C_{1}=C_{2}$ leads to an equivalent capacitance of approximately zero.

\subsection{Non-zero Capacitance}

In the scenario where we reverse the polarity of the voltage applied to the same circuit, as illustrated in Fig. \eqref{fig:Unidirectional-Capacitor-using}(b), positive capacitance can be achieved. Utilizing the same analysis as before, we consider $i_{+}=i_{-}=0$ and $V_{f}=V_{2}$. Consequently, the current flowing through $C_{1}$ is given by:

\begin{eqnarray*}
	i_{C} & = & \frac{V_{1}-V_{2}}{Z_{C_{1}}}\\
	\frac{V_{1}-V_{2}}{i_{C}} & = & Z_{C_{1}}
\end{eqnarray*}

In this case, we obtain a positive capacitive reactance $Z_{C_{1}}$. Introducing a parallel capacitor $C_{2}$ results in a total capacitance of $C_{1}+C_{2}$.

\subsection{Testing Unidirectional Capacitance}

\begin{figure}[htp]
\centering
\includegraphics[width=0.45\textwidth]{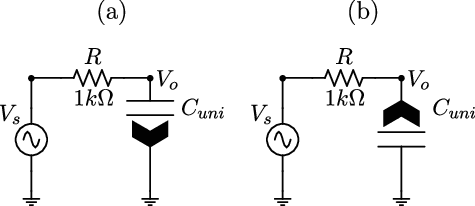} 
\caption{ \label{fig:INIC_test} Unidirectional Capacitor testing setup (a) zero-capacitance test (b) non-zero capacitance test}
\end{figure}

To test the unidirectional capacitance, we setup a typical $RC$ low-pass filter and used our unidirectional capacitor as a test capacitance. In Fig. \ref{fig:INIC_test}(a) we tested the zero capacitance and by reversing the polarity of the capacitor in Fig. \ref{fig:INIC_test}(b) we test non-zero capacitance.

\begin{figure}[h]
\centering
\includegraphics[width=0.45\textwidth]{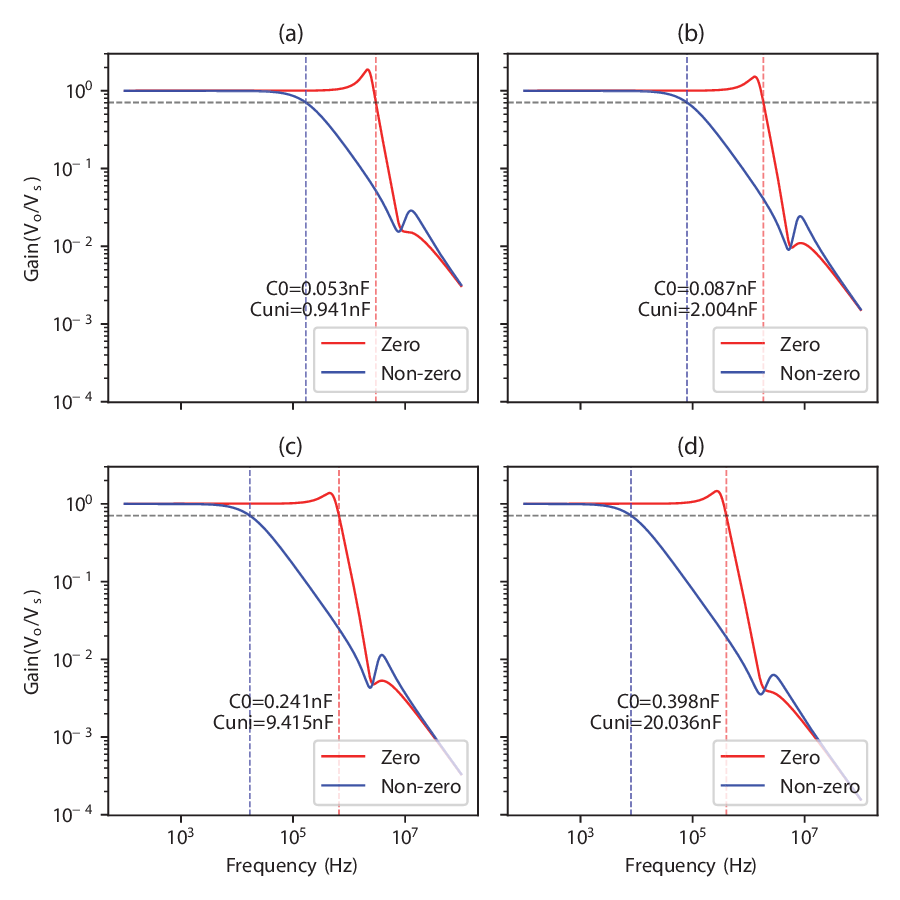}
\caption{\label{fig:INIC_test_sim}Low pass filter response using unidirectional capacitor (a) $0.47nF$ (b) $1nF$ (c) $4.7nF$ (d) $10nF$}
\end{figure}

Fig. \ref{fig:INIC_test_sim} shows LTSpice simulation results of
testing our unidirectional capacitors in a low pass filter configuration. From the voltage gain($=\frac{V_{o}}{V_{s}}$), we calculated the 3dB frequency of a low pass $RC$ filter as $f_{3dB}=\frac{1}{2\pi RC}$. We used $1k\Omega$ as a test resistor. For \textit{zero} (blue line) and \textit{non-zero}(red line) configurations, we found non-zero capacitances are 17, 23, 39 and 50 times higher than the zero capacitances. 

\section{Constraints on Knot Types Achievable in Our Setup}
In our model, not all types of knot configurations can be realized; only knots with an unknotting number of one can be achieved. The unknotting number, a concept from knot theory, measures the minimum number of crossing changes needed to transform a given knot into the simplest possible knot, known as the unknot (a simple loop). In the context of our model with two bands in a 1D setup containing long-range coupling, it is indeed true that only twist knots with an unknotting number of one can be realized. For example, in the case of a Hopf link or a trefoil knot configuration, only a single crossing needs to be changed to transform the knot configuration into an unknot configuration.
Furthermore, it is true that knots with larger unknotting numbers (such as the Cinquefoil knot with an unknotting number of two) are not possible in our current two-band model, as each band must have a unique complex value at each $k$-point.

\section{Robustness of the knot configurations against disorder}
\begin{figure}[htp]
\centering
\includegraphics[width=0.48\textwidth]{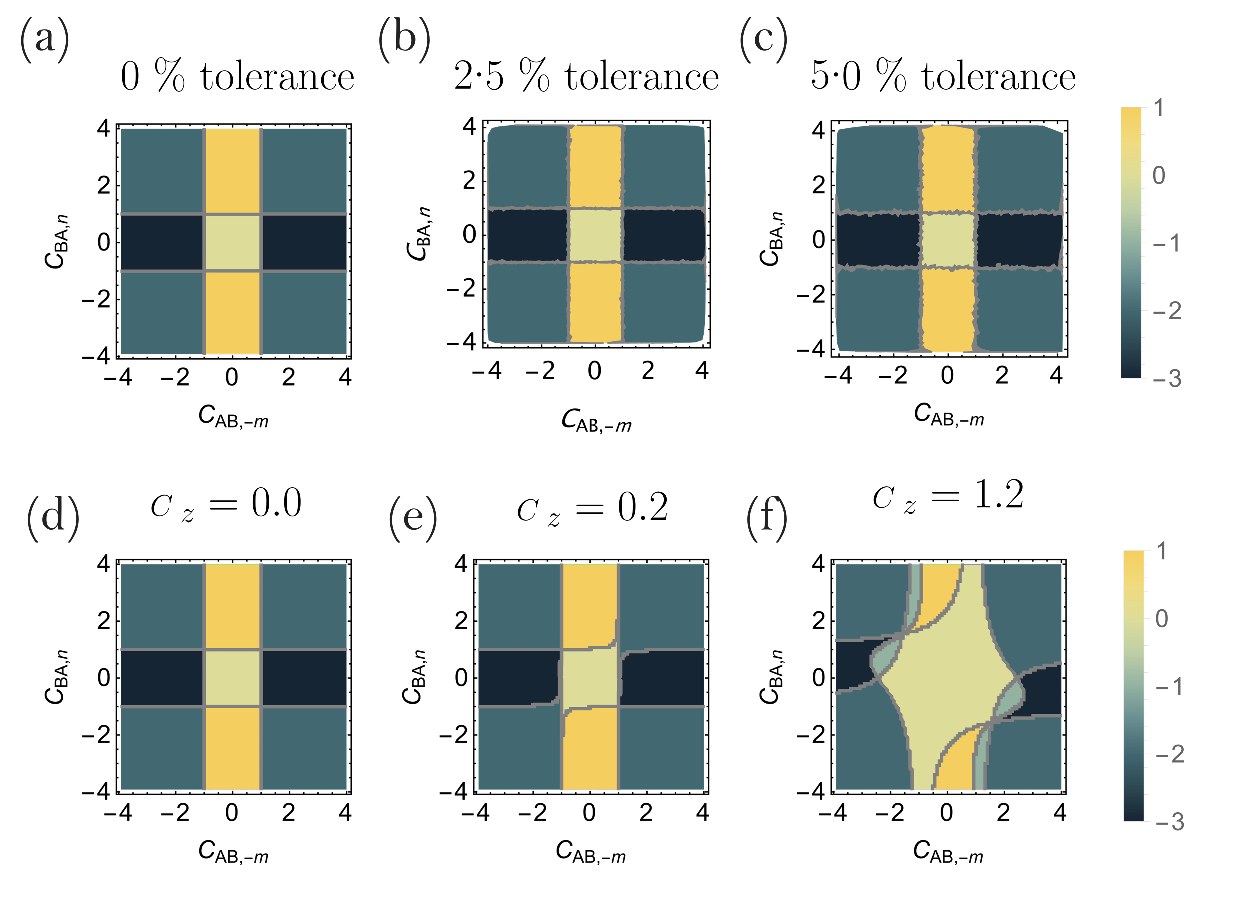} 
\caption{ \label{figr2} Braiding index phase diagram for the system described by Eq. \ref{H1} of the main text with $m = 3$ and$ n = 1$, showing the effects of various tolerances in the coupling parameters: (a) 0\% tolerance, (b) 2.5\% tolerance, and (c) 5.0\% tolerance. (d-f) Braiding index phase diagram in the presence of a mass-like term proportional to $\sigma_z$ at different $C_z$ values, (d) $C_z=0$,\ (e) $C_z=0.2$, and (f) $C_z=1.2$. Common parameter used: $C_{AB,0}=1$.}
\end{figure}

Our system demonstrates robustness against small parasitic effects. To verify this, we varied the components within a 5\% tolerance of Eq. \ref{H1} and plotted the topological knot phase diagram in Fig.\ref{figr2}a-c, which showed a profile almost identical to the original one. Thus, the knot configuration is quite protected against small parasitic variations.
However, the situation changes when a $\sigma_z $ term is included in the Hamiltonian. If we rewrite the Hamiltonian in Eq. \ref{H1} with a mass term as
\begin{equation}
  H_2 \left( k \right) = H_1 \left( k \right) + C_z \sigma_z;
  \label{ap4}
\end{equation}

where, $\sigma_z$ is the third Pauli matrix and $C_z$ corresponds to the mass-like term in the Hamiltonian, this term breaks the chiral symmetry of the Hamiltonian. In the presence of a small mass term (i.e., $C_z \ll C_i$), the braiding index phase diagram does not change significantly (see  Fig.\ref{figr2}d). Therefore, the complex braiding/knot configuration remains protected.
However, with a larger mass-like term (i.e., $C_z\approx C_i$),   the braiding index phase diagram changes a lot and the region with zero knot index (unknotting phase) expands in the parameter plane, as shown in Fig.\ref{figr2}f, while retaining other knot configurations with non-zero knot index in smaller range of parameters. As a result, knot structures convert to an unknot configuration across the wider parameter regime. Thus, knot structures are still protected in the presence of a large chiral symmetry-breaking ${C_z\sigma}_z$ term but at the different coupling parameters range. 

\section{Stability of the Proposed Electric Circuit Network}

In this section, we discuss some important points regarding the stability of the proposed electric circuit network. Indeed, the use of active elements such as operational amplifiers (OpAmps) in a negative impedance converter configuration to achieve unidirectional couplings can introduce stability concerns.

To address this, we conducted additional analyses beyond the steady-state AC simulations originally performed with LTSpice. Specifically, we examined the time dynamics of the system to assess the stability of the circuit. Our findings indicate that the eigenvalues of the Laplacian with a negative imaginary part can indeed lead to exponentially increasing time evolution of the corresponding eigenstates, potentially causing instability.
For systems with periodic boundary conditions, this issue is particularly critical as even slightly unstable modes can diverge due to the loop feedback of the boundary conditions. To mitigate this risk, careful design of the OpAmp circuit is essential, including the implementation of feedback mechanisms to ensure stability.

Additionally, we have outlined the steps taken to ensure stability in the possible experimental setup, including the design parameters and feedback mechanisms used to counteract potential instabilities. To prevent the system from going unstable due to non-causal instability, we added an additional resistor $R_0$ to each node to shift the entire admittance spectrum upward such that no complex eigenvalues have negative imaginary parts. This adjustment does not alter the eigenstate profile, allowing us to still observe the same braiding patterns. With these changes in the circuit design, time domain analysis of the system converges, and the system becomes stable.

Furthermore, to achieve almost real-like outputs from circuit simulations, we added a parasitic resistance of $ 100\ m\Omega$ to each inductor. While this inclusion of parasitic resistance results in slight mismatches from the numerical results, as shown in Fig. 8 of the main manuscript, the circuit now provides almost real outputs that are stable.


\end{document}